\documentclass[graybox]{svmult}

\usepackage{mathptmx}       
\usepackage{helvet}         
\usepackage{courier}        
\usepackage{type1cm}        
\usepackage{makeidx}         
\usepackage{graphicx}        
\usepackage{multicol}        
\usepackage[bottom]{footmisc}

\usepackage{bm}
\usepackage{color}
\usepackage{amsmath}

\makeindex             

\begin{document}

\title*{Many-Body Perturbation Theory (MBPT) and Time-Dependent Density-Functional Theory (TD-DFT): MBPT Insights About What is Missing in, and Corrections to, the TD-DFT Adiabatic Approximation}
\titlerunning{MBPT and TD-DFT}

\author{Mark E. Casida and Miquel Huix-Rotllant}
\institute{Mark E. Casida \at D\'epartement de Chimie Mol\'eculaire, 
Institut de Chimie Mol\'eculaire de Grenoble, Universit'e Joseph Fourier (Grenoble I),
301 rue de la Chimie, BP 53, 38041 Grenoble Cedex 9, France
\email{mark.casida@ujf-grenoble.fr}
\and Miquel Huix-Rotllant \at Institut f\"ur Physikalische und Theoretische Chimie,
Universit\"at Frankfurt am Main, Germany \email{miquel.huix@gmail.com}}
%
%
\maketitle


\abstract*{Each chapter should be preceded by an abstract (10--15 lines long) that summarizes the content. The abstract will appear \textit{online} at \url{www.SpringerLink.com} and be available with unrestricted access. This allows unregistered users to read the abstract as a teaser for the complete chapter. As a general rule the abstracts will not appear in the printed version of your book unless it is the style of your particular book or that of the series to which your book belongs.
Please use the 'starred' version of the new Springer \texttt{abstract} command for typesetting the text of the online abstracts (cf. source file of this chapter template \texttt{abstract}) and include them with the source files of your manuscript. Use the plain \texttt{abstract} command if the abstract is also to appear in the printed version of the book.}

\abstract{ 
In their famous paper Kohn and Sham formulated a formally exact 
density-functional theory (DFT) for the ground-state energy and density
of a system of $N$ interacting electrons, albeit limited at the time by 
certain troubling representability questions.  As no practical exact form
of the exchange-correlation (xc) energy functional was known, the xc-functional
had to be approximated, ideally by a local or semilocal functional.  Nowadays
however the realization that Nature is not always so nearsighted has driven us
up Perdew's Jacob's ladder to find increasingly nonlocal density/wavefunction
hybrid functionals.  Time-dependent (TD-) DFT is a younger development
which allows DFT concepts to be used to describe the temporal evolution of 
the density in the presence of a perturbing field.  Linear response (LR)
theory then allows spectra and other information about excited states to be
extracted from TD-DFT.  Once again the exact TD-DFT xc-functional must be
approximated in practical calculations and this has historically been done
using the TD-DFT adiabatic approximation (AA) which is to TD-DFT very much like
what the local density approximation (LDA) is to conventional ground-state
DFT.  While some of the recent advances in TD-DFT focus on what can be
done within the AA, others explore ways around the AA.  After giving
an overview of DFT, TD-DFT, and LR-TD-DFT, this article will focus on
many-body corrections to LR-TD-DFT as one way to building hybrid 
density-functional/wavefunction methodology for incorporating aspects
of nonlocality in time not present in the AA.
}  

\section{Introduction}
\label{sec:intro}

\begin{quote}
\noindent
``I have not included chemistry in my list [of the physical sciences] because,
  though Dynamical Science is continually reclaiming large tracts of good ground from
  one side of Chemistry, Chemistry is extending with still greater rapidity on the
  other side, into region where the dynamics of the present day must put her hand on her
  mouth.  But Chemistry is a Physical Science...''\\ --- James Clerk Maxwell,
  {\em Encyclopaedia Britannica}, ca.\ 1873 \cite{R09}
\end{quote}

Much has changed since when Maxwell first defended chemistry as a physical science.
The physics applied to chemical systems now involves as much, if not more, quantum
mechanics than classical dynamics.  However some things have not changed.  Chemistry
still seems to extend too rapidly for first principles modeling to keep up.
Fortunately density-functional theory (DFT) has established itself as a 
computationally-simple way to extend {\em ab initio}\footnote{The term {\em ab initio}
is used here as it is typically used in quantum chemistry.  That is, {\em ab initio}
refers to first-principles Hartree-Fock-based theory, excluding DFT.  In contrast, the
term {\em ab initio} used in the solid state physics literature usually encompasses
DFT.} accuracy to larger systems than where 
{\em ab initio} quantum chemical methods can traditionally be applied. The reticence
to use DFT for describing excited states has even given way as linear response (LR-)
time-dependent (TD-) DFT has become an established way to calculate excited-state
properties of medium- and large-size molecules.  
One of the strengths of TD-DFT is that it is formally-exact theory.  However, as in
traditional DFT, problems arise in practice because of the need to make approximations.
Of course, from the point of view
of a developer of new methods, when people are given a little then they immediately
want more. As soon as LR-TD-DFT was shown to give reasonably promising results in 
one context, then many people in the modeling community immediately wanted to apply 
LR-TD-DFT in a whole range of more challenging contexts.  
It then became urgent to explore the limits of applicability of {\em approximate} TD-DFT
and to improve approximations in order to extend these limits.  Much work has been done
on this problem and there are many success stories to tell about LR-TD-DFT.
Indeed many of the articles in this book describe some of these challenging contexts where 
conventional LR-TD-DFT approximations do work.  In this chapter, however we want to focus on the cutting
edge where LR-TD-DFT finds itself seriously challenged and yet progress is being
made.  In particular, what we have in mind are photochemical applications where
interacting excited states of fundamentally different character need to be described 
with similar accuracy and where bonds may be in the process of breaking or forming.
The approach we will take is to introduce a hybrid method where many-body perturbation
theory (MBPT) corrections are added on top of LR-TD-DFT.  We will also use the tools
we have developed to gain some insight into what needs to be included in the TD-DFT
exchange-correlation (xc) functional in order for it to better describe photochemical
problems.

Applications of LR-TD-DFT to photochemistry are no longer rare.  Perhaps the earliest 
attempt to apply LR-TD-DFT to photochemistry was the demonstration that avoided 
crossings between formaldehyde excited-state curves could indeed be described with this 
method \cite{CJCS98}.  Further hope for photochemistry from LR-TD-DFT was
raised again only a few years later \cite{C02,DM02}, with an example application to
the photochemistry of oxirane appearing in another five years time \cite{CJI+07,TTR+08}.
Ref.~\cite{CND11} provides a recent review of the present state of LR-TD-DFT applied
to photochemistry and where some of the difficulties lie.

\begin{figure}
  \caption{Typical curves for the singlet photochemical isomerization of ethylene.
  \label{fig:ethylene_singlet_photochem}
       }
      \includegraphics[angle=0,scale=0.3]{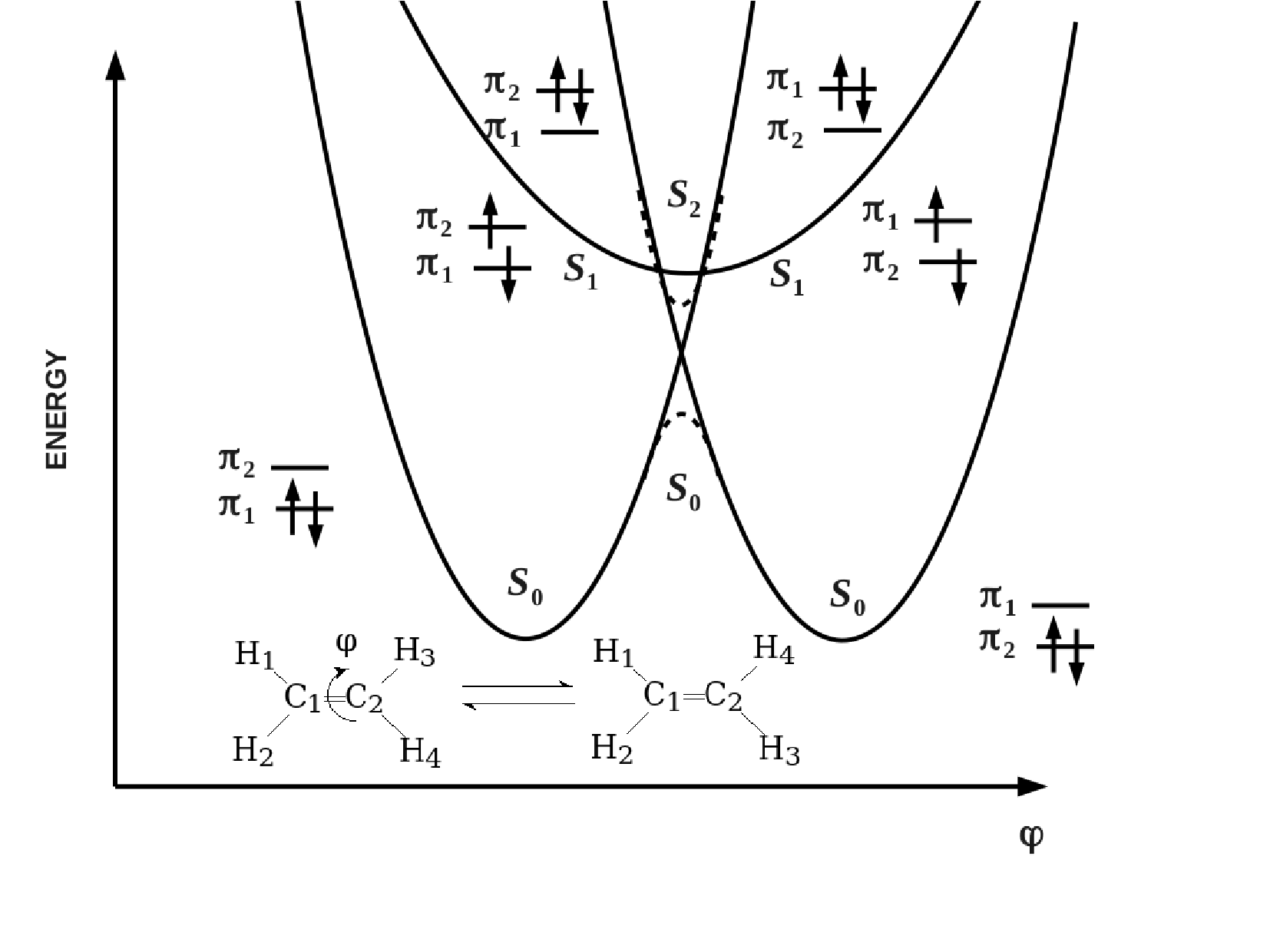}
\end{figure}
Let us try to focus on some of key problems.  Photophenomena are frequently divided
into photophysics, when the photoprocess ends with the same molecules with which it 
started, and photochemistry, when the photoprocess ends with different molecules.
This is illustrated by the cartoon in Fig.~\ref{fig:ethylene_singlet_photochem}.
An example of a typical photophysical process would be beginning at one $S_0$ minimum, 
exciting to the singly-excited $S_1$ state, and reverting to the same $S_0$ minimum. 
In contrast, an example of a typical photochemical process would be exciting from 
one $S_0$ minimum to an $S_1$ excited state, followed by moving along the $S_1$ surface, 
through avoided crossings, conical intersections, and other photochemical funnels, 
to finally end up at the other $S_0$ minimum.  State-of-the-art LR-TD-DFT does
a reasonable job modeling photophysical processes but has much more difficulty
with photochemical processes.  The main reason is easily seen in 
Fig.~\ref{fig:ethylene_singlet_photochem} --- namely that photochemical processes 
often require an explicit treatment of doubly excited states and these are beyond 
the scope of conventional LR-TD-DFT.  There are several ways to remedy this problem
which have been discussed in a previous review article \cite{CH12}.  In this article,
we will concentrate on one way to explore and correct the double excitation problem
using a hybrid MBPT/LR-TD-DFT approach.

The rest of this chapter is organized as follows.  The next section 
(Sec.~\ref{sec:review}) provides a small review of the current state of DFT,
TD-DFT, and LR-DFT.  Section~\ref{sec:MBPT} begins with an introduction to
the key notions of MBPT needed to derive corrections to approximate LR-TD-DFT
and derives some basic equations.  Section~\ref{sec:dressed} shows these corrections
can be used in practical applications through an exploration of dressed LR-TD-DFT.
Ideally it would be nice to be able to use these corrections to improve the xc
functional of TD-DFT.  However this involves an additional localization step which
is examined in Sec.~\ref{sec:local}.  Section~\ref{sec:conclude} sums up with some
perspectives.

\section{Some Review}
\label{sec:review}

This section reviews a few concepts which in some sense are hoary
with age: DFT is about 50 years old, TD-DFT is about 30 years old, 
and LR-TD-DFT (in the form of the Casida equations) is about 20 years old.  
Thus many of the basic concepts are now well known.  However this section 
is both necessary to define some notation and because some aspects of these 
subjects have continued to evolve and so need to be updated.

\subsection{Density-Functional Theory (DFT)}

Hohenberg and Kohn \cite{HK64} and Kohn and Sham \cite{KS65} defined 
DFT in the mid-1960s when they gave formal rigor to earlier work by Thomas, 
Fermi, Dirac, Slater, and others.  This initial work has been nicely reviewed
in well-known texts \cite{PY89,DG90,KH00} and so we shall not dwell on details here
but rather concentrate on what is essential in the present context.
Hartree atomic units ($\hbar = m_e = e = 1$) will be used throughout unless
otherwise specified.

Kohn and Sham introduced orthonormal auxiliary functions (Kohn-Sham orbitals) $\psi_i(1)$
and corresponding occupation numbers $n_i$ which allow the density to be expressed as,
\begin{equation}
  \rho(1) = \sum_i n_i \vert \psi_i(1) \vert^2 \, ,
  \label{eq:review.1}
\end{equation}
and the electronic energy to be expressed as,
\begin{equation}
  E = \sum_i n_i \langle \psi_i \vert \hat{t}_s + v \vert \psi_i \rangle 
    + E_H[\rho] + E_{xc}[\rho] \, .
  \label{eq:review.2}
\end{equation}
Here we use a notation where $i = (\vec{r}_i, \sigma_i)$ stands for the space 
$\vec{r}_i$ and spin $\sigma_i$ coordinates of electron $i$,
$\hat{t}_s = -(1/2) \nabla^2$ is the noninteracting kinetic energy operator,
$v$ is the external potential which represents the attraction of the electron to the
nuclei as well as any applied electric fields, 
$E_H[\rho] = \int \int \rho(1)\rho(2)/r_{12} \, d1 d2$ is the Hartree (or Coulomb) energy,
and $E_{xc}[\rho]$ is the xc-energy which includes everything not included in the other
terms (i.e., exchange, correlation, and the difference between the interacting and
noninteracting kinetic energies).  Minimizing the energy [Eq.~(\ref{eq:review.2})]
subject to the constraint of orthonormal orbitals gives the Kohn-Sham orbital equation,
\begin{equation}
  \hat{h}_s[\rho] \psi_i  =  \epsilon_i \psi_i \, ,
  \label{eq:review.3}
\end{equation}
where the Kohn-Sham Hamiltonian, $\hat{h}_s[\rho](1)$, is the sum of 
$\hat{t}_s(1)+v(1)$, the Hartree (or Coulomb) potential
$v_H[\rho](1) = \int \rho(2)/r_{12} \, d2$, and the xc-potential
$v_{xc}[\rho](1) = \delta E_{xc}[\rho]/\delta \rho(1)$.

An important but subtle point is that the Kohn-Sham equation should be solved
self-consistently with lower energy orbitals filled before higher energy orbitals
({\em Aufbau} principle) as befits a system of noninteracting electrons.  
If this can be done with integer occupancy, then the system is said to be
noninteracting $v$-representable (NVR).  Most programs try to enforce NVR,
but it now seems likely that NVR fails for many systems even in exact 
Kohn-Sham DFT.   The alternative is to consider fractional occupation within
an ensemble formalism.  An important theorem then states that only the last occupied
degenerate orbitals may be fractionally occupied (see, e.g., Ref.~\cite{DG90}, pp.\ 55-56).  
Suitable algorithms are rare
as maintaining this condition can lead to degenerate orbitals having different occupation
numbers which, in turn, may require minimizing the energy with respect to unitary 
transformations within the space spanned by the degenerate occupied orbitals with 
different occupation numbers.  These points have been previously discussed in somewhat
greater detail in Ref.~\cite{CH12}.  Most programs show at least an effective failure
of NVR when using approximate functionals, in particular around regions of strong electron
correlation such as where bonds are being made or broken (e.g., avoided crossing of the 
$S_0$ surfaces in Fig.~\ref{fig:ethylene_singlet_photochem}) which often shows
up as self-consistent field (SCF) convergence failures.

\begin{table}
\caption{Jacob’s ladder for functionals \cite{PS01}. (An updated version is given in 
Ref.~\cite{PRC+09}.) 
\label{tab:jacob}
}
\begin{tabular}{lcl}
\hline 
 & \multicolumn{1}{c}{\em Quantum Chemical Heaven} & \\
Double-hybrid & \line(1,0){100} & $\rho(1)$, $x(1)$, $\tau(1)$, $\psi_i(1)$, $\psi_a(1)^h$ \\
Hybrid & \line(1,0){100} & $\rho(1)$, $x(1)$, $\tau(1)$, $\psi_i(1)^g$ \\
mGGA$^c$ & \line(1,0){100} & $\rho(1)$, $x(1)$, $\tau(1)^e$, $\nabla^2 \rho(1)^f$ \\
GGA$^b$ & \line(1,0){100} & $\rho(1)$, $x(1)^d$ \\
LDA$^a$ & \line(1,0){100} & $\rho(1)$ \\
 & \multicolumn{1}{c}{\em Hartree World} & \\
\hline 
\end{tabular}
\\
$^a$ Local density approximation.\\
$^b$ Generalized gradient approximation.\\
$^c$ Meta generalized gradient approximation.\\
$^d$ The reduced gradient $x(1) = \vert \vec{\nabla} \rho(1) \vert / \rho^{4/3}(1)$.\\
$^e$ The local kinetic energy $\tau(1) = \sum_p n_p \psi_p(1) \nabla^2 \psi_p(1)$.\\
$^f$ There is some indication that the local kinetic energy density $\tau(1)$ and the
Laplacian of the charge density, $\nabla^2 \rho(1)$, contain comparable information
\cite{PC07}.\\
$^g$ Occupied orbitals.\\
$^h$ Unoccupied orbitals.
\end{table}

As no practical exact form of $E_{xc}$ is known, it must be approximated in practice.
In the original papers, $E_{xc}$ should depend only upon the charge density.  However
our notation already reflects the modern tendency to allow a spin-dependence in $E_{xc}$ 
(spin-DFT).  This additional degree of freedom makes it easier to develop improved
density-functional approximations (DFAs).  In recent years, this tendency to add
additional functional dependencies into $E_{xc}$ has lead to generalized Kohn-Sham theories
corresponding to different levels of what Perdew has refered to Jacob's ladder 
\footnote{
``Jacob set out from Beersheba and went on his way towards Harran.  He came to a
certain place and stopped there for the night, because the sun had set; and, taking
one of the stones there, he made it a pillow for his head and lay down to sleep.
He dreamt that he saw a ladder, which rested on the ground with its top reaching
to heaven, and angles of God were going up and down it.''
--- The Bible, Genesis 28:10-13}
for 
functionals (Table~\ref{tab:jacob}).  The LDA and GGA are pure DFAs.  Higher levels
are no longer fall within the pure DFT formalism \cite{G01} and in particular are subject 
to a different interpretation of orbital energies.

Of particular importance to us is the hybrid level which incorporates some amount
of Hartree-Fock exchange.  Inspired by the adiabatic connection formalism in DFT and
seeking functionals with thermodynamic accuracy, Becke suggested a functional of 
roughly the form \cite{B93}, 
\begin{equation}
  E_{xc}^{hybrid} = E_x^{GGA} + a \left(E_x^{HF} - E_x^{GGA}\right) + E_c^{GGA} \, .
  \label{eq:review.4}
\end{equation}
The $a$ parameter was intially determined semi-empirically but a choice of $a = 0.25$
was later justified on the basis of MBPT \cite{PEB96}.  This is a global hybrid (GH), to 
distinguish it from yet another type of hybrid, namely the range-separated hybrid (RSH).
Initially proposed by Savin \cite{S95}, RSHs separate the $1/r_{12}$ interelectronic 
repulsion into a short-range (SR) part to be treated by density-functional theory and 
a long-range (LR) part to be treated by wavefunction 
methodology.  A convenient choice
uses the complementary error function for the short-range part,
$(1/r_{12})_{SR} = \mbox{erfc}(\gamma r_{12})/r_{12}$, and the error function for the
long-range part, $(1/r_{12})_{LR} = \mbox{erf}(\gamma r_{12})/r_{12}$.
In this case, $\gamma=0$ corresponds to pure DFT while $\gamma=\infty$ corresponds to 
Hartree-Fock. See Ref.~\cite{BLS10} for a recent review of one type of RSH.

\subsection{Time-Dependent (TD-) DFT}


Conventional Hohenberg-Kohn-Sham DFT is limited to the ground stationary state,  
but chemistry is also concerned with linear and nonlinear optics and molecules in
excited states.  Time-dependent DFT has been developed to address these issues.
This subsection first reviews formal TD-DFT and then briefly discusses TD-DFAs.
There are now a number of review articles on TD-DFT (some of which are cited in 
this chapter), two summer school multi-author texts \cite{MUN+06,MNN+11}, and 
now a single-author textbook \cite{U12}.  Our review of formal TD-DFT roughly 
follows Ref.~\cite{U12} pp.~50--58 which the reader may wish to consult for further details.
Our comments about the Frenkel-Dirac variational principle and TD-DFAs comes
from our own synthesis of the subject.

A great deal of effort has been put into making formal TD-DFT as rigorous as
possible and firming up the formal underpinnings of TD-DFT remains an area of
active research.  At the present time, formal TD-DFT is based upon two theorems, 
namely the Runge-Gross theorem \cite{RG84} and the van~Leeuwen theorem \cite{L99}.
They remind one of us (MEC) of some wise words from his thesis director (John E.\ Harriman)
at the time of his (MEC's) PhD studies: ``Mathematicians always seem to know more than 
they can prove.''\footnote{This is formalized in mathematical logic theory 
by G\"odel's incompleteness theorem which basically says that there are always 
more things that are true than can be proven to be true.} The Runge-Gross and 
van~Leeuwen theorems are true for specific cases where they can be proven, but 
we believe them to hold more generally and efforts continue to find more general proofs.  

\paragraph{Runge-Gross theorem}
This theorem states, with two caveats, that the time-dependent external potential 
$v({\bf 1})$ is determined up to an arbitrary function of time by the initial wavefunction
$\Psi_0=\Psi(t_0)$ at some time $t_0$ and by the time-dependent charge density 
$\rho({\bf 1})$.  Here we have enriched our notation to include time, 
${\bf i} = (i,t_i) = (\vec{r}_i, \sigma_i, t_i)$.  The statement that the 
external potential is only determined up to an arbitrary function of time simply 
means that the phase of the associated wave function is only determined up to
a spatially-constant time-dependent constant.  
This is because two external potentials differing by an additive function of
time $\tilde{v}(1) = v(1) + c(t_1)$ lead to associated wave functions 
$\tilde{\Psi}(t) = e^{-i\alpha(t)} \Psi(t)$ where $d\alpha(t)/dt = c(t)$.
A consequence of the Runge-Gross theorem is that expectation values of observables 
$\hat{A}(t)$
are functionals of the initial wavefunction and of the time-dependent charge density,
\begin{equation}
  A[\rho,\Psi_0](t)  = \langle \Psi[\rho,\Psi_0](t) \vert \hat{A}(t) 
    \vert \Psi[\rho,\Psi_0](t) \rangle
  \, .
  \label{eq:review.5}
\end{equation}
The proof of the theorem assumes (caveat 1) that the external potential is expandable
in a Taylor series in time in order to show that the time-dependent current density
determines the time-dependent external potential up to an additive function of time.
The proof then goes on to make a second assumption (caveat 2) that the external 
potential goes to zero at large $r$ at least as fast as $1/r$ in order to prove
that the time-dependent charge density determines the time-dependent current density.

\paragraph{van~Leeuwen theorem}
Given a system with an electron-electron interaction $w(1,2)$, external potential 
$v({\bf 1})$, and initial wavefunction $\Psi_0$, and another system with the same
time-dependent charge density $\rho({\bf 1})$, possibly different electron-electron 
interaction $\tilde{w}(1,2)$, and initial wavefunction $\tilde{\Psi}_0$, then
the external potential of the second system $\tilde{v}({\bf 1})$ is uniquely determined
up to an additive function of time.  Notice that we recover the Runge-Gross theorem
when $w(1,2)=\tilde{w}(1,2)$ and $\Psi_0=\tilde{\Psi}_0$.  However the most interesting
result is perhaps when $\tilde{w}(1,2)=0$ because this corresponds to a Kohn-Sham-like
system of noninteracting electrons, showing us that the external potential of such
a system is unique and ultimately justifying the time-dependent Kohn-Sham equation,
\begin{equation}
  \hat{h}[\rho,\Psi_0,\tilde{\Psi}_0]({\bf 1}) \psi_i({\bf 1}) = i\frac{\partial}{\partial t} \psi_i({\bf 1}) \, ,
  \label{eq:review.6}
\end{equation}
where,
\begin{equation}
  \hat{h}[\rho,\Psi_0,\tilde{\Psi}_0]({\bf 1}) = \hat{t}_s + v({\bf 1}) 
   + v_H[\rho]({\bf 1}) + v_{xc}[\rho,\Psi_0,\tilde{\Psi}_0]({\bf 1}) \, .
  \label{eq:review.7}
\end{equation}
The proof of the theorem assumes (caveat 1) that the external potential is expandable 
in a Taylor series in time {\em and} (caveat 2) that the charge density is expandable 
in a Taylor series in time.  Work on removing these caveats is ongoing 
\cite{MTWB10,RL11,RGPL12,RNL13} (Ref.~\cite{U12} pp.~57--58 provides a brief, but dated, summary).

\paragraph{Frenkel-Dirac action}
This is a powerful and wide-spread action principle used to derive time-dependent
equations within approximate formalisms.  Making the action
\begin{equation}
  A = \int_{t_0}^{t_1} \langle \Psi(t') \vert i\frac{\partial}{\partial t'} - \hat{H}(t')
      \vert \Psi(t') \rangle \, dt' \, ,
  \label{eq:review.8}
\end{equation}
stationary subject to the conditions that $\delta \Psi(t_0) = \delta \Psi(t_1) = 0$ 
leads to the time-dependent Schr\"odinger equation 
$\hat{H}(t)\Psi(t) = i \partial \Psi(t)/\partial t$.  Runge and Gross initially suggested
that $A=A[\rho,\Psi_0]$ and used this to derive a more explicit formula for the TD-DFT
xc-potential as a functional derivative of an xc-action, however this led to causality
problems.  A simple explanation and way around these contradictions was presented
by Vignale \cite{V08} who noted that, as the time-dependent Schr\"odinger equation is
a first-order partial differential equation in time, $\Psi(t_1)$ is determined by 
$\Psi(t_0)$ so that, while $\delta \Psi(t_0)$ may be imposed, $\delta \Psi(t_1)$ may
not be imposed.  The proper Frenkel-Dirac-Vignale action principle is then,
\begin{equation}
  \delta A = i \langle \Psi(t_1) \vert \delta \Psi(t_1)\rangle \, .
  \label{eq:review.9}
\end{equation}
In many cases, the original Frenkel-Dirac action principle gives the same results
as the more sophisticated Frenkel-Dirac-Vignale action principle.  Ref.~\cite{MDRS11}
gives one example of where this action principle has been used to derive an
xc-potential within a TD-DFA.  Other solutions to the Dirac-Frenkel causality problem
in TD-DFT may also be found in the literature \cite{R96,L98,L01,M05,M13}.

\paragraph{Time-dependent density-functional approximations (TD-DFAs)}
As the exact TD-DFT xc-functional is unknown, it must be approximated.  In most
cases, we can ignore the initial state dependences because we
are treating a system initially in its ground stationary state exposed to
a time-dependent perturbation.   This is because, if the initial 
state is the ground stationary state, then according to the first
Hohenberg-Kohn theorem of conventional DFT $\Psi_0=\Psi_0[\rho]$ and 
$\tilde{\Psi}_0=\tilde{\Psi}_0[\rho]$.  

The simplest and most successful TD-DFA is the TD-DFT adiabatic approximation 
(AA) which states that the xc-potential reacts instantaneously and without 
memory to any temporal change in the time-dependent density,
\begin{equation}
  v_{xc}^{AA}[\rho]({\bf 1}) = \frac{\delta E_{xc}[\rho_{t_1}(1)]}
               {\delta \rho_{t_1}(1)}
  \label{eq:review.10}
\end{equation}
The notation is a bit subtle here: $\rho_{t_1}(1)$ is $\rho({\bf 1})=\rho(1,t_1)$
at a fixed value of time, meaning that $\rho_{t_1}(1)$ is uniquely a function of
the space and spin coordinates albeit at fixed time $t_1$.  The AA approximation 
has been remarkably successful and effectively defines conventional TD-DFT.

\begin{table}
\caption{Jacob’s ladder for memory functionals \cite{PS01}. 
\label{tab:TDjacob}
}
\begin{tabular}{lcl}
\hline 
 & \multicolumn{1}{c}{\em Quantum Chemical Heaven} & \\
TD-RDMT$^d$ & \line(1,0){100} & $\gamma(1,2,t)^g$, $\theta_i(t)^h$\\
TD-OEP$^c$ & \line(1,0){100} & $\psi_i(\bf 1)^f$ \\
L-TD-DFT$^b$ & \line(1,0){100} & fluid position and deformation tensor \\
TD-CDFT$^a$ & \line(1,0){100} & $\rho({\bf 1})$, $\vec{j}({\bf 1})^e$ \\
TD-DFT & \line(1,0){100} & $\rho({\bf 1})$ \\
 & \multicolumn{1}{c}{\em Hartree World} & \\
\hline 
\end{tabular}
\\
$^a$ TD current-density-functional theory.\\
$^b$ Lagrangian TD-DFT.\\
$^c$ TD optimized effective potential.\\
$^d$ TD reduced-density-matrix theory.\\
$^e$ The current density.\\
$^f$ TD occupied orbitals.\\
$^g$ TD reduced-density matrix.\\
$^h$ Natural orbital phases.
\end{table}
Going beyond the TD-DFT AA is subject of ongoing work.  Defining new Jacob's ladders
for TD-DFT may be helpful here.  The first attempt to do so was the definition by
one of us (MEC) of a ``Jacob's jungle gym'' consisting of parallel Jacob's ladders
for $E_{xc}$, $v_{xc}({\bf 1})$, $f_{xc}({\bf 1},{\bf 2}) = \delta v_{xc}({\bf 1})/
\delta \rho({\bf 2})$, etc.~\cite{C02}.  This permitted the use of simultaneous
use of different functionals on the different ladders on the grounds that accurate
lower derivatives did not necessarily mean accurate higher derivatives.  Of course,
being able to use a consistent level of approximation across all ladders could
be important for some types of applications (e.g., those involving analytical 
derivatives).  With this in mind, the authors recently suggested a new Jacob's
ladder for TD-DFT (Table~\ref{tab:TDjacob}).

\subsection{Linear Response (LR-) TD-DFT}

As originally formulated TD-DFT seems ideal for the calculation of
nonlinear optical (NLO) properties from the dynamical response of the molecular 
dipole moment $\vec{\mu}(t)$ to an applied electric field 
$\vec{\varepsilon}(t)=\vec{\varepsilon}\cos (\omega t)$,
\begin{equation}
  \Delta \vec{\mu}(t) = \int {\bf \alpha}(t-t') \vec{\varepsilon}(t') dt' 
  + \mbox{HOT} \, ,
  \label{eq:review.11a}
\end{equation}
using real-time numerical integration of the TD Kohn-Sham equation, but it
may also be used to calculate electronic absorption spectra.  This subsection
explains how.
  
In Eq.~(\ref{eq:review.11}) ``HOT'' stands for ``higher-order terms'' and the 
quantity ${\bf \alpha}$ is the dynamic dipole polarizability.  After Fourier 
transform, Eq.~(\ref{eq:review.11}) becomes,
\begin{equation}
  \Delta \vec{\mu}(\omega) =  {\bf \alpha}(\omega) \vec{\varepsilon}(\omega) 
  + \mbox{HOT} \, ,
  \label{eq:review.11}
\end{equation}
If the applied field is suffiently small then we are in the LR regime where
we may neglect the HOT and calculate the dipole polarizability as 
$\alpha_{i,j}(\omega)=\Delta \mu_i(\omega)/\varepsilon_j(\omega)$.
Electrical absorption spectra may be calculated from this because of the
sum-over-states theorem in optical physics,
\begin{equation}
  \alpha(\omega) = \sum_{I \neq 0} \frac{f_I}{\omega_I^2-\omega^2} \, ,
  \label{eq:review.12}
\end{equation}
where $\alpha = (1/3) (\alpha_{xx}+\alpha_{yy}+\alpha_{zz})$.  Here
\begin{equation}
  \omega_I = E_I-E_0 \, ,
  \label{eq:review.13}
\end{equation}
is the excitation energy
\footnote{Remember that $\hbar = 1$ in the atomic units used here.}
and 
\begin{equation}
  f_I = \frac{2}{3} \omega_I \vert \langle 0 \vert \vec{r} \vert I \rangle \vert^2
  \, 
  \label{eq:review.14}
\end{equation}
is the corresponding oscillator strength.  This sum-over-states
theorem makes good physical sense because we expect the response of the
charge density and dipole moment to become infinite (i.e., to jump suddenly)
when the photon frequency corresponds to an electronic excitation energy.
Usually in real-time TD-DFT programs, the spectral function is calculated
as,
\begin{equation}
  S(\omega) = \frac{2\omega}{\pi} \Im \alpha(\omega+i\eta) \, ,
  \label{eq:review.15}
\end{equation}
which generates a Lorentzian broadened spectrum with broadening controlled by
the $\eta$ parameter.  The connection with the experimentally observed molar
extinction coefficient as a function of $\nu=\omega/(2\pi)$ is,
\begin{equation}
  \epsilon(\nu) = \frac{\pi N_A e^2}{m_e c (4\pi \epsilon_0) \ln (10)} S(2\pi \nu) \, ,
  \label{eq:review.16}
\end{equation}
in SI units.

So far this is fine for calculating spectra, but not for assigning and studying
individual states.  For that, it is better to take another approach using
the susceptibility,
\begin{equation}
  \chi({\bf 1},{\bf 2}) = \frac{\delta \rho({\bf 1})}{\delta v_{appl}({\bf 2})} \, ,
  \label{eq:review.17}
\end{equation}
which describes the response of the density to the applied pertubation 
$v_{appl}$,
\begin{equation}
  \delta \rho({\bf 1}) = \int \chi({\bf 1},{\bf 2}) \delta v_{appl}({\bf 2}) \, d{\bf 2}
  \, .
  \label{eq:review.18}
\end{equation}
The response of the density of the Kohn-Sham fictitious system of noninteracting electrons is
identical but the potential is now the Kohn-Sham single-particle potential,
\begin{equation}
  \delta \rho({\bf 1}) = \int \chi_s({\bf 1},{\bf 2}) \delta v_s({\bf
    2}) \, d{\bf 2} \, .
  \label{eq:review.19}
\end{equation}
In contrast to the interacting susceptibility of Eq.~(\ref{eq:review.17}), the noninteracting susceptibility,
\begin{eqnarray}
  \chi_s({\bf 1},{\bf 2}) = \frac{\delta \rho({\bf 1})}{\delta
    v_s({\bf 2})} \, ,
  \label{eq:review.20}
\end{eqnarray}
is known exactly from MBPT.  Of course the effective potential is the sum of the applied potential and the potential due to the response of the self-consistent field, $v_{Hxc}$,
\begin{equation}
  \delta v_s({\bf 1}) = \delta v_{appl}({\bf 1}) + \int f_{Hxc}({\bf
    1},{\bf 2}) \delta \rho({\bf 2}) \, d{\bf 2} \, ,
  \label{eq:review.21a}
\end{equation}
where $f_{Hxc}({\bf 1},{\bf 2}) = \delta v_{Hxc}({\bf 1})/\delta \rho({\bf 2})$ is the functional derivative of the Hartree plus exchange-correlation self-consistent field.  Manipulating these equations is facilitated by a matrix representation in which the integration is interpreted as a sum over a continuous index.  Thus,
\begin{equation}
  \delta {\bm \rho} = {\bm \chi} \delta {\bm v}_{appl} = {\bm \chi}_s
  \left( \delta {\bm v}_{appl} + {\bm f}_{Hxc} \delta {\bm \rho}
  \right) \, ,
  \label{eq:review.21}
\end{equation}
is easily manipulated to give a Bethe-Salpeter-like equation (Sec.~\ref{sec:MBPT}),
\begin{equation}
  {\bm \chi} = {\bm \chi}_s + {\bm \chi}_s {\bm f}_{Hxc} {\bm \chi} \,
  ,
  \label{eq:review.22}
\end{equation}
or, written out more explicitly,
\begin{equation}
  \chi({\bf 1},{\bf 4}) = \chi_s({\bf 1},{\bf 4}) + \int \chi_s({\bf
    1},{\bf 2}) f_{Hxc}({\bf 2},{\bf 3}) \chi({\bf 3},{\bf 4}) \,
  d{\bf 2} d{\bf 3} \, .
  \label{eq:review.23}
\end{equation}
Equation~(\ref{eq:review.21}) may be solved iteratively for $\delta {\bm \rho}$.
Alternatively $\delta {\bm \rho}$ may be obtained by solving,
\begin{equation}
  \left( {\bm \chi}_s^{-1} - {\bm f}_{Hxc} \right) \delta {\bm \rho} =
  \delta {\bm v}_{appl} \, ,
  \label{eq:review.24}
\end{equation}
which typically involves iterative Krylov space techniques because of the large
size of the matrices involved.

This last equation may be manipulated to make the most common form of LR-TD-DFT
used in quantum chemistry \cite{C95} \footnote{This equation is not infrequently 
called the ``Casida equation'' in the TD-DFT literature (e.g., as in Ref.~\cite{U12} 
pp.~145--153.)}.  This is a pseudoeigenvalue problem,
\begin{equation}
  \left[ \begin{array}{cc} {\bf A}(\omega) & {\bf B}(\omega) \\ {\bf
        B}^*(\omega) & {\bf A}^*(\omega) \end{array} \right]
  \left( \begin{array}{c} {\bf X} \\ {\bf Y} \end{array} \right) =
  \omega \left[ \begin{array}{cc} {\bm 1} & {\bm 0} \\ {\bm 0} & -{\bm
        1} \end{array} \right] \left( \begin{array}{c} {\bf X} \\ {\bf
      Y} \end{array} \right) \, ,
  \label{eq:review.25}
\end{equation}
where,
\begin{eqnarray}
  A_{ia,jb}(\omega) & = & \delta_{i,j} \delta_{a,b} \epsilon_{a,i} + (ia|f_{Hxc}(\omega)|jb) \nonumber \\ B_{ia,bj}(\omega) & = & (ia|f_{Hxc}(\omega)|bj) \, .
  \label{eq:review.26}
\end{eqnarray}
Here,
\begin{equation}
  (pq|f|rs) = \int \int \psi_p^*(1) \psi_q(1) f(1,2) \psi_r^*(2) \psi_s(2) \, 
  d1 d2 \, , \label{eq:review.26a}
\end{equation}
is a two electron integral in Mulliken ``charge-cloud'' notation over 
the kernel $f$ which may either be the Hartree kernel [$f_H(1,2)=\delta_{\sigma_1,\sigma_2} / r_{12}$] or the xc-kernel or the sum of the two (Hxc).
The index notation is $i,j,...$ for occupied spin-orbitals, $a,b,...$ for virtual 
spin-orbitals and $p,q,...$ for unspecified spin-orbitals (either occupied or unoccupied)
\footnote{Sometimes we call this the {\sc FORTRAN} index convention in reference to the
default variable names for integers in that computer language}. 
Also we have introduced the compact notation,
\begin{equation}
  \epsilon_{rs\cdots,uv\cdots} = (\epsilon_r + \epsilon_s + \cdots) - (\epsilon_u + \epsilon_v + \cdots) \, .
  \label{eq:review.28}
\end{equation}
Equation~(\ref{eq:review.26}) has paired excitation and de-excitation solutions. 
Its eigenvalues are (de-)excitation energies the vectors ${\bf X}$ and ${\bf Y}$ 
provide information about transition moments.  In particular the oscillator strength,
of the transition with excitation energy $\omega_I$ may be calculated from ${\bf X}_I$ 
and ${\bf Y}_I$.~\cite{C95}  When the adiabatic approximation (AA) to the xc-kernel is 
made, the ${\bf A}$ and ${\bf B}$ matrices become independent of frequency. As a 
consequence, the number of solutions is equal to the number of 1-electron excitations, 
albeit dressed to include electron correlation effects.  Allowing the ${\bf A}$ and 
${\bf B}$ matrices to have a frequency dependence allows the explicit inclusion of 
2-electron (and higher) excited states.

The easiest way to understand what is missing in the AA is within the 
so-called Tamm-Dancoff approximation (TDA). The usual AA TDA equation,
\begin{equation}
  {\bf A} {\bf X} = \omega {\bf X} \, ,
  \label{eq:review.29}
\end{equation}
is restricted to single excitations. The configuration interaction (CI) 
equation~\cite{C96},
\begin{equation}
  \left( {\bf H} - E_0 {\bm 1} \right) {\bf C} = \omega {\bf C} \, ,
  \label{eq:review.30}
\end{equation}
which includes all excitations of the system, can be put into the form of 
Eq.~(\ref{eq:review.29}), but with a frequency-dependent ${\bf A}(\omega)$ matrix. 
This can be simply done by partitioning the full CI Hamiltonian into a singles 
excitations part (${\bf A}_{1,1}$) and multiple-excitations part (${\bf A}_{2+,2+}$) as,
\begin{equation}
  \left[ \begin{array}{cc} {\bf A}^{CI}_{1,1} & {\bf A}^{CI}_{1,2+}
      \\ {\bf A}^{CI}_{2+,1} & {\bf A}^{CI}_{2+,2+} \end{array}
    \right] \left( \begin{array}{c} {\bf C}_1 \\ {\bf
      C}_{2+} \end{array} \right) = \omega \left( \begin{array}{c}
    {\bf C}_1 \\ {\bf C}_{2+} \end{array} \right) \, ,
  \label{eq:review.31}
\end{equation}
provided we can ignore any coupling between the ground state and excited states.  
Applying the standard L\"owdin-Feshbach partitioning technique to 
Eq.~(\ref{eq:review.31})~\cite{L64}, we obtain
\begin{equation}
  \left[ {\bf A}^{CI}_{1,1} + {\bf A}^{CI}_{1,2+} \left( \omega {\bm
      1}_{2+,2+} - {\bf A}^{CI}_{2+,2+} \right)^{-1} {\bf
      A}^{CI}_{2+,1} \right] {\bf C}_1 = \omega {\bf C}_1 \, ,
  \label{eq:review.32}
\end{equation}
in which it is clearly seen that multiple-excitation states arise from a 
frequency-dependent term missing in the AA xc-kernel~\cite{C96}. 

In the remainder of this chapter, we will first show how MBPT may be used to
derive expressions for the ${\bf A}^{CI}_{1,2+}$, ${\bf A}^{CI}_{2+,1}$,
and ${\bf A}^{CI}_{2+,2+}$ blocks and show how this may be used in the form
of dressed TD-DFT to correct the AA.  Then we will discuss localization of the
terms beyond the AA in order to obtain some insight into the analytic behavior
of the xc-kernel.

\section{Many-Body Perturbation Theory (MBPT)}
\label{sec:MBPT}

This section elaborates on the polarization propagator (PP) approach.  As the PP was originally
inspired by the Bethe-Salpeter equation (BSE) and as the BSE often crops up in articles from
the solid-state physics community which are concerned with both TD-DFT and 
MBPT~\cite{ORR02,RORO02,SOR03,MSR03,STP04,BDLS05,RSB+09}, we
will try to make the connection between the PP and BSE approaches as clear as possible.
Although the two  MBPT approaches are formally equivalent,
differences emerge because the BSE approach emphasizes the time representation while the PP 
approach emphasizes the frequency representation.  This can and typically does lead to 
different approximations.  In particular it seems to be easier to derive pole 
structure-conserving approximations needed for treating 2-electron and higher excitations 
in the frequency representation than in the time representation.  This and prior experience 
with the PP approach in the quantum chemistry community~\cite{OJ77,NJO80,NJO81,JS81,S82,TSS99} 
have lead us to favor the PP approach.  We shall make extensive use of diagrams in order to 
give an overview of our manipulations.  Whenever possible, more elaborate mathematical
manipulations will be relegated to the appendix.

\subsection{Green's Functions}

Perhaps the most common and
arguably the most basic quantity in MBPT is the 1-electron Green's
function defined by,
\begin{equation}
  iG({\bf 1},{\bf 2}) = \langle 0 \vert {\cal T} \{ \hat{\psi}_H({\bf
    1}) \hat{\psi}^\dagger_H({\bf 2}) \} \vert 0 \rangle \, .
  \label{eq:MBPT.1}
\end{equation}
Here, the subscript $H$ indicates that the field operators are understood to
be in the Heisenberg representation.  Also ${\cal T}$ is the usual
time-ordering operator, which includes anticommutation in our case
(i.e., for fermions),
\begin{eqnarray}
  {\cal T} \{ \hat{\psi}_H({\bf 1}) \hat{\psi}^\dagger_H({\bf 2}) \} & = &
  \theta(t_1-t_2) \hat{\psi}_H({\bf 1}) \hat{\psi}_H^\dagger({\bf 2})
  \nonumber \\ & - & \theta(t_2-t_1) \hat{\psi}_H^\dagger({\bf 2})
  \hat{\psi}_H({\bf 1}) \, .  
  \label{eq:MBPT.2}
\end{eqnarray}
The 2-electron Green's function is (see p.~116 of Ref.~\cite{FW71}),
\begin{equation}
  G({\bf 1},{\bf 2};{\bf 3},{\bf 4}) = (-i)^2 \langle 0 \vert {\cal T}
  \{ \hat{\psi}_H({\bf 1}) \hat{\psi}_H({\bf 2})
  \hat{\psi}_H^\dagger({\bf 4}) \hat{\psi}_H^\dagger({\bf 3}) \} \vert
  0 \rangle \, .
  \label{eq:MBPT.3}
\end{equation}

The usual MBPT approach to evaluating the susceptibility, $\chi$, uses
the fact that it is the retarded form,
\begin{equation}
  i \chi({\bf 1},{\bf 2}) = \theta(t_1-t_2) \langle 0 \vert [
    \tilde{\rho}_H({\bf 1}) , \tilde{\rho}_H({\bf 2}) ] \vert 0
  \rangle \, ,
  \label{eq:MBPT.4}
\end{equation}
of the time-ordered correlation function,
\begin{equation}
  i \chi({\bf 1},{\bf 2}) = \langle 0 \vert {\cal T} \{
  \tilde{\rho}_H({\bf 1}) \tilde{\rho}_H({\bf 2}) \} \vert 0 \rangle
  \, ,
  \label{eq:MBPT.5}
\end{equation}
where,
\begin{equation}
  \tilde{\rho}_H({\bf 1}) = \hat{\psi}_H^\dagger({\bf 1})
  \hat{\psi}_H({\bf 1}) - \langle 0 \vert \hat{\psi}_H^\dagger({\bf
    1}) \hat{\psi}_H({\bf 1}) \vert 0 \rangle \, ,
  \label{eq:MBPT.6}
\end{equation}
is the density fluctuation operator.
(See for example Ref.~\cite{FW71} pp.\ 172-175 and p.\ 151.)

We will also need several generalizations of the susceptibility
and the density fluctuation operator.  The first is the 
particle-hole (ph) propagator~\cite{S82}, which we chose to write
as,
\begin{equation}
  i L({\bf 1},{\bf 2};{\bf 3},{\bf 4}) = \langle 0 \vert {\cal T} \{
  \tilde{\gamma}({\bf 1},{\bf 2}) \tilde{\gamma}({\bf 4},{\bf 3}) \}
  \vert 0 \rangle \, ,
  \label{eq:MBPT.7}
\end{equation}
where,
\begin{equation}
  \tilde{\gamma}({\bf 1},{\bf 2}) = \hat{\psi}_H^\dagger({\bf 2})
  \hat{\psi}_H({\bf 1}) - \langle 0 \vert {\cal T}
  \{\hat{\psi}_H^\dagger({\bf 2}) \hat{\psi}_H({\bf 1}) \} \vert 0
  \rangle \, ,
  \label{eq:MBPT.8}
\end{equation}
is a sort of density matrix fluctuation operator (or would be if we
constrained $t_1=t_2$ and $t_3=t_4$).  Notice that the ph-propagator
is a four-time quantity.

[It may be useful to try to place $L$ in the context of other
2-electron propagators: The particle-hole response function,~\cite{S82}
\begin{equation}
  R({\bf 1},{\bf 2};{\bf 3},{\bf 4}) = G({\bf 1},{\bf 2};{\bf 3},{\bf
    4}) - G({\bf 1},{\bf 3}) G({\bf 2},{\bf 4}) \, .
  \label{eq:MBPT.9a}
\end{equation}
Then $L$ is related to $R$ by the relation,
\begin{equation}
  L({\bf 1},{\bf 2};{\bf 3},{\bf 4}) = i R({\bf 1},{\bf 4};{\bf
    2},{\bf 3}) \, \mbox{.]}
  \label{eq:MBPT.9}
\end{equation}

We will also need the polarization propagator (PP) which is the two-time
quantity,
\begin{equation}
  \Pi(1,2;3,4;t-t') = L(1t,2t;3t',4t') \, .
  \label{eq:MBPT.10}
\end{equation}
Written out explicitly,
\begin{eqnarray}
  & & i\Pi(1,2;3,4;t-t') \nonumber \\ & = & \langle 0 \vert {\cal T}\{
  \hat{\psi}_H^\dagger(2t^+) \hat{\psi}_H(1t)
  \hat{\psi}^\dagger_H(3t'^+) \hat{\psi}_H(4t') \}\vert 0 \rangle
  \nonumber \\ & - & \langle 0 \vert {\cal T}\{
  \hat{\psi}_H^\dagger(2t^+) \hat{\psi}_H(1t) \} \vert 0 \rangle
  \langle 0 \vert {\cal T}\{ \hat{\psi}_H^\dagger(3t'^+)
  \hat{\psi}_H(4t') \} \vert 0 \rangle \, . 
  \label{eq:MBPT.11}
\end{eqnarray}
[The second term is often dropped in the definition of the PP.  It is
  there to remove $\omega=0$ excitations in the Lehmann
  representation. (See for example pp.\ 559-560 of
  Ref.~\cite{FW71}.)]  The retarded version of the PP is the
susceptibility describing the response of the 1-electron density
matrix,
\begin{equation}
  \gamma(1,2;t) = \langle 0 \vert \hat{\psi}^\dagger(2t)
  \hat{\psi}(1t) \vert 0 \rangle \, ,
  \label{eq:MBPT.12}
\end{equation}
to a general (not necessarily local) applied perturbation,
\begin{equation}
  \Pi(1,2;3,4;t-t') = \frac{\delta \gamma(1,2;t)}{\delta
    w_{appl}(3,4;t')} \, ,
  \label{eq:MBPT.13}
\end{equation}
which is a convolution.  After Fourier transform,
\begin{equation}
  \delta \gamma(1,2;\omega) = \int \Pi(1,2;3,4;\omega) \delta
  w_{appl}(3,4;\omega) \, d3 d4 \, ,
  \label{eq:MBPT.14}
\end{equation}
or,
\begin{equation}
  \delta {\bm \gamma}(\omega) = {\bm \Pi}(\omega) \delta {\bm
    w}_{appl}(\omega) \, ,
  \label{eq:MBPT.15}
\end{equation}
in matrix form.

\subsection{Diagram Rules}

The representation of MBPT expansions in terms of diagrams is very
convenient for bookkeeping purposes.  Indeed certain ideas such as the
linked-cluster theorem \cite{K66} or the
concept of a ladder approximation (see e.g., Ref.~\cite{FW71} p.\ 136)
are most naturally expressed in terms of diagrams.  Also
diagrams drawn according to systematic rules allow an easy
way to check algebraic expressions.  This is how we have used diagrams
in our research.  However we introduce diagrams here for a different reason,
namely because they provide a concise way to explain our work.

Several types of MBPT diagrams exist in the literature.  These divide into 
four main classes which we call Feynman, Abrikosov, Goldstone, and Hugenholtz.  
Such diagrams can be distinguished by whether they are time-ordered (Goldstone and
Hugenholtz) or not (Feynman and Abrikosov) and by whether they treat
the electron repulsion interaction as a wavy or dotted line with an
incoming and an outgoing arrow at each end (Feynman and Goldstone) or
in a symmetrized way as a point with two incoming and two outgoing
arrows (Abrikosov and Hugenholtz).  These differences affect how they
are to be translated into algebraic expressions as does the nature of
the quantity being expanded (wave function, one-electron Green's
function, self-energy, polarization propagator, etc.)  Given this
plethora of types of diagrams and the difficulty of finding a clear
explanation of how to read polarization propagator diagrams, we have
chosen to present rules for how our diagrams should be
translated into algebraic expressions.  This is perhaps especially
necessary because while the usual practice in the solid-state
literature is to use time-unordered diagrams with electron repulsions
represented as wavy or dotted lines (i.e., Feynman diagrams), while
the usual practice in the quantum chemistry literature of using
time-ordered diagrams with electron repulsions represented as points
(i.e., Hugenholtz diagrams).  


\begin{figure}
  \includegraphics[width=0.75 \columnwidth]{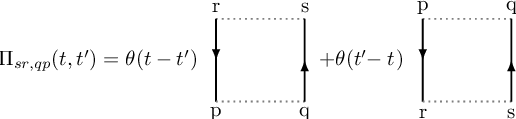}
  \caption{Basic time-ordered finite basis set representation PP
    diagram.}
   \label{fig:rule1}
\end{figure}
\begin{figure}
  \includegraphics[width=0.75 \columnwidth]{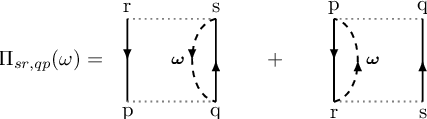}
  \caption{Basic frequency and finite basis set representation PP
    diagram.}
   \label{fig:rule2}
\end{figure}
\begin{figure}
  \includegraphics[width=0.75 \columnwidth]{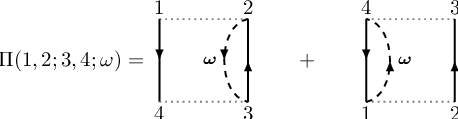}
  \caption{Basic frequency and real space representation PP diagram.}
   \label{fig:rule3}
\end{figure}
\begin{figure}
  \includegraphics[width=0.25 \columnwidth]{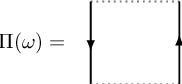}
  \caption{Time-unordered representation PP diagram.}
   \label{fig:rule4}
\end{figure}

We will limit ourselves to giving precise rules for the polarization
propagator (PP) since these rules are difficult to find in the literature.
The PP expressed in an orbital basis is,
\begin{equation}
  \Pi(1,2,3,4;t-t') = \sum_{pqrs} {\Pi_{sr,qp}(t-t')
    \psi_r(1)\psi^*_s(2)\psi_q^*(3)\psi_p(4)} \, ,
  \label{eq:MBPT.16}
\end{equation}
where,
\begin{eqnarray}
  \Pi_{sr,qp}(t-t') &=& -i\theta(t-t') \langle0\vert {\hat
    r}_H^\dagger(t) {\hat s}_H(t) {\hat q}_H^\dagger(t') {\hat
    p}_H(t') \vert0\rangle \nonumber \\ &-& i \theta(t'-t)
  \langle0\vert {\hat q}_H^\dagger(t') {\hat p}_H(t') {\hat
    r}_H^\dagger(t) {\hat s}_H(t) \vert0\rangle \, .\nonumber \\
  \label{eq:MBPT.17}
\end{eqnarray}
This makes it clear that the PP is a two time particle-hole propagator
which either propagates forward in time or backward in time.  To
represent it we introduce the following rules:
\begin{itemize}
  \item[(1)] Time increases vertically from bottom to top.  This is in
    contrast to a common convention in the solid-state literature
    where time increases horizontally from right to left.
  \item[(2)] A PP is a two time quantity. Each of these two times is
    indicated by a horizontal dotted line.  This is one type of
    ``event'' (representing the creation/destruction of an
    excitation).
  \item[(3)] Time-ordered diagrams use directed lines (arrows).
    Down-going arrows correspond to holes running backward in time,
    that is, to occupied orbitals.  Up-going arrows correspond to
    particles running forward in time, that is, unoccupied orbitals.
\end{itemize}
At this point, the PP diagrams look something like
Fig.~\ref{fig:rule1}.  Fourier transforming leads us to the
representation shown in Fig.~\ref{fig:rule2}.  An additional rule has
been introduced:
\begin{itemize}
  \item[(4)] A downward $\omega$ arrow on the left indicates forward
    ph-propagation.  An upward $\omega$ arrow on the right indicates
    backward ph-propagation.
\end{itemize}
Diagrams for the corresponding position space representation are shown
in Fig.~\ref{fig:rule3}.  Usually the labels ($p$, $q$, $r$, and $s$
or $1$, $2$, $3$, and $4$) are suppressed.  If the $\omega$ arrows are
also suppressed then there is no information about time-ordering and
both diagrams may be then written as a single time-unordered diagram
as in Fig.~\ref{fig:rule4}.  Typical Feynman diagrams are unordered
in time.

\begin{figure}
  \includegraphics[width=0.75 \columnwidth]{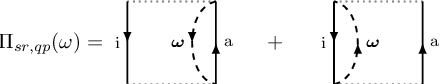}
  \caption{Zero-order PP diagrams.}
   \label{fig:rule5}
\end{figure}

Perturbation theory introduces certain denominators in the algebraic
expressions corresponding to the diagrams.  These may be represented
as cuts between events.
\begin{itemize}

  \item[(5)] Each horizontal cut between events contributes a factor
    $(\pm\omega + \sum_p{\epsilon_{p}}-\sum_{h}{\epsilon_{h}})^{-1}$,
    where $\sum_p$ ($\sum_h$) stands for the sum over all particle
    (hole) lines that are cut.  The omega line only appears in the sum
    if it is also cut.  It enters with a $+$ sign if it is directed
    upwards and with a $-$ sign if it is directed downwards.

  \item[(6)] There is also an overall sign given by the formula
    $(-1)^{h+l}$, where $h$ is the number of hole lines and $l$ is the
    number of closed loops, including the horizontal dotted event
    lines but ignoring the $\omega$ lines.

\end{itemize}
Diagrams are shown for the independent particle approximation in
Fig.~\ref{fig:rule5}.  The first diagram reads,
\begin{equation}
  \Pi_{ai,ai}(\omega) = \frac{1}{\omega+\epsilon_i-\epsilon_a} \, .
  \label{eq:MBPT.18}
\end{equation}
The second diagram reads,
\begin{equation}
  \Pi_{ia,ia}(\omega) = \frac{1}{-\omega+\epsilon_i-\epsilon_a} =
  \frac{-1}{\omega+\epsilon_a-\epsilon_i} \, .
  \label{eq:MBPT.19}
\end{equation}
These two equations are often condensed in the literature as,
\begin{equation}
  \Pi_{pq,rs}(\omega) = \delta_{p,r} \delta_{q,s}
  \frac{n_q-n_p}{\omega+\epsilon_q-\epsilon_p} \, .
  \label{eq:MBPT.20}
\end{equation}
\begin{figure}
\includegraphics[width=\columnwidth]{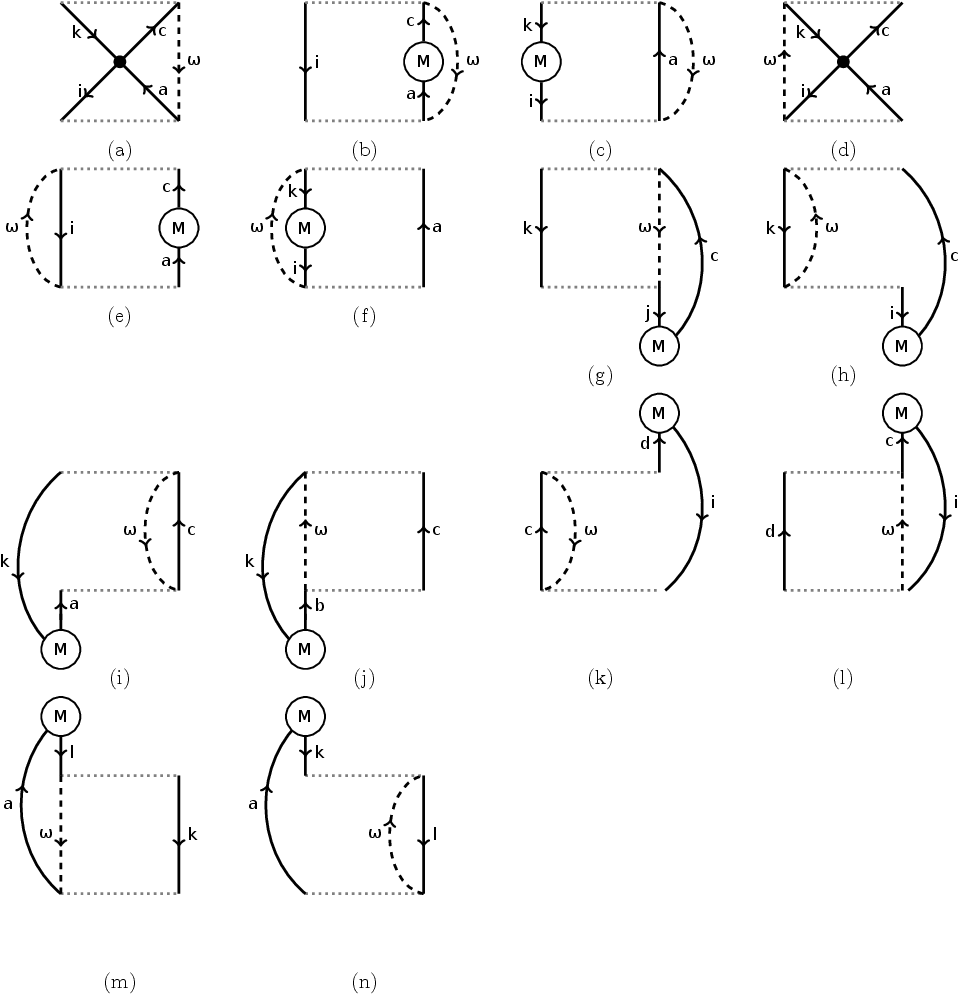}
\caption{First-order time-ordered diagrams Hugenholtz for 
    ${\bm \Pi}(\omega) - {\bm \Pi}_s(\omega)$ ({\em vide infra}). 
    Diagrams (a)--(i) involve coupling between the particle-hole space, diagrams (g), (h),
    (m), and (n) involve coupling between particle-hole space and particle-particle, and diagrams 
    (i)--(l) couple the particle-hole space with the hole-hole space.
  \label{fig:SOPPA1}
  }
\end{figure}
Let us now introduce one-electron perturbations in the form of M
circles.
\begin{itemize}

  \item[(7)] Each M circle in a diagram contributes a factor of
    $\langle p \vert \hat{M}_{xc} \vert q \rangle$, where $p$ is an
    incoming arrow and $q$ is an outgoing arrow and $\hat{M}_{xc}$ is 
    ``xc-mass operator'' which is the difference between the Hartree-Fock
    exchange self-energy and the xc-potential [Eq.~(\ref{eq:MBPT.30})].  
    (Thus $\langle \mbox{in} \vert \hat{M}_{xc} \vert \mbox{out} \rangle$.)
    For example, the
    term corresponding to Fig.~\ref{fig:SOPPA1} (b) contains a factor
    of $\langle a \vert \hat{M}_{xc} \vert c \rangle$, while the term
    corresponding to Fig.~\ref{fig:SOPPA1} (f) contains a factor of
    $\langle k \vert \hat{M}_{xc} \vert i \rangle$.  This is a second
    type of ``event'' (representing ``collision'' with the quantity
    $M_{xc}$).

\end{itemize}
For example, the term corresponding to Fig.~\ref{fig:SOPPA1} (j) is,
\begin{equation}
  \Pi_{ck,cb}(\omega) = \frac{\langle k \vert \hat{M}_{xc} \vert
    b \rangle} {(\omega-\epsilon_k+\epsilon_c)(\epsilon_k-\epsilon_b)}
  \, .
  \label{eq:MBPT.21}
\end{equation}

\begin{figure}
  \includegraphics[width=0.75 \columnwidth]{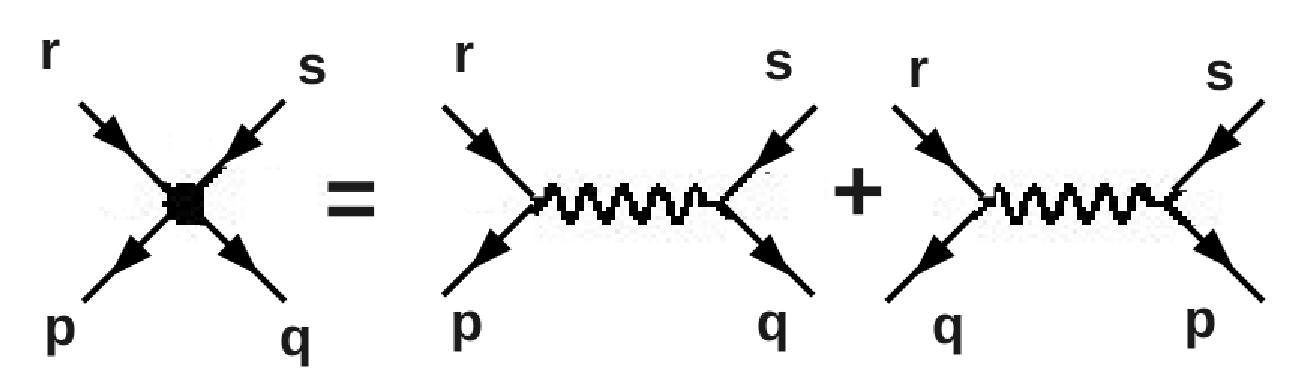}
  \caption{Electron repulsion integral diagrams.}
   \label{fig:rule6}
\end{figure}
This brings us to the slightly more difficult treatment of electron
repulsions.
\begin{itemize}

\item[(6)] When electron repulsion integrals are represented by dotted
  lines (Feynman and Goldstone diagrams), each end of the line
  corresponds to the labels corresponding to the same spatial point.
  The dotted line representation may be condensed into points
  (Abrikosov and Hugenholtz diagrams) as in Fig.~\ref{fig:rule6}.  A
  point with two incoming arrows, labeled $r$ and $s$, and two
  outgoing arrows, labeled $p$ and $q$ contributes a factor of
  $(rs\vert \vert pq) = (rp \vert f_H \vert sq) - (rq \vert f_H \vert
  sp)$.  (Thus $( \mbox{in, in} \vert \vert \mbox{out,out} ) =
  (\mbox{left in, right in} \vert \mbox{left in, right in})
  - (\mbox{left in, right in} \vert \mbox{left in, right in})$.
  The minus sign is not part of the diagram as it is taken into account
  by other rules.)  The integral notation is established in 
  Eq.~(\ref{eq:review.26a}) and the integral,
  \begin{equation}
    (pq\vert \vert rs) = \int \psi_p^*(1) \psi_r^*(2) \frac{1}{r_{12}}
    (1-{\cal P}_{12}) \psi_q(1) \psi_s(2) \, d1 d2 \, .
    \label{eq:MBPT.30b} 
  \end{equation}

\item[(7)] To determine the number of loops and hence the overall sign
  of a diagram in which electron repulsion integrals are expanded as
  dots, then write each dot as a dotted line (it does not matter which
  one of the two in Fig.~\ref{fig:rule6} is chosen) and apply rule
  (6).  The order of indices in each integral $(rs\vert \vert pq)$
  should correspond to the expanded diagrams.  (When Goldstone
  diagrams are interpreted in this way, we call them Brandow
  diagrams.)

\item[(8)] An additional factor of $1/2$ must be added for each pair
  of equivalent lines.  These are directed lines whose interchange, in
  the absence of further labeling, leaves the Hugenholtz diagram
  unchanged.

\end{itemize}
For example, the term corresponding to Fig.~\ref{fig:SOPPA1} (a) is,
\begin{eqnarray}
  \Pi_{ck,ai}(\omega) & = & -\frac{(ka||ic)}{
    (-\omega+\epsilon_k-\epsilon_c) (-\omega+\epsilon_i-\epsilon_a)}
  \nonumber \\ & = & \frac{(ak||ic)}{(-\omega+\epsilon_k-\epsilon_c)
    (-\omega+\epsilon_i-\epsilon_a)} \, .
  \label{eq:MBPT.22}
\end{eqnarray}
Additional information about Hugenholtz and other diagrams may be
found, for example, in Ref.~\cite{W84}.

\subsection{Dyson's equation and the Bethe-Salpeter equation (BSE)}

Two of the most basic equations of diagrammatic MBPT are Dyson's equation
for the 1-electron Green's function and the BSE for the ph-propagator.
Both require the choice of a zero-order picture which we take here to be the
exact or approximate Kohn-Sham system of noninteracting electrons.  We will
denote the zero-order quantities by the subscript $s$ (for single particle).

\begin{figure}
  \includegraphics[width=0.75 \columnwidth]{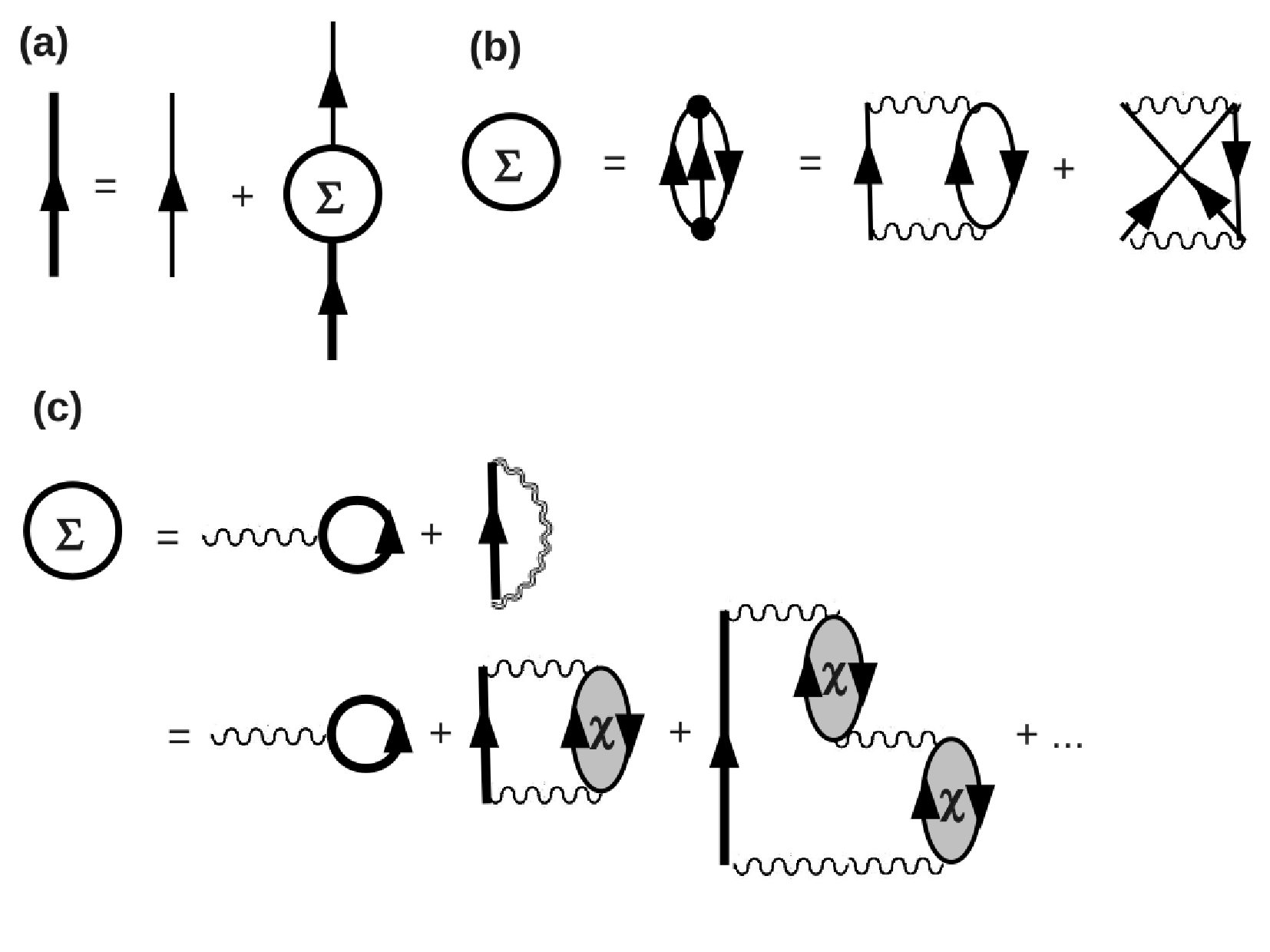}
  \caption{Time-unordered (Feynman and Abrikosov) 1-electron Green's function
    diagrams: (a) Dyson's equation, (b) second-order self-energy quantum chemistry
    approximation, (c) $GW$ self-energy solid-state physics approximation.
   \label{fig:Dyson}
   } 
\end{figure}
Dyson's equation relates the true 1-electron Green's function ${\bm G}$ to the
zero-order Green's function ${\bm G}_s$ via the (proper) self-energy ${\bm \Sigma}$,
\begin{equation}
  G({\bf 1},{\bf 2}) = G_s({\bf 1},{\bf 2}) + \int G_s({\bf 1},{\bf 3})
  \Sigma({\bf 3},{\bf 4}) G({\bf 4},{\bf 2}) \, d{\bf 3} d{\bf 4} \, ,
  \label{eq:MBPT.23}
\end{equation}
or more concisely,
\begin{equation}
  {\bm G} = {\bf G}_s + {\bm G}_s {\bm \Sigma} {\bm G} \, .
  \label{eq:MBPT.24}
\end{equation}
This is shown diagrammatically in Fig.~\ref{fig:Dyson}.  
It is to be emphasized that these diagrams are {\em un}ordered in time as it is not possible
to write a Dyson equation for time-ordered diagrams.  Also shown in Fig.~\ref{fig:Dyson} 
are typical low-order self-energy approximations.  Typical quantum chemistry approximations 
(b) involve explicit antisymmetrization of electron-repulsion integrals while solid-state 
physics approximations (c) emphasize dynamical screening.  Each approach has its strength 
and its weaknesses and so far the two approaches have defied any rigorous attempts at merger.

\begin{figure}
  \includegraphics[width=0.75 \columnwidth]{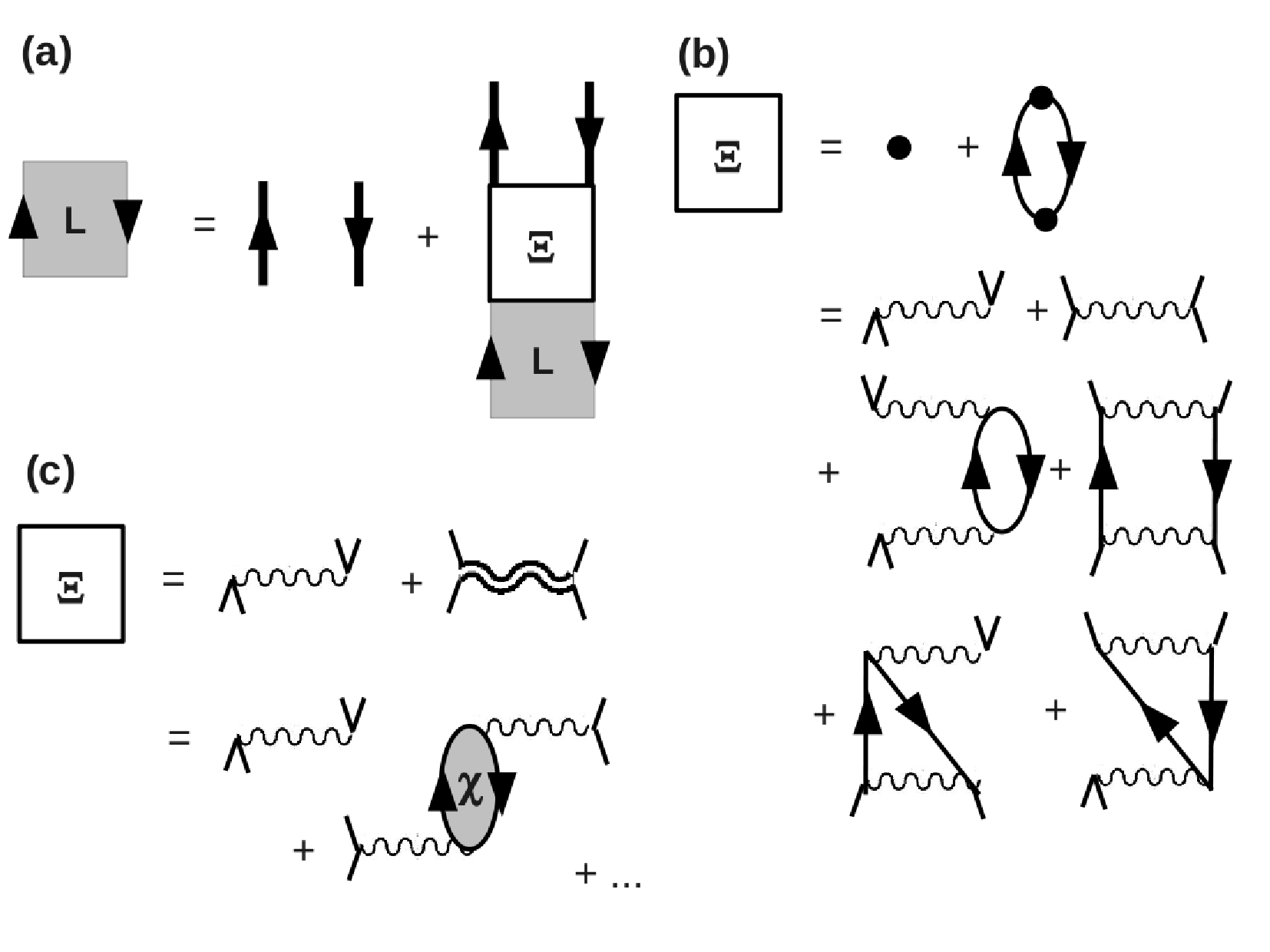}
  \caption{Time-unordered (Feynman and Abrikosov) ph-propagator
    diagrams: (a) BSE, (b) second-order self-energy quantum chemistry
    approximation, (c) $GW$ self-energy solid-state physics approximation.
    Note in part (c) that the solid-state physics literature will often
    turn the $v$ and $w$ wiggly lines at right angles to each other to
    indicate the same thing that we have indicated here by adding tab lines.
   \label{fig:BSE}
   } 
\end{figure}
The BSE is ``Dyson's equation'' for the ph-propagator,
\begin{eqnarray}
 & & L({\bf 1},{\bf 2};{\bf 7},{\bf 8}) = L_s({\bf 1},{\bf 2};{\bf
    7},{\bf 8}) \nonumber \\ & + & \int L_s({\bf 1},{\bf 2};{\bf
    3},{\bf 4}) \Xi_{Hxc}({\bf 3},{\bf 4};{\bf 5},{\bf 6}) L({\bf
    5},{\bf 6};{\bf 7},{\bf 8}) \, d{\bf 3} d{\bf 4} d{\bf 5} d{\bf 6}
  \, , 
  \label{eq:MBPT.25}
\end{eqnarray}
or
\begin{equation}
  {\bm L} = {\bm L}_s + {\bm L}_s {\bm \Xi}_{Hxc} {\bm L} \, ,
  \label{eq:MBPT.26a}
\end{equation}
in matrix notation.  Here 
\begin{equation}
  i L_s({\bf 1},{\bf 2};{\bf 3},{\bf 4}) = G_s({\bf 1},{\bf
    4})G_s({\bf 2},{\bf 3}) \, ,
  \label{eq:MBPT.26}
\end{equation}
is the ph-propagator for the zero-order picture (in our case, the exact or 
approximate Kohn-Sham fictitious system of noninteracting electrons), and 
the 4-point quantity, ${\bm \Xi}_{Hxc}$, may be deduced from a Feynman diagram expansion as 
the proper part of the ph-response function ``self-energy''.
This is shown diagrammatically in Fig.~\ref{fig:BSE}.
Again the quantum chemical approximations emphasize antisymetrization of the electron
repulsion integrals which is needed for proper inclusion of double excitations
while solid-state physics emphasizes use of a screened interaction.
While no rigorous way is yet known for combining screening and antisymmetrization, an
interesting pragmatic suggestion may be found in Ref.~\cite{SRC+11}.

\subsection{Superoperator equation-of-motion (EOM) polarization propagator (PP) approach}

We will now specialize to the PP and show how to obtain a ``Casida-like'' equation for
excitation energies and transition moments.  This will not yet give us correction terms
to AA LR-TD-DFT but it will give us some important tools to help us build
correction terms.  The basic idea in this section is to take the
exact or approximate Kohn-Sham system of independent electrons as the zero-order picture,
\begin{equation}
  {\hat H}^{(0)} = {\hat h}_{KS} \, ,
  \label{eq:MBPT.27}
\end{equation}
to add the perturbation,
\begin{equation}
  {\hat H}^{(1)} = {\hat V} + {\hat M}_{xc} \, .
  \label{eq:MBPT.28}
\end{equation}
and to do MBPT.  Here ${\hat V}$ is the fluctuation operator,
\begin{equation}
  {\hat V} = \frac{1}{4} \sum_{pqrs}{(pq\vert\vert rs){\hat p}^\dagger
    {\hat r}^\dagger {\hat s}{\hat q}} - \sum_{pqr}{(pr\vert \vert rq)
    {\hat p}^\dagger {\hat q}} \, ,
  \label{eq:MBPT.29}
\end{equation}
\begin{equation}
  {\hat M}_{xc} = \sum_{pq}{ (p \vert {\hat \Sigma}_x^{HF} - {\hat
      v}_{xc} \vert q) {\hat p}^{\dagger} {\hat q}} \, ,
  \label{eq:MBPT.30} 
\end{equation}
and ${\hat \Sigma}_x^{HF}$ is the HF exchange operator defined in terms 
of the occupied Kohn-Sham orbitals and the integral of Eq.~\ref{eq:MBPT.30b}.
{\em Heuristically} this will give us a series of diagrams which we must resum 
to have the proper analytic structure of the exact PP so we can take 
advantage of this analytic structure to produce the desired ``Casida-like'' 
equation.  {\em Rigorously} we actually first begin with some exact equations
in the superoperator equation-of-motion (EOM) formalism to deduce the
analytic structure of the PP.  This exact structure is then developed
in a perturbation expansion so that we can perform an order analysis of each
of the terms entering into basic ``Casida-like'' equation.  As we shall see,
not every diagram is generated by this procedure, either because they are not
needed or because of approximations which we have chosen to make.

Our MBPT expansions are in terms of the bare electron repulsion 
[or more exactly the ``fluctuation potential'' Eq.~(\ref{eq:MBPT.29})], 
rather than the screened interaction used in solid-state 
physics~\cite{ORR02,RSB+09}.
The main advantage of working with the bare interaction is a balanced treatment of direct 
and exchange diagrams, which is especially important for treating two- and higher-electron 
excitations.  While we will 
automatically include what the solid state community refers to as vertex effects, the 
disadvantage of our approach is that it is likely to break down in solids when screening 
becomes important.  The specific approach we will take is the now well-established
second-order polarization propagator approximation (SOPPA) of Nielsen, J{\o}rgensen, 
and Oddershede~\cite{OJ77,NJO80,NJO81,JS81}.  The usual presentation of the SOPPA approach 
is based upon the superoperator equation-of-motion (EOM) approach previously used by 
one of us~\cite{C05}. However the SOPPA approach is very similar in many ways to the 
second-order algebraic diagrammatic construction [ADC(2)] approach of 
Schirmer~\cite{S82,TSS99} and we will not hesitate to refer to this approach as needed 
(particularly with regard to the inclusion of various diagrammatic contributions.)  
The only thing really new here is the change from a Hartree-Fock to a Kohn-Sham zero-order 
picture and the concomitant inclusion of (many) additional terms.  Nevertheless it will 
be seen that the final working expressions are fairly compact.

\begin{figure}
  \includegraphics[width=0.75 \columnwidth]{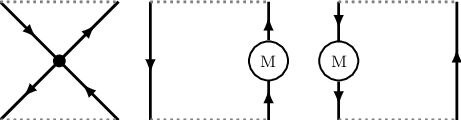}
  \caption{Topologically different first-order time-unordered Abrikosov diagrams for 
    the PP.
   \label{fig:tdia}
    }
\end{figure}
Before going into the details of the superoperator EOM approach, let us anticipate some
of the results by looking at some of the diagrams which emerge from this analysis.
We have seen [Eq.~(\ref{eq:MBPT.10})] that the PP is just the restriction of the ph-propagator
to two, rather than four, times.  Thus heuristically it suffices to take the ph-propagator diagrams,
fix two times, and then take all possible time orderings.
Defining order as the order in the number of times $\hat{V}$ and/or ${\hat M}_{xc}$ 
appear, then all of the time-unordered first-order terms are shown in Fig.~\ref{fig:tdia}.
Fixing two times and restricting ourselves to an exchange-only theory
gives the 14 time-ordered diagrams shown in Fig.~\ref{fig:SOPPA1}.
As we shall see below in a very precise mathematical way, dangling parts below or above
the horizontal dotted lines correspond respectively to Hugenholtz diagrams for 
initial-time and final-time perturbed wavefunctions.  
(Two other first-order Goldstone diagrams are found in Ref.~\cite{S82} 
with the electron repulsion dot above or below the two dotted lines, however
a more detailed analysis shows that these terms neatly cancel out in the 
final analysis.)
The area between the dotted lines corresponds to time propagation.  In this case,
there are only one-hole/one-particle excitations between the two horizontal dotted lines.
Our final results are in perfect agreement with diagrams appearing in the exact exchange
(EXX) theory as obtained by Hirata {\em et al.}~\cite{HIBG05} which are equivalent to the more
condensed form given by G\"orling~\cite{G98}.

\begin{figure}
\includegraphics[width=0.75 \columnwidth]{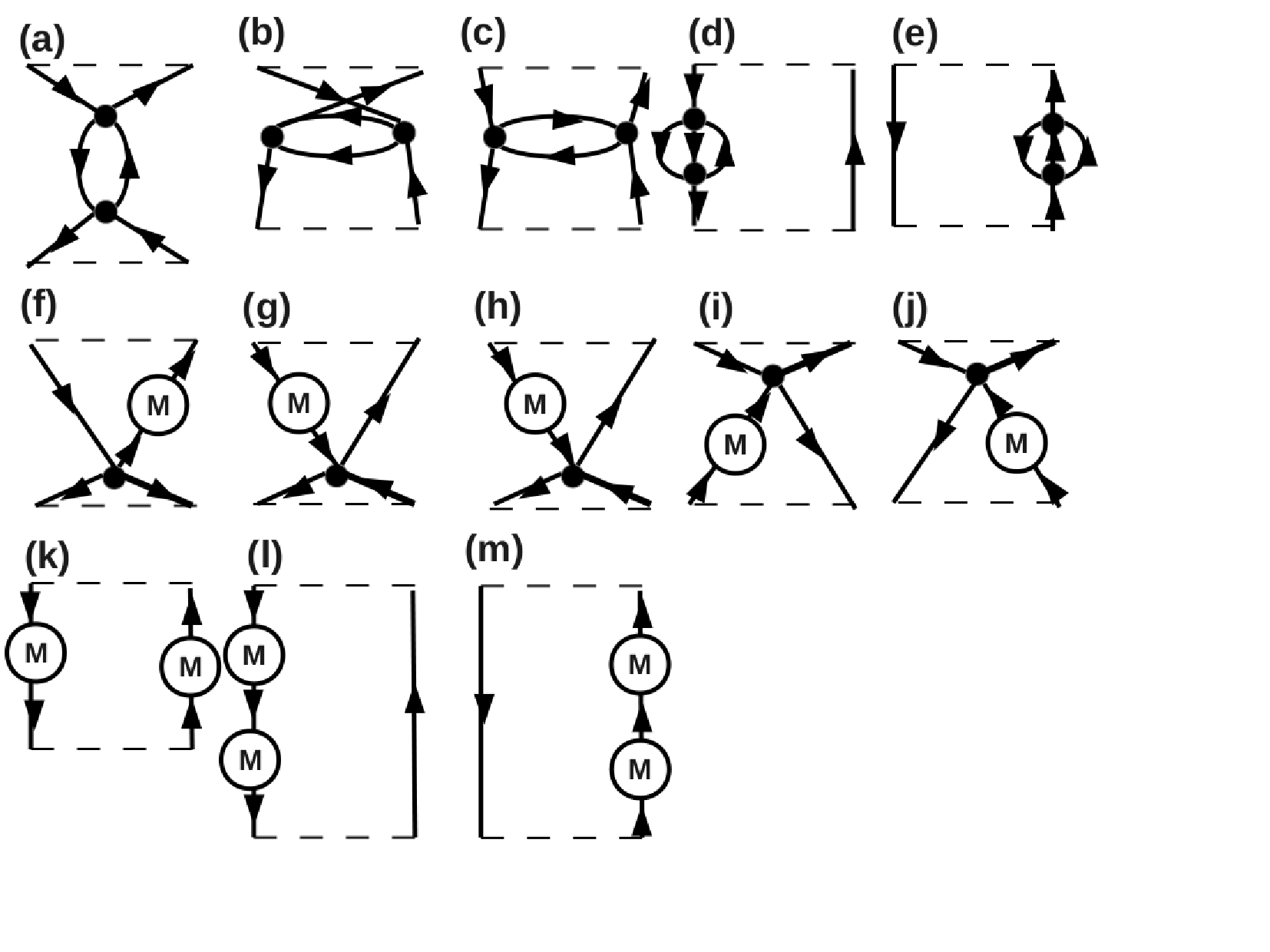}
\caption{Second-order time-unordered Abrikosov PP diagrams.  Not all of the
  time-ordered Hugenholtz diagrams are generated by our procedure---only about 140 
  Hugenholtz diagrams.
  \label{fig:secondorderAbrikosov}
  }
\end{figure}
\begin{figure}
\includegraphics[width=0.5 \columnwidth]{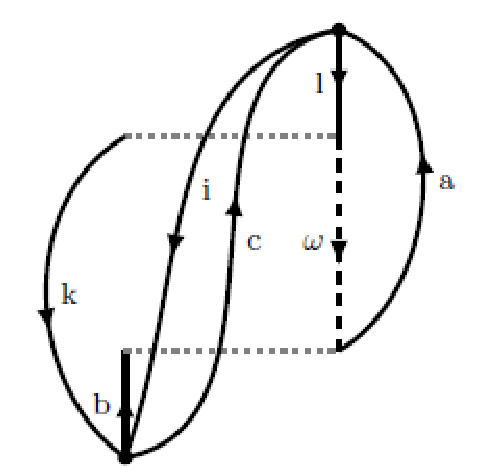}
\caption{An example of a second-order time-ordered Hugenholtz PP diagrams.  
  \label{fig:secondorderHugenholtz}
  }
\end{figure}

Figure~\ref{fig:secondorderAbrikosov} shows all 13 second-order time-unordered diagrams.
While this may not seem like very many, our procedure generates about 140 
time-ordered Hugenholtz diagrams (and even more Feynman diagrams).
A typical time-ordered Hugenholtz diagram is shown
in Fig.~\ref{fig:secondorderHugenholtz}.  The corresponding equation, 
\begin{equation}
   \Pi_{sr,qp}^{diag}(\omega) = \sum_{a,b,c,i,k,l} 
     \frac{(pq\vert \vert ba)(kl \vert \vert rs)}
      {\epsilon_{ik,bc}(\omega-\epsilon_{ik,ca})\epsilon_{il,ac}} \, ,
\end{equation}
shows that this diagrams has poles at the double excitations $\epsilon_{ik,ca}$.
Thus we see that the polarization propagator does have poles at double
excitations, but we are not really ready to do calculations yet.  There
are two main reasons: (i) we need a more sophisticated formalism which will
allow the single and double excitations to mix with each other and (ii)
we would like a (pseudo)eigenvalue equation to solve.  Thus we still have 
to do quite a bit more work to arrive at a ``Casida-like'' equation with 
explicit double excitations, but the basic idea is already present in what
we have done so far.

To do so, it is first convenient to express the PP in a molecular orbital basis
as,
\begin{eqnarray}
  &&\Pi(1,2,3,4;t-t') = \nonumber \\
  &&\sum_{pqrs}
  {\Pi_{sr,qp}(t-t') \psi_r(1)\psi^*_s(2)\psi_q^*(3)\psi_p(4)} \, ,
  \label{eq:MBPT.31} 
\end{eqnarray}
where
\begin{eqnarray}
  -\Pi_{sr,qp}(t-t') &=&  i\theta(t-t') \langle0\vert {\hat
  r}_H^\dagger(t) {\hat s}_H(t) {\hat q}_H^\dagger(t') {\hat
  p}_H(t') \vert0\rangle  \nonumber \\
  &+&i
   \theta(t'-t) \langle0\vert {\hat q}_H^\dagger(t') {\hat p}_H(t')
  {\hat r}_H^\dagger(t) {\hat s}_H(t) \vert0\rangle \, .
  \label{eq:MBPT.32} 
\end{eqnarray}
As explained in Ref.~\cite{FW71}, this change of convention with respect to 
that of Eq.~(\ref{eq:MBPT.11}) turns out to be more convenient.  Also 
note that, since the PP depends only upon the time difference, $t-t'$, we 
can shift the origin of the time-scale so that $t'=0$ without lose of 
generality.  

Equation~(\ref{eq:MBPT.32}) can be more easily manipulated by making use 
of the superoperator formalism.  A (Liouville-space) superoperator 
${\breve X}$ is defined by its action on an (Hilbert-space) operator 
${\hat A}$ as
\begin{equation}
  {\breve X}{\hat A} = [{\hat X},{\hat A}] = {\hat X}{\hat A} - {\hat
    A}{\hat X} \, .
  \label{eq:MBPT.33} 
\end{equation}
When $\breve X$ is the Hamiltonian operator, $\breve H$, one often speaks 
of the Liouvillian.  An exception is the identity superoperator, 
${\breve 1}$, whose action is simply given by,
\begin{equation}
  {\breve 1}{\hat A} = {\hat A} \, .
  \label{eq:MBPT.34} 
\end{equation}
The Heisenberg form of orbital creation and annihilation operators is 
easily expressed in terms of the Liouvillian superoperator,
\begin{equation}
  {\hat p}_H(t) = e^{i{\hat H}t} {\hat p} e^{-i{\hat H}t} =
  e^{i{\breve H}t}{\hat p} \, .
  \label{eq:MBPT.35} 
\end{equation}
Then
\begin{eqnarray}
   -\Pi_{sr,qp}(t) &=& i\theta(t) \langle0\vert \left[ e^{i{\breve
         H}t} \left( {\hat r}^\dagger {\hat s} \right) \right] {\hat
     q}^\dagger {\hat p} \vert0\rangle \nonumber \\ &+&i \theta(-t)
   \langle0\vert {\hat q}^\dagger {\hat p} \left[ e^{i{\breve H}t}
     \left( {\hat r}^\dagger {\hat s} \right) \right] \vert0\rangle \,
  \label{eq:MBPT.36} 
\end{eqnarray}
Taking the Fourier transform [with appropriate convergence factors (not 
shown)] gives,
\begin{equation}
  -\Pi_{sr,qp}(\omega) = ( {\hat p}^\dagger{\hat q} \vert
  (\omega{\breve 1}+{\breve H})^{-1} \vert {\hat r}^\dagger{\hat s} )
  \, ,
  \label{eq:MBPT.37} 
\end{equation}
where we have introduced the superoperator metric\footnote{Technically this
is not a metric, because the overlap matrix is symplectic rather than 
positive definite.  Howevever we will call it a metric as it can be used 
in much the same way as a true metric.},
\begin{equation}
  ({\hat A} \vert {\breve X} \vert {\hat B}) = \langle 0 \vert [{\hat
      A^\dagger},[{\hat X},{\hat B}]] \vert 0 \rangle \, .
  \label{eq:MBPT.38} 
\end{equation}
[It may be useful to note that,
\begin{equation}
  -\Pi_{sr,qp}(\omega) = \Pi_{rs,pq}(\omega) \, ,
  \label{eq:MBPT.39a} 
\end{equation}
follows as an easy consequence of the above definitions.  Moreover 
since we typically use real orbitals and a finite basis set, the PP 
is a real symmetric matrix.  This allows us 
to simply identify $\Pi$ as the superoperator resolvant,
\begin{equation}
  \Pi_{pq,rs}(\omega) = ( {\hat p}^\dagger{\hat q} \vert
  (\omega{\breve 1}+{\breve H})^{-1} \vert {\hat r}^\dagger{\hat s} )
  \, \text{.]}
  \label{eq:MBPT.39} 
\end{equation}

Since matrix elements of a resolvant superoperator are harder to manipulate than resolvants of a 
superoperator matrix, we will transform Eq.~(\ref{eq:MBPT.37}) into the later form by introducing 
a complete set of excitation operators.  The complete set
\begin{equation}
  \{ {\bf T}^\dagger \} = \{ {\bf T}^\dagger_1\ ;\ {\bf
    T}^\dagger_2\ ;\ ... \} = \{ {\hat a}^\dagger {\hat
    i}^{\vphantom{\dagger}}\ ,\ {\hat i}^\dagger {\hat
    a}^{\vphantom{\dagger}}\ ;\ {\hat a}^\dagger {\hat
    i}^{\vphantom{\dagger}}{\hat b}^{\dagger} {\hat
    j}^{\vphantom{\dagger}}\ ,\ {\hat i}^\dagger {\hat
    a}^{\vphantom{\dagger}}{\hat j}^{\dagger} {\hat
    b}^{\vphantom{\dagger}}\ ; \ ... \} \, ,
  \label{eq:MBPT.40} 
\end{equation}
leads to the resolution of the identity (RI),
\begin{equation}
  {\bf \breve 1} =  { \vert{\bf T}^\dagger )({\bf
      T}^\dagger\vert{\bf T}^\dagger)^{-1}({\bf T}^\dagger\vert} \, .
  \label{eq:MBPT.41} 
\end{equation}
We have defined the operator space differently from the previous work of 
one of us~\cite{C95} to be more consistent with the literature on the 
field of PP calculations. The difference is actually the commutation of 
two operators which introduces one sign change.
Insertion into Eq.~(\ref{eq:MBPT.37}) and use of the relation,
\begin{equation}
  ({\bf T}^\dagger \vert (\omega{\breve 1}+{\breve H})^{-1} \vert {\bf
    T}^\dagger ) = ({\bf T}^\dagger\vert{\bf T}^\dagger)({\bf
    T}^\dagger\vert\omega{\breve 1}+{\breve H} \vert{\bf
    T}^\dagger)^{-1}({\bf T}^\dagger\vert{\bf T}^\dagger)
  \label{eq:MBPT.42} 
\end{equation}
then gives,
\begin{equation}
  -\Pi_{sr,qp}(\omega) = ({\hat p}^\dagger{\hat
    q}^{\vphantom{\dagger}}\vert {\bf T}^\dagger)({\bf T}^\dagger
  \vert \omega{\breve 1}+{\breve H} \vert {\bf T}^\dagger)^{-1}({\bf
    T}^\dagger \vert {\hat r}^\dagger{\hat s}^{\vphantom{\dagger}}) \,
  .
  \label{eq:MBPT.43} 
\end{equation}
This shows us the analytical form of the exact polarization propagator.

The corresponding ``Casida-like'' pseudoeigenvalue equation is,
\begin{equation}
  ({\bm T}^\dagger \vert \breve{H} \vert {\bm T}^\dagger) \vec{Z}_I
  =  \omega_I ({\bm T}^\dagger \vert {\bm T}^\dagger) \vec{Z}_I \, ,
  \label{eq:MBPT.44}
\end{equation}
with normalization,
\begin{equation}
  \vec{Z}^\dagger_I ({\bm T}^\dagger \vert {\bm T}^\dagger) \vec{Z}_J = \delta_{I,J} \, .
  \label{eq:MBPT.45}
\end{equation}
Let us also seek a sum-over-states expression for the polarization
propagator.  








Spectral expansion tells us that,
\begin{equation}
  {\bm \Gamma}(\omega) = \omega ({\bm T}^\dagger \vert {\bm T}^\dagger)
        + ({\bm T}^\dagger \vert \breve{H} \vert {\bm T}^\dagger)
   = \sum_I ({\bm T}^\dagger \vert {\bm T}^\dagger) \vec{Z}_I 
     \left( \omega + \omega_I \right) \vec{Z}_I^\dagger 
     ({\bm T}^\dagger \vert {\bm T}^\dagger) \, ,
    \label{eq:MBPT.47}
\end{equation}
and,
\begin{equation}
  {\bm \Gamma}^{-1}(\omega) = \left[ \omega ({\bm T}^\dagger \vert {\bm T}^\dagger)
        + ({\bm T}^\dagger \vert \breve{H} \vert {\bm T}^\dagger) \right]^{-1}
   = \sum_I \vec{Z}_I \left( \omega + \omega_I \right)^{-1} \vec{Z}_I^\dagger 
    \, .
    \label{eq:MBPT.48}
\end{equation}
So Eq.~(\ref{eq:MBPT.43}) reads,
\begin{equation}
  -\Pi_{sr,qp}(\omega) = \sum_I ({\hat p}^\dagger{\hat
    q}^{\vphantom{\dagger}}\vert {\bf T}^\dagger) \vec{Z}_I \left( \omega + \omega_I \right)^{-1} \vec{Z}_I^\dagger 
   ({\bf
    T}^\dagger \vert {\hat r}^\dagger{\hat s}^{\vphantom{\dagger}}) \, .
  \label{eq:MBPT.49}
\end{equation}
This means that the PP has poles given at the pseudoeigenvalues of 
Eq.~(\ref{eq:MBPT.44}) and that the eigenvectors may be used to calculate
oscillator strengths via Eq.~(\ref{eq:MBPT.49}).

As the ``Casida-like'' equation [Eq.~(\ref{eq:MBPT.44})] is so important,
let us rewrite it as, 
\begin{equation}
  \left[ \begin{array}{cc} {\bm A} & {\bm B} \\
         {\bm B}^* &  {\bm A}^* \end{array} \right]
  \left( \begin{array}{c} \vec{X} \\ \vec{Y} \end{array} \right) = \omega
   \left[ \begin{array}{cc} {\bm S}_{A,A} & {\bm S}_{A,B} \\
         {\bm S}_{B,A} &  {\bm S}_{B,B} \end{array} \right]
  \left( \begin{array}{c} \vec{X} \\ \vec{Y} \end{array} \right) \, ,
  \label{eq:MBPT.50}
\end{equation}
which is roughly,
\begin{equation}
  \left[ \begin{array}{cc} {\bm A} & {\bm B} \\
         {\bm B}^* &  {\bm A}^* \end{array} \right]
  \left( \begin{array}{c} \vec{X} \\ \vec{Y} \end{array} \right) = \omega
   \left[ \begin{array}{cc} {\bm 1} & {\bm 0} \\
         {\bm 0} &  -{\bm 1} \end{array} \right]
  \left( \begin{array}{c} \vec{X} \\ \vec{Y} \end{array} \right) \, ,
  \label{eq:MBPT.51}
\end{equation}
The ${\bm A}$ and ${\bm B}$ matrices, as well as the $\vec{X}$ and
$\vec{Y}$ partition according to whether they refer to one-electron
excitations or two-electron excitations.   In the Tamm-Dancoff
approximation the ${\bm B}$ matrices are neglected so we can write,
\begin{equation}
  \left[ \begin{array}{cc} {\bm A}_{1,1}^{(0+1+2)} & {\bm A}_{1,2}^{(1)} \\
         {\bm A}_{2,1}^{(1)} & {\bm A}_{2,2} \end{array} \right]
  \left( \begin{array}{c} \vec{C}_1 \\ \vec{C}_2 \end{array} \right)
  = \omega \left( \begin{array}{c} \vec{C}_1 \\ \vec{C}_2 \end{array} \right)
  \label{eq:MBPT.52}
\end{equation}
Here $\vec{X}$ has been replaced by $\vec{C}$ as is traditional and to 
reflect the normalization $\vec{C}^\dagger \vec{C} = 1$.

The superscripts in Eq.~(\ref{eq:MBPT.53}) reflect a somewhat difficult
order analysis which is carried out in the Appendix.  
This analysis consists of expanding the polarization propagator algebraically
and then matching each term to a set of diagrams to see what order of 
each EOM matrix is needed to get a given order of polarization propagator.

The result in the case of the ${\bm A}$ matrices is,
\begin{eqnarray}
  \left( {\bm A}_{1,1}^{(0+1+2)}\right)_{kc,ia} & = & \delta_{i,k} F^{(0+1+2)}_{a,c} - \delta_{a,c} F^{(0+1+2)}_{i,k}
          + (ai \vert \vert kc) \nonumber \\
  \left( {\bm A}_{2,1}^{(1)} \right)_{kc,jbia} & = & -\delta_{i,k} (bc \vert \vert aj ) + \delta_{j,k} (bc \vert \vert ai )
         \nonumber \\
         & - & \delta_{b,c} (ai \vert \vert kj) + \delta_{k,j} (bi \vert \vert kj) \nonumber \\
  \left( {\bm A}_{2,2}^{(0)} \right)_{ldkc,jbia} & = & \delta_{i,k} \delta_{c,a} \delta_{d,b} 
        \epsilon_{ab,ij}
  \, , 
  \label{eq:MBPT.53} 
\end{eqnarray}
where $ F^{(0+1)}_{r,s}=\delta_{r,s} \epsilon_r +  M^{xc}_{r,s}$ is the matrix of the Hartree-Fock operator constructed with Kohn-Sham orbitals and
\begin{eqnarray}
  F^{(0+1+2)}_{a,c} & = & F^{(0+1)}_{a,c} 
                   + \sum_l \frac{M_{l,a} M_{l,c}}{\epsilon_{l,a}}
   \nonumber \\
     & - & \frac{1}{2} \sum_{l,m,d} \frac{(ld \vert \vert mc) (dl\vert \vert am)} {\epsilon_{lm,ad}}  \nonumber \\
  F^{(0+1+2)}_{i,k} & = & F^{(0+1)}_{i,k} + \sum_d \frac{M_{k,d} M_{d,i}}{\epsilon_{i,d}} 
     \nonumber \\
     & - & \frac{1}{2} \sum_{l,d,e} \frac{(le\vert \vert kd) (dl \vert \vert ei)}{\epsilon_{im,de}} \, ,
  \label{eq:MBPT.54} 
\end{eqnarray}
include second-order corrections.  (Note that extra factors of 1/2 will occur 
in these expressions when spin is taken explicitly into account.)  
In practice a zero-order approximation to ${\bm A}_{2,2}$ is insufficient
and we must use an expression correct through first order,
\begin{eqnarray}
  \left( {\bm A}_{2,2}^{(0+1)} \right)_{aibj,ckdl} & = & 
  \delta_{i,k} \delta_{j,l} \left(\delta_{a,c} F^{(0+1)}_{b,d} + 
  \delta_{b,d} F_{a,c}^{(0+1)} \right) 
   -\delta_{a,c}\delta_{b,d} \left( \delta_{j,l} F_{i,k}^{(0+1)}
   -\delta_{i,k} F_{d,l}^{(0+1)} \right) \nonumber \\
  & -& \delta_{a,c} f_{i,j,k,l}(b,d) - \delta_{b,d}f_{i,j,k,l}(a,c)
   + \delta_{a,d}f_{i,j,k,l}(b,c) + \delta_{b,c}f_{i,j,k,l}(a,d)
   \nonumber \\
  & - & \delta_{a,c} \delta_{b,d}(kj \vert \vert li) 
  - \delta_{j,l} \delta_{k,i}(ad \vert \vert bc) \, ,
  \label{eq:MBPT.55}
\end{eqnarray}
where,
\begin{equation}
  f_{i,j,k,l}(p,q) = \delta_{i,k} (lj\vert \vert pq) + \delta_{j,l} (ki \vert \vert pq)
         - \delta_{k,j} (li\vert \vert pq) - \delta_{i,l} (kj \vert \vert pq)
  \, .
  \label{eq:MBPT.56}
\end{equation}

We will refer to the resultant method as extended SOPPA/ADC(2).
It is 
immediately seen that truncating to first order recovers the usual 
configuration interaction singles (CIS) equations in a noncanonical basis set.
We now have the essential tools to proceed with the rest of this chapter.

\section{Dressed LR-TD-DFT}
\label{sec:dressed}

We now give one answer to the problem raised in the introduction of how to 
include explicit double excitations in LR-TD-DFT.   This answer goes by 
the name dressed LR-TD-DFT and consists of a hybrid MBPT/AA LR-TD-DFT
method.  We will first give the basic idea and comment on some of the
early developments.  We will then go into the practical details which
are needed to make a useful implementation of dressed LR-TD-DFT.  Finally
we will introduce the notion of Brillouin corrections which are undoubtedly
important for photochemistry.

\subsection{Basic Idea}

\begin{figure}
      \includegraphics[width=0.75 \columnwidth]{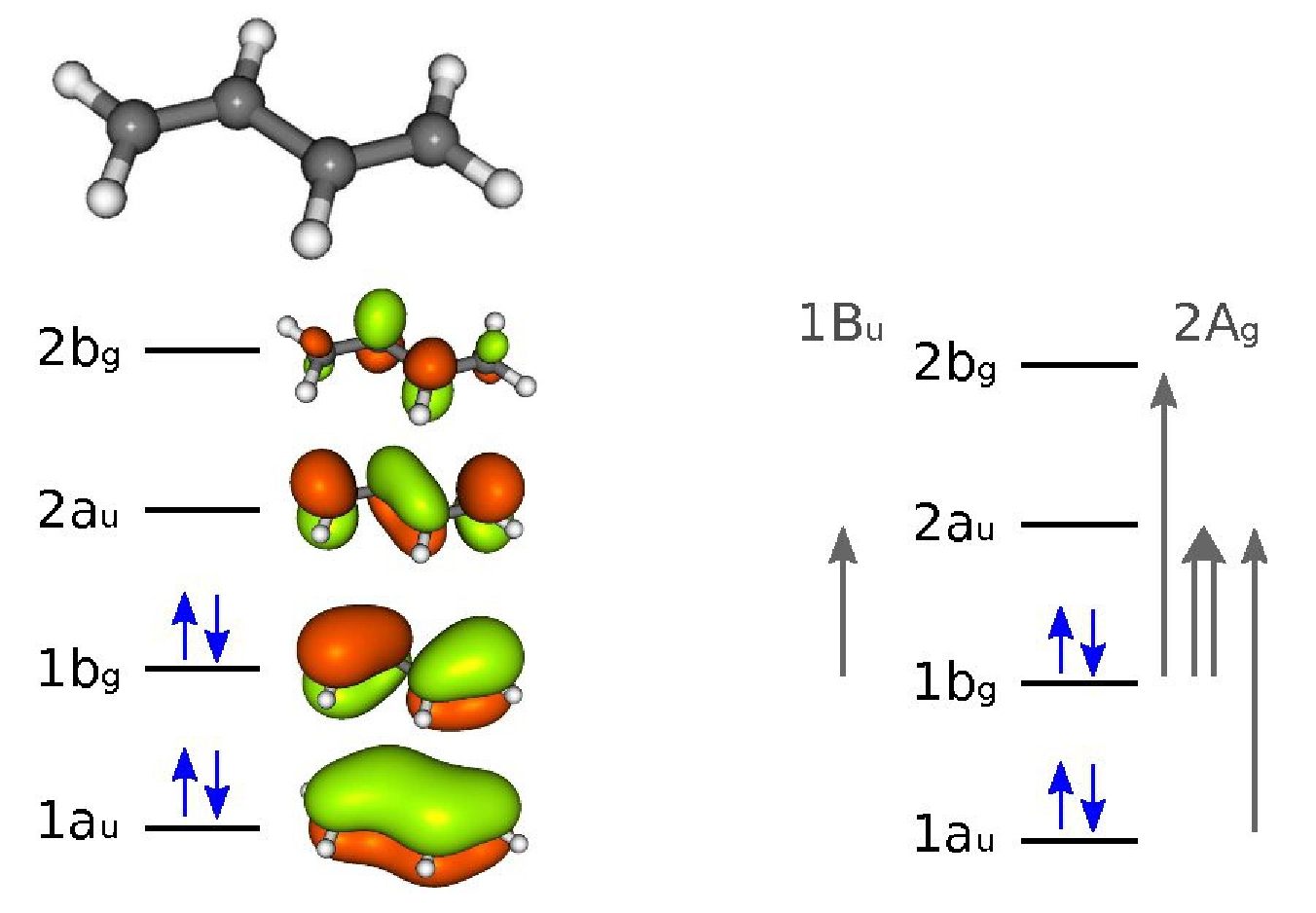}
  \caption{Doubles contribution to the $^1A_g$ excited state of butadiene.
  Since the obvious two lowest singly-excited singlets $^1(1b_g,2b_g)$ and
  $^1(1a_u,2a_u)$ are quasidegenerate in energy, they mix to form new 
  singly-excited singlets $(1/\sqrt(2))[^1(1b_g,2b_g) \pm ^1(1a_u,2a_u)]$.
  One of these is quasidegenerate with the doubly-excited singlet dark state
  $^1(1b_g^2,2a_u^2)$.  The resultant mixing modifies the energy and intensity
  of the observed $^1A_g$ excited state.
  \label{fig:butadiene}
       }
\end{figure}
As emphasized in Sec.~\ref{sec:review}, simple counting arguments show that
the AA limits LR-TD-DFT to single excitations, albeit dressed to include some
electron correlation.  However explicit double excitations are sometimes 
needed when describing excited states.  This was discussed in the introduction
in the context of photochemistry (Fig.~\ref{fig:ethylene_singlet_photochem}).
It is well-known in {\em ab initio} quantum chemistry that double excitations 
can be important when describing vertical excitations and the best known 
example is briefly discussed in the caption of Fig.~\ref{fig:butadiene}.  

\begin{figure}
      \includegraphics[width=0.5 \columnwidth]{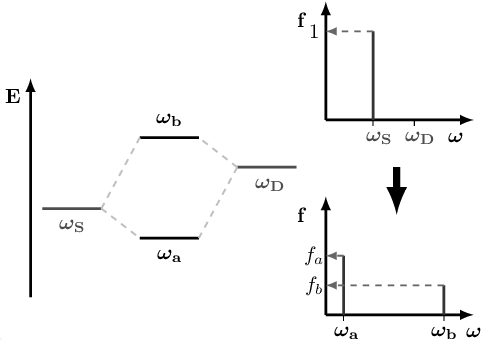}
  \caption{Two-level model used by Maitra {\em et al}. in their {\em
      heuristic} derivation of dressed TDDFT.  See explanation in text.
      \label{fig:BurkeModel}
      }
\end{figure}
At first this may seem a little perplexing because the fact that the 
oscillator strength is the transition matrix element of a one-electron operator
[Eq.~(\ref{eq:review.14})] means that the oscillator strength of a double
excitation relative to a single-determinantal ground-state wavefunction
should be zero---that is, the doubly excited state should be spectroscopically
dark.  What happens is easily explained by the two-level model shown in
Fig.~\ref{fig:BurkeModel} which is sufficient to give a first
explanation of the butadiene case for example.  (In the butadiene case, the
singly-excited state to be used is already a mixture of two different 
one-hole/one-particle states.)
Figure~\ref{fig:BurkeModel} shows a bright singly-excited state with 
excitation energy $\omega_S$ and oscillator strength $f_S=1$ interacting 
with a dark doubly-excited state with excitation energy $\omega_D$ and 
oscillator strength $f_D=0$ via a coupling matrix element $x$. 
The CI problem is simply,
\begin{equation}
  \left[ \begin{array}{cc} \omega_S & x \\ x & \omega_D \end{array}
    \right] \left( \begin{array}{c} C_S \\ C_D \end{array} \right) =
  \omega \left( \begin{array}{c} C_S \\ C_D \end{array} \right) \, ,
  \label{eq:dressed.1}
\end{equation}
which can be formally solved obtaining
\begin{eqnarray}
  \omega_S & = & \omega_a \cos^2 \theta + \omega_b \sin^2 \theta
  \nonumber \\ \omega_D & = & \omega_a \sin^2 \theta + \omega_b \cos^2
  \theta \, ,
  \label{eq:dressed.2}
\end{eqnarray}
for some value of $\theta$.
Notice that the average excitation energy is 
conserved in the coupled problem ($\omega_a + \omega_b = \omega_S + \omega_D$) 
and that something similar occurs with the oscillator strengths. This leads to 
the common interpretation that the coupling ``shatters the singly-excited 
peaks into two satellite peaks.'' 

Now let us see how this wavefunction theory compares with LR-TD-DFT and
how Maitra {\em et al}.~\cite{MZCB04} decided to combine the two into a 
hybrid method.  Of course, the proper comparison with CI is LR-TD-DFT
within the TDA.
Applying the partitioning technique to Eq.~(\ref{eq:dressed.1}), we obtain
\begin{equation}
  \left( \omega_S + \frac{x^2}{\omega - \omega_D} \right) C_S = \omega C_S \, .
  \label{eq:dressed.4}
\end{equation}
Comparing with the diagonal TDA LR-TD-DFT within the two-orbital model,
\begin{equation}
  \omega = \epsilon_{a,i} + ( ia \vert f_{Hxc}(\omega) \vert ia )  \, ,
  \label{eq:dressed.5}
\end{equation}
shows that,
\begin{equation}
  ( ia \vert f_{Hxc}(\omega) \vert ia ) = \left( \omega_S -
  \epsilon_{a,i} \right) + \frac{x^2}{\omega - \omega_D} \, .
  \label{eq:dressed.6}
\end{equation}
Maitra {\em et al}.\ \cite{MZCB04} interpreted the first term as the adiabatic part,
\begin{equation}
  f_{Hxc}^{\text{AA}} = \omega_S - \epsilon_{a,i} \, ,
  \label{eq:dressed.7}
\end{equation}
and second term as the nonadiabatic correction,
\begin{equation}
  f_{Hxc}^{\text{NA}}(\omega) = \frac{x^2}{\omega - \omega_D} \, .
  \label{eq:dressed.8}
\end{equation}
Additionally, it is easy to show that
\begin{equation}
  x^2 = \omega_S \omega_D - \omega_a \omega_b \, .
  \label{eq:dressed.9}
\end{equation}
which is the form of the numerator used by Maitra {\em et al}.~\cite{MZCB04}. 
The suggestion of Maitra {\em et al}., which defines dressed LR-TD-DFT,
is to calculate the nonadiabatic correction terms [Eq.~(\ref{eq:dressed.8})]
from MBPT~\cite{MZCB04}.  Thus $x$ and $\omega_D$ in Eq.~(\ref{eq:dressed.1}) 
are to be calculated using MBPT rather than using DFT.  

\subsection{Practical Details and Applications}

Applications of dressed LR-TD-DFT to the butadiene and related problems 
have proven to be very encouraging~\cite{MZCB04,CZMB04,MW09,GB09}.
Nevertheless several things were missing in these seminal papers.
In the first place, they did not always use exactly the same formalism
for dressed LR-TD-DFT and not always the same DFAs.  Moreover, while the
formalism showed encouraging results for a few molecules for those 
excitations which were thought to be most affected explicit inclusion 
of double excitations, the same references failed to show that predominantly 
single excitations were left largely unaffected by the dressing of 
AA LR-TD-DFT.  These questions were carefully addressed in Ref.~\cite{HIRC11},
with some surprising answers.

The implementation of dressed LR-TD-DFT considered in Ref.~\cite{HIRC11}
was to add just a few double excitations to AA LR-TD-DFT and solve the
TDA equation,
\begin{equation}
  \left[ \begin{array}{cc} {\bm A}_{1,1}^{(\text{AA})} & {\bm A}_{1,2}^{(1)} \\
         {\bm A}_{2,1}^{(1)} & {\bm A}_{2,2}^{(0+1)} \end{array} \right]
  \left( \begin{array}{c} \vec{C}_1 \\ \vec{C}_2 \end{array} \right)
  = \omega \left( \begin{array}{c} \vec{C}_1 \\ \vec{C}_2 \end{array} \right)
   \, .
  \label{eq:dressed.10}
\end{equation}
Thus the calculation of the ${\bm A}_{1,1}$ block which is one of the most 
difficult to calculate in the extended SOPPA/ADC(2) theory is very much
simplified by using AA LR-TD-DFT.  The ${\bm A}_{2,2}$ block must however be
calculated through first order in practice.  It was confirmed that adding
only a few (e.g., 100) double excitations led to little difference in 
calculated eigenvalues unless the double excitation were quasidegenerate
with a single excitation.  There is thus no significant problem in practice
of double counting electron correlation effects when using this hybrid
MBPT/LR-TD-DFT method.  Tests were carried out on the test set of Schreiber
{\em et al.} consisting of 28 organic chromophores with 116 well-characterized
singlet excitation energies \cite{SSST08a}.

Note that the form of Eq.~(\ref{eq:dressed.10}) was chosen instead of the
form,
\begin{eqnarray}
  \left( {\bm A}_{1,1}^{(\text{AA})} + {\bm K}_{1,1}^{\text{NA}}(\omega) \right)
   \vec{C}_1 & = & \omega \vec{C}_1 \nonumber \\
  {\bm K}_{1,1}^{\text{NA}}(\omega) & = & {\bm A}_{1,2}^{(1)}
  \left( \omega {\bm 1} - {\bm A}_{2,2}^{(0+1)} \right)^{-1} {\bm A}_{2,1}^{(1)}
  \, , \label{eq:dressed.10A}
\end{eqnarray}
for computational simplicity.  However Eq.~(\ref{eq:dressed.10A}) is
the straightforward extension of the dressed kernel given at the end of the 
previous subsection and is easy to generalize to the full response theory 
case (i.e., without making the TDA).

We confirm the previous report that using the LDA for the AA LR-TD-DFT 
part of the calculation often gives good agreement with vertical excitation 
energies having significant double excitation contributions
\cite{HHH01}. However most excitations are dominated by a singles and these
are significantly underestimated by the AA LDA.  Inclusion of double excitations
tended to decrease the typically already too low AA LDA excitation energy.
The AA LR-TD-DFT block was then modified to behave like a global hybrid 
functional with 20\% Hartree-Fock exchange.  The excitations with significant
doubles character were then found to be overestimated but the addition of
the doubles MBPT contribution again gave good agreement with benchmark
{\em ab initio} results.  This was consistent with previous experience with
dressed LR-TD-DFT \cite{MZCB04,CZMB04,MW09,GB09}. {\em The real surprise was 
the discovery that adding the MBPT to the hybrid functional made very little 
difference for the majority of excitations which are dominated by single 
excitation character.}  It thus seems that a dressed LR-TD-DFT requires
the use of hybrid functional.

\subsection{Brillouin Corrections} 

So far dressed LR-TD-DFT allows us to include explicit double excitations 
and so to describe photochemical funnels between excited states.  However
a worrisome point remains, namely how to include doubles contributions to
the ground state in the same way that we include doubles contributions to
excited states so that we may describe, for example, the photochemical funnel
between $S_1$ and $S_0$ in Fig.~\ref{fig:ethylene_singlet_photochem}.
It is not clear how to do this in LR-TD-DFT where the excited-state 
potential energy surfaces are just obtained by adding the excitation energies
at each geometry to the ground-state DFT energies.  Not only does such a 
procedure lead to the excited states inheriting the convergence difficulties 
of the ground state surface coming from places with noninteracting 
$v$-representability difficulties, but there is no coupling between
the ground state and singly excited states.  This is similar to what
happens with Brillouin's theorem in CIS calculations and leads to problems
describing conical intersections.  However adding in the missing nonzero
terms (which we call Brillouin corrections) to dressed LR-TD-DFT is easy
in the TDA.

It is good to emphasize at this point that we are making an {\em ad hoc}
correction, albeit one which is eminently reasonable from a wavefunction
point of view.  Formally correct approaches might include: (i) acknowledging
that part of the problem may lie in the fact that non-interacting 
$v$-representability in Kohn-Sham DFT often breaks down at key places
on ground-state potential energy surfaces when bonds are formed or broken, 
so that conventional Kohn-Sham DFT may no longer be a good starting point;
(ii) examining nonadiabatic xc-kernels which seem to include some degree of
multideterminantal ground-state character in their response such as that of
Maitra and Tempel \cite{MT06}; (iii) introducing explicit 
multideterminantal character into the description of the Kohn-Sham 
DFT ground state.  We will come back to this again in our final section,
but for now we will just try the {\em ad hoc} approach of adding
Brillouin corrections to TDA dressed LR-TD-DFT.  Note that this will also
have an indirect effect on interactions between excited states, though the
primary effect will be between excited states and the ground state.

\begin{figure}
\begin{tabular}{lll}
(a) Adiabatic & (b) Dressed & (c) Brillouin dressed \\
&&\\
\includegraphics[width=0.30 \columnwidth]{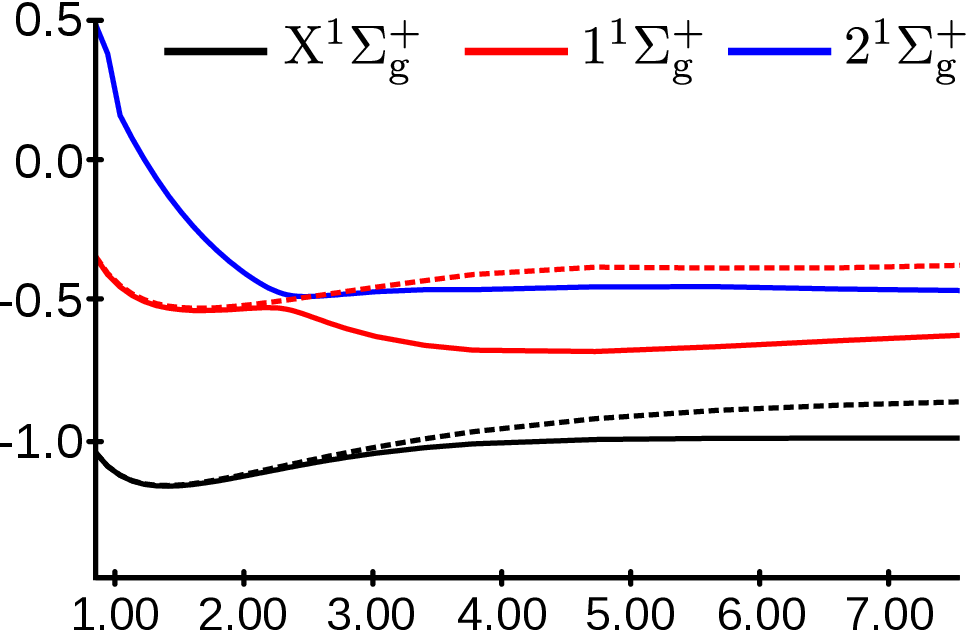}&
\includegraphics[width=0.30 \columnwidth]{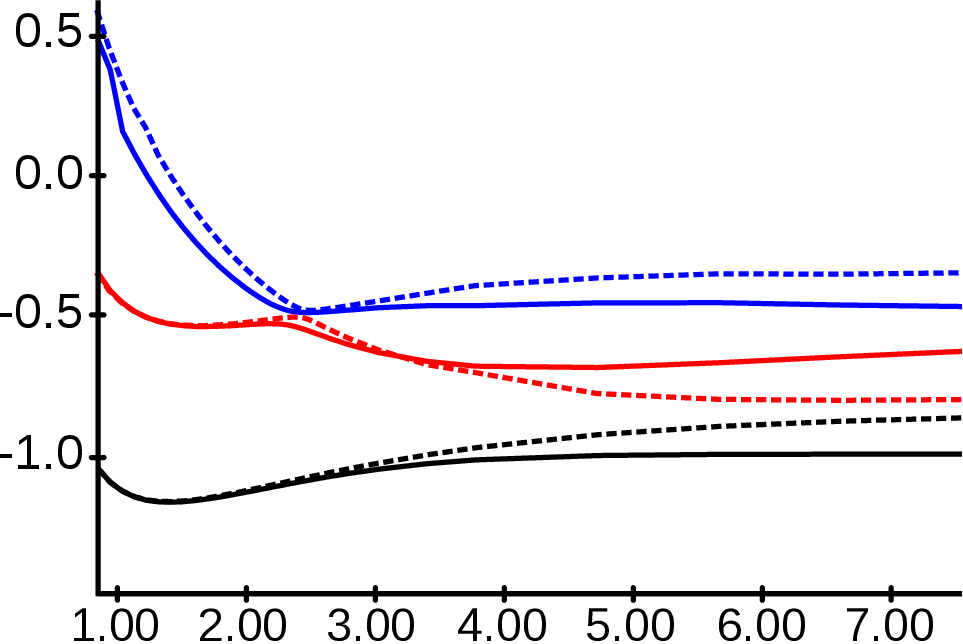}&
\includegraphics[width=0.30 \columnwidth]{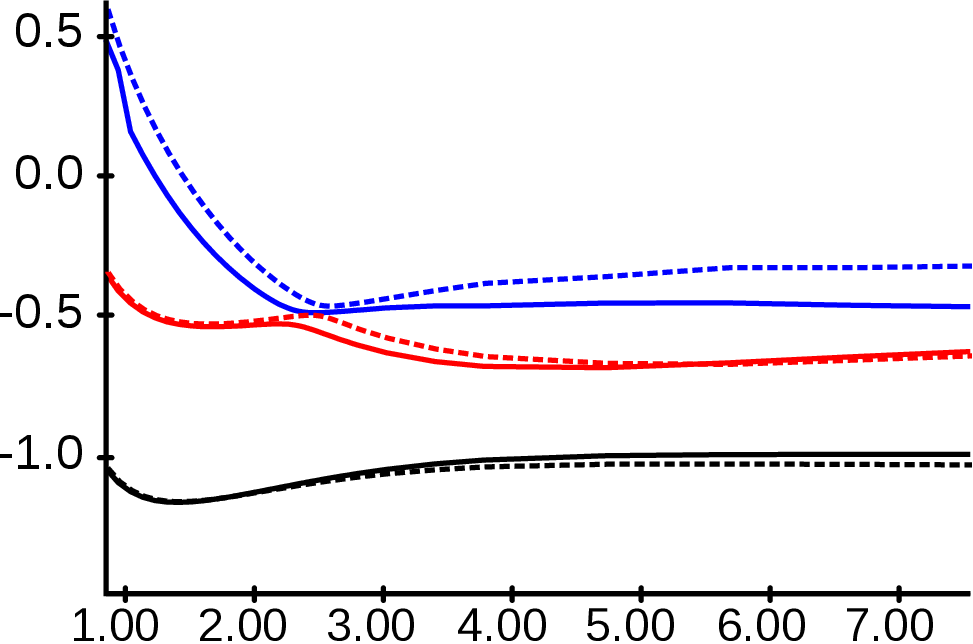}\\
\end{tabular}
\vspace{0.3cm}
\caption{Potential energy surfaces of the ground- and two-lowest 
excited states of $\Sigma_\mathrm{g}^+$ symmetry. Comparison of CISD 
(solid lines) with adiabatic, dressed and hybrid LR-TD-BH\&HLYP/TDA 
(dashed lines). All calculations have been performed with a cc-pVTZ basis set. 
All axes are in Hartree atomic units (bohr for the $x$-axis and hartree
for the $y$-axis).  Unlike the ethylene potential energy curves ({\em vide
infra}), no shift has been made in the potential energy curves.
\label{fig:h2disso}
}
\end{figure}
It suffices to add an extra column and row to the TDA problem to take into
account the ground-state determinant in hybrid DFT.  This gives,
\begin{equation}
  \left[ \begin{array}{ccc} 0 & {\bm A}_{0,1} & {\bm A}_{0,2} \\
        {\bm A}_{1,0} & {\bm A}_{1,1}^{(\text{AA})} & {\bm A}_{1,2}^{(1)} \\
        {\bm A}_{2,0} & {\bm A}_{2,1}^{(1)} & {\bm A}_{2,2}^{(0+1)} \end{array} \right]
  \left( \begin{array}{c} C_0 \\ \vec{C}_1 \\ \vec{C}_2 \end{array} \right)
  = \omega \left( \begin{array}{c} C_0 \\ \vec{C}_1 \\ \vec{C}_2 \end{array} \right)
   \, .
  \label{eq:dressed.11}
\end{equation}
where the extra matrix elements are calculated as,
\begin{equation}
  \left( {\bm A}_{0,1} \right)_{jb}  
  =  \langle j \vert \hat{M}_{xc} \vert b \rangle \, ,
  \label{eq:dressed.12}
\end{equation}
and,
\begin{equation}
  \left( {\bm A}_{0,2} \right)_{kcld}  
  =  2\left[ (kc \vert \vert ld) - (kd \vert \vert lc) \right] \, .
  \label{eq:dressed.13}
\end{equation}

Of course, we can also derive a corresponding nonadiabatic correction to the 
xc-coupling matrix,
\begin{eqnarray}
  \left( {\bm A}_{1,1}^{(\text{AA})} + {\bm K}_{1,1}^{\text{NA}}(\omega) \right)
   \vec{C}_1 & = & \omega \vec{C}_1 \nonumber \\
  {\bm K}_{1,1}^{\text{NA}}(\omega) & = & 
  \left( \begin{array}{cc} {\bm A}_{1,0} & {\bm A}_{1,2}^{(1)} \end{array} \right)
  \left[ \begin{array}{cc} \omega 1 & -{\bm A}_{0,2} \\
  -{\bm A}_{2,0} &
   \omega {\bm 1} - {\bm A}_{2,2}^{(0+1)} \end{array} \right]^{-1} 
  \left( \begin{array}{c} {\bm A}_{0,1} \\ {\bm A}_{2,1}^{(1)} 
  \end{array} \right)
  \, . \label{eq:dressed.13A}
\end{eqnarray}
The extension beyond the TDA is not obvious in this case.

\paragraph{Dissociation of molecular hydrogen}


Molecular hydrogen dissociation is a prototypical case where doubly-excited 
configurations are essential for describing the potential energy surfaces of 
the lowest-lying excited states. The three lowest singlet states of 
$\Sigma_g^+$ symmetry can be essentially described by three CI 
configurations, namely $(1\sigma_g^21\sigma_u^02\sigma_g^0)$, 
$(1\sigma_g^11\sigma_u^02\sigma_g^1)$ and 
$(1\sigma_g^01\sigma_u^22\sigma_g^0)$, referred as ground, single, and 
double configuration respectively. 

Obviously, the double configuration plays an essential role when a 
restricted single-determinant is used as reference. On the one hand, the 
mixing of ground and double configurations is necessary for describing the 
correct -1 hartree dissociation energy of H$_2$. On the other hand, the single 
and double configurations mix at around 2.3 Bohr, thus producing an 
avoided crossing. These features are shown in Fig.~\ref{fig:h2disso}, 
where we compare different flavors of TD-DFT  with the CISD 
benchmark (shown as solid lines in all graphs). 

Adiabatic TD-DFT (shown in Figure \ref{fig:h2disso}~(a)) misses completely the double configuration, and so neither the avoided crossing nor the dissociation limit are described correctly. It is noteworthy, however, that CISD and adiabatic TD-DFT curves are superimposed for states X$^1\Sigma^+_g$ and 1$^1\Sigma^+_g$ at distances lower than 2.3 bohr, where the KS assumption is fully satisfied. At distances larger than 2.3 bohr, the 1$^1\Sigma^+_g$ state corresponds to the CISD 2$^1\Sigma_g^+$ state. This is because the  1$^1\Sigma^+_g$ in TD-DFT is diabatic, as it does not contain the doubly-excited configuration. The dissociation limit is also overestimated as it is usual from RKS with common xc functionals.

Dressed TD-DFT (shown in panel b) includes the double configuration. On the one hand, the avoided crossing is represented correctly. However, the gap between the $1^1\Sigma_g^+$ and the $2^1\Sigma_g^+$ is smaller than the CISD crossing. The dissociation limit, however, is not correctly represented, as dressed TD-DFT does not include the ground- to excited-state interaction. Therefore, the double configuration dissociates at the same limit as the ground configuration.

Brioullin dressed TD-DFT (shown in panel b) includes also the ground- and double 
configuration mixture additional to the single- and double mixing of 
dressed TD-DFT. On the one hand, the avoided crossing is represented 
more precisely, with a gap closer to that of CISD. Now, the dissociation 
limit is more correctly described. Still, there is a slight error in the 
dissociation energy limit, probably due to the double counting of correlation. 
This could be alleviated by a parameterization of the Brillouin-corrected 
dressed TD-DFT functional.

\paragraph{Ethylene torsion}

 \begin{figure}
\begin{tabular}{lll}
(a) Adiabatic & (b) Dressed & (c) Brillouin dressed \\
&&\\
\includegraphics[width=0.30 \columnwidth]{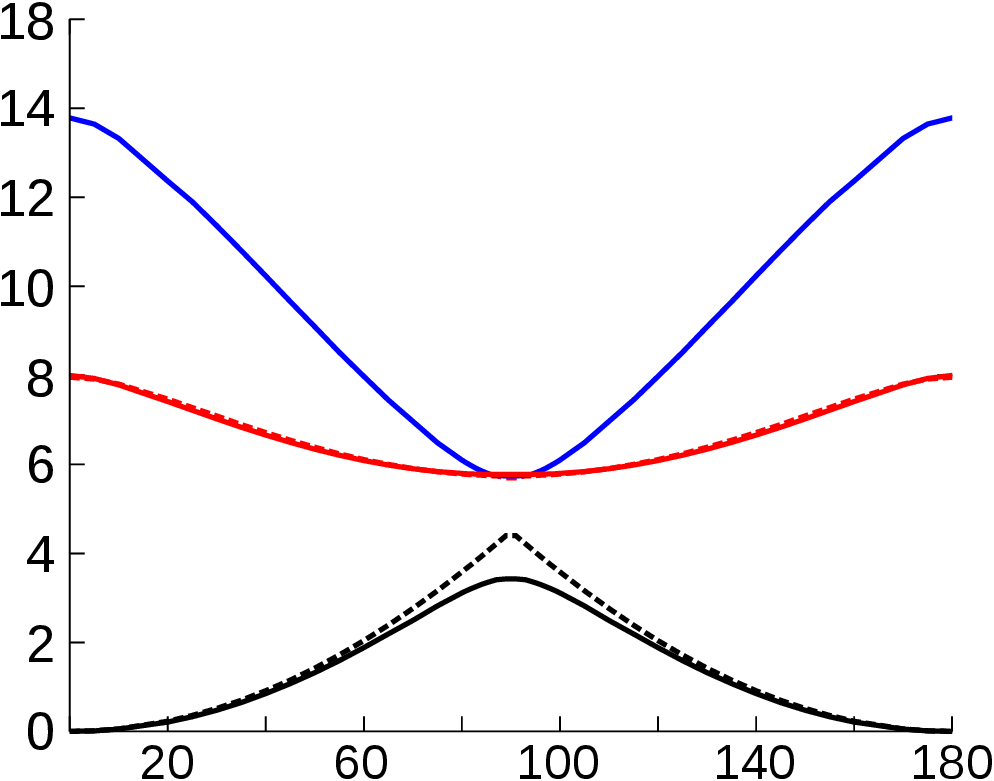}&
\includegraphics[width=0.30 \columnwidth]{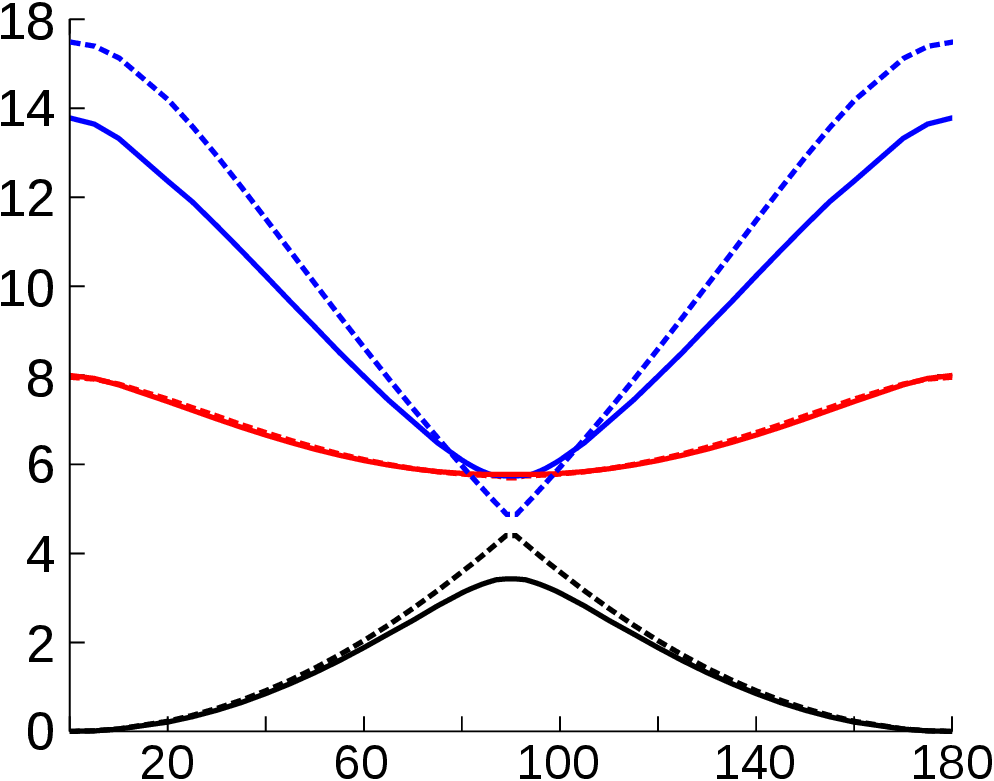}&
\includegraphics[width=0.30 \columnwidth]{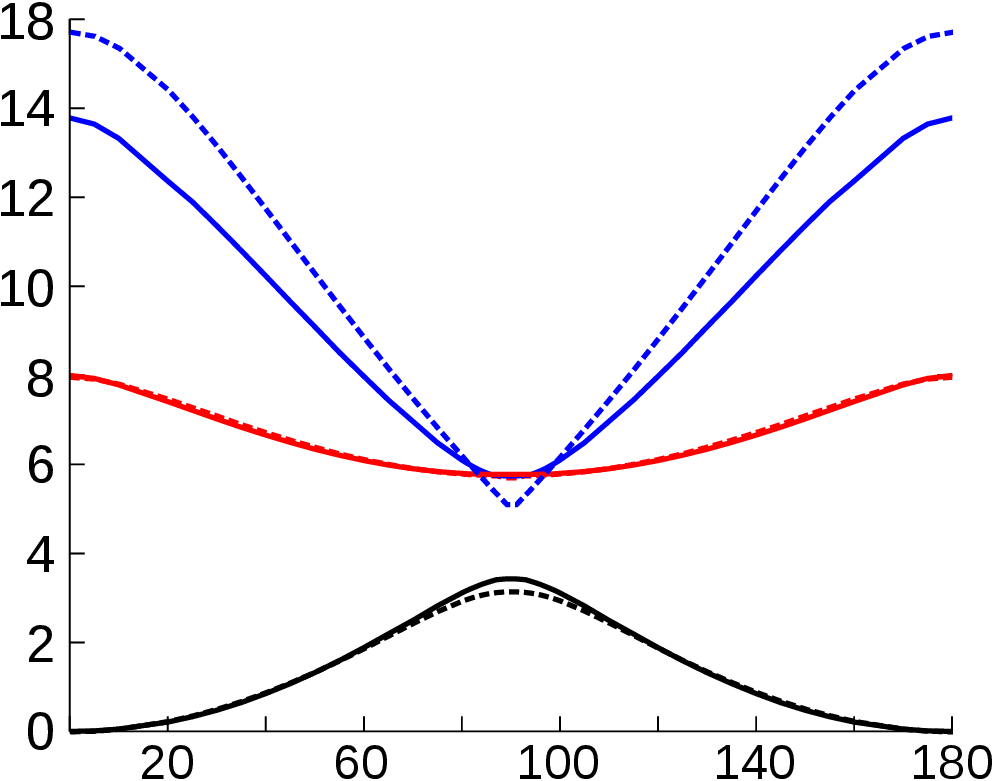}\\
\end{tabular}
\vspace{0.3cm}
   \caption{
Potential energy cuts of the S$_0$, S$_1$ and S$_2$ states of ethylene along 
the twisting coordinate: $x$-axis in degrees, $y$-axis in eV. 
All the curves have been shifted so that the ground-state curve at 0$^\circ$
corresponds to 0 eV.  The solid lines correspond to a CASSCF(2,2)/MCQDPT2 
calculation, and the dashed lines to the different models using the BH\&HLYP 
functional and the Tamm-Dancoff approximation. The 6-31++G(d,p) basis set have 
been employed in all calculations.  (Note that these curves are in good 
agreement with similar calculations previously reported in Fig.~7.3 of 
Chapter 7 of Ref.~\cite{H11}, albeit with a different functional.)
   \label{fig:ethylene}
        }
 \end{figure}

In Figure \ref{fig:ethylene}, we show the potential energy surfaces of 
S$_0$, S$_1$ and S$_2$ of ethylene along the torsional coordinate. The static 
correlation of these three states can be essentially represented by three 
configurations, namely the ground-state configuration $(\pi^2\pi^{*,0})$, 
the singly-excited configuration $(\pi^1\pi^{*,1})$ and the doubly-excited 
configuration $(\pi^0\pi^{*,2})$.

From the CASSCF(2,2)/MCQDPT2, we observe that the ground- and doubly-excited
configurations are heavily mixed at 90$^\mathrm{o}$, forming an avoided 
crossing. At this angle, the S$_1$ and S$_2$ states are degenerate. These 
features are not captured by adiabatic TD-DFT (see panel a). 
Indeed the doubly-excited configuration is missing, and so the ground-state 
features a cusp at the perpendicular conformation. The S$_1$, which is 
essentially represented by a single excitation, is virtually superimposed 
with the CASSCF(2,2)/MCQDPT2 result. The dressed TD-DFT (see panel b) 
includes the double excitation, but the surfaces of S$_0$ and S$_2$ appear 
as diabatic states, due to the fact that the ground- to excited-state coupling 
term is missing. This is largely fixed by introducing the Brillouin corrections
(see panel c). The ground-state is now in very good agreement with the 
CASSCF(2,2)/MCQDPT2 S$_0$ state, although the degeneracy of S$_1$ and S$_2$ 
at 90$^\mathrm{o}$ is still not fully captured. Thus the picture given by 
Brillouin-corrected LR-TD-DFT is qualitatively correct with respect to 
the multi-reference results.

\section{Effective Exchange-Correlation (xc) Kernel}
\label{sec:local}

We now have the tools to deduce a MBPT expression for the TD-DFT xc-kernel. 
It should be emphasized that this is not a new exercise but that we seem to be
the only ones to do so within the PP formalism.  We think this may have the
advantage of making a rather complicated subject more accessible to Quantum
Chemists already familiar with the PP formalism.

The problem of constructing xc-correlation objects such as the xc-potential
$v_{xc}$ and the xc-kernel $f_{xc}(\omega)$ from MBPT for use in 
DFT was been termed ``{\em ab initio} DFT'' by Bartlet \cite{BSB05,BB06}.  At
the exchange-only level, the terms optimized effective potential (OEP) 
\cite{TS76,T89}
or exact exchange \cite{G99,IHB99} are also used and OEP is also used to
include the correlated case \cite{C95a,C99a}.  At first glance, nothing much is
gained.  For example, the calculated excitation energies and oscillators 
strengths in {\em ab initio} TD-DFT must be, by construction, exactly the 
same as those from MBPT. Nor does this approach give explicit functionals
of the density (though it may be thought of as giving implicit functionals).
However it does allow us to formulate expressions for and to calculate purely
(TD-) DFT objects and hence can provide insight into, and computational
checks of, the behavior of such illusive objects as $v_{xc}$ and 
$f_{xc}(\omega)$.  

Here we will concentrate on the latter, namely the
xc-kernel.  Previous work along these lines has been carried out for the
kernel by directly taking the derivative of the OEP energy expression
with the constraint that the orbitals come from a local potential.  This was
first done by G\"orling in 1998 \cite{G98} for the full time-dependent 
exchange-only problem.  In 2002, Hirata {\em et al.} redid the derivation
for the static case \cite{HIGB02}.  Later, in 2006, a diagrammatic derivation of
the static result was given by Bokhan and Bartlett \cite{BB06}, and the 
functional derivative of the kernel $g_x$ has been treated by Bokhan and 
Bartlett in the static exchange-only case \cite{BB07}.  

In this section, we
will take a somewhat different and arguably more direct approach than that 
used in the aforementionned articles, in that we will make direct use of 
the fundamental relation,
\begin{equation}
  \chi({\bf 1}, {\bf 2}) = L({\bf 1}, {\bf 1}^+, {\bf 2},{\bf 2}^+)
  = \Pi(1,1,2,2,t_1-t_2)
  \label{eq:local.1}
\end{equation}
where ${\bf i}^+$ is infinitessimally later than ${\bf i}$.  This approach
has been used by Totkatly, Stubner, and Pankaratov to develop a diagrammatic
expression for $f_{xc}(\omega)$ \cite{TP01,TSP02}.
It also leads to the ``Nanoquanta approximation,''
so named by Lucia Reining because it was simultaneiously derived by several
different people \cite{ORR02,RORO02,SOR03,STP04,BDLS05,MSR03} involved in the 
so-called Nanoquanta group.  (See also pp.~318-329 of Ref.~\cite{U12}.)

The work presented here differs from previous work
in two respects, namely (i) we make a direct connection with the PP formalism
which is more common in quantum chemistry than is the full BSE approach
(they are formally equivalent but differ in practice through the approximations
used) and (ii) we introduce a matrix formulation based upon Harriman's
contraction $\hat{\Upsilon}$ and expansion operators $\hat{\Upsilon}^\dagger$.  
This allows us to introduce the concept of the localiser $\Lambda(\omega)$
which shows explicitly how localization in space results requires the 
introduction of additional frequency dependence.  Finally we 
recover the formulae of G\"orling and Hirata {\em et al.} and produce
a rather trivial proof of the Gonze and Scheffler result \cite{GS99} that
this additional frequency dependence ``undoes'' the spatial localization 
procedure in particular cases.

We first seek a compact notation for Eq.~(\ref{eq:local.1}).  Harriman 
considered the relation between the space of kernels of operators and the
space of functions \cite{H83,H86}.  In order to main consistency with the
rest of this paper, we will generalize Harriman's notion from space-only to
space and spin coordinates.  Then the collapse operator is defined by,
\begin{equation}
  \hat{\Upsilon} A(1,2) = A(1,1) \, ,
  \label{eq:local.2}
\end{equation}
for an arbitrary operator kernel.  The adjoint of the collapse operator is 
the so-called expansion operator,
\begin{equation}
  \hat{\Upsilon}^\dagger f(1) = f(1) \delta(1-2) \, ,
  \label{eq:local.3}
\end{equation}
for an arbitrary function $f(1)$.  Clearly $\hat{\Upsilon}^\dagger \hat{\Upsilon} A(1,2) = A(1,1) \delta(1-2) \neq A(1,2)$.
The ability to express these operators as matrices (${\bm \Upsilon}$ and 
${\bm \Upsilon}^\dagger$) facilitates finite basis set applications.

We may now rewrite Eq.~(\ref{eq:local.1}) as,
\begin{equation}
  {\bm \chi}(t_1-t_2) = {\bm \Upsilon} {\bm L}(t_1,t_1^+,t_2,t_2^+) {\bm \Upsilon}^\dagger
  = {\bm \Upsilon} {\bm \Pi}(t_1-t_2) {\bm \Upsilon}^\dagger
  \label{eq:local.4}
\end{equation}
Comparing,
\begin{equation}
  {\bm \chi}(t_1-t_2) = {\bm \chi}_s(t_1-t_2) + \int {\bm \chi}_s(t_1-t_3)
  {\bm f}_{Hxc}(t_3-t_4) {\bm \chi}(t_4-t_2) \, dt_3 dt_4 \, ,
  \label{eq:local.5}
\end{equation}
with the BSE,
\begin{eqnarray}
  {\bm L}(t_1,t_2,t_3,t_4) & = & {\bm L}_s(t_1,t_2,t_3,t_4) \nonumber \\
  & + & \int {\bm L}_s(t_1,t_2,t_5,t_6) {\bm \Xi}_{Hxc}(t_5,t_6,t_7,t_8) 
    {\bm L}(t_7,t_8,t_3,t_4) \, dt_5 dt_6 dt_7 dt_8 \, , \nonumber \\
  \label{eq:local.6}
\end{eqnarray}
or, more precisely with,
\begin{eqnarray}
  {\bm \chi}(t_1-t_2) & = & {\bm \Upsilon} {\bm L}(t_1,t_1^+,t_2,t_2^+) {\bm \Upsilon}^\dagger \nonumber \\
  & = & {\bm \Upsilon} {\bm L}_s(t_1,t_1^+,t_2,t_2^+) {\bm \Upsilon}^\dagger
  \nonumber \\
  & + & \int {\bm \Upsilon} {\bm L}_s(t_1,t_1^+,t_5,t_6) {\bm \Xi}_{Hxc}(t_5,t_6,t_7,t_8) 
    {\bm L}(t_7,t_8,t_2,t_2^+) \, dt_5 dt_6 dt_7 dt_8  \nonumber \\
  & = & {\bm \chi}_s(t_1-t_2) 
  \nonumber \\
  & + & \int {\bm \Upsilon} {\bm L}_s(t_1,t_1^+,t_5,t_6) {\bm \Xi}_{Hxc}(t_5,t_6,t_7,t_8) 
    {\bm L}(t_7,t_8,t_2,t_2^+) \, dt_5 dt_6 dt_7 dt_8  
    \, , \nonumber \\
  \label{eq:local.7}
\end{eqnarray}
then shows that,
\begin{eqnarray}
  \int {\bm \Upsilon} {\bm L}(t_1,t_1^+,t_3,t_3^+) {\bm \Upsilon}^\dagger
  {\bm f}_{Hxc}(t_3-t_4) 
  {\bm \Upsilon} {\bm L}(t_4,t_4^+,t_2,t_2^+) {\bm \Upsilon}^\dagger
  \, dt_3 dt_4 & = & 
  \nonumber \\
    \int {\bm \Upsilon} {\bm L}_s(t_1,t_1^+,t_5,t_6) {\bm \Xi}_{Hxc}(t_5,t_6,t_7,t_8) 
    {\bm L}(t_7,t_8,t_2,t_2^+) \, dt_5 dt_6 dt_7 dt_8 \, .
   \nonumber \\
  \label{eq:local.8}
\end{eqnarray}
If we take advantage of the Kohn-Sham reference giving us the exact density,
then the Hartree part cancels out so that we actually get,
\begin{eqnarray}
  \int {\bm \Upsilon} {\bm L}(t_1,t_1^+,t_3,t_3^+) {\bm \Upsilon}^\dagger
  {\bm f}_{xc}(t_3-t_4) 
  {\bm \Upsilon} {\bm L}(t_4,t_4^+,t_2,t_2^+) {\bm \Upsilon}^\dagger
  \, dt_3 dt_4 & = & 
  \nonumber \\
    \int {\bm \Upsilon} {\bm L}_s(t_1,t_1^+,t_5,t_6) {\bm \Xi}_{xc}(t_5,t_6,t_7,t_8) 
    {\bm L}(t_7,t_8,t_2,t_2^+) \, dt_5 dt_6 dt_7 dt_8 \, .
   \nonumber \\
  \label{eq:local.9}
\end{eqnarray}

While this is certainly a beautiful result, it is nevertheless plagued with
four-time quantities which may be eliminated by using the PP.
\begin{equation}
  {\bm \Pi}(t_1-t_2) 
  = {\bm \Pi}_s(t_1-t_2) + \int {\bm \Pi}_s(t_1-t_3)
  {\bm K}_{Hxc}(t_3-t_4) {\bm \Pi}(t_4-t_2) \, dt_3 dt_4 \, ,
  \label{eq:local.10}
\end{equation}
where we have introduced the coupling matrix defined by,
\begin{equation}
  {\bm K}_{Hxc} = {\bm \Pi}_s^{-1} - {\bm \Pi}^{-1} \, .
  \label{eq:local.11}
\end{equation}
The price we have to pay is that the coupling matrix can not be easily
expanded in Feynman diagrams, but that in no way prevents us from determining
appropriate algebraic expressions for it.
We may then write,
\begin{eqnarray}
   & &  \int {\bm \Upsilon} {\bm \Pi}_s(t_1-t_3) {\bm \Upsilon}^\dagger
  {\bm f}_{xc}(t_3-t_4) {\bm \Upsilon} {\bm \chi}(t_4-t_2) {\bm \Upsilon}^\dagger \, dt_3 dt_4
 \nonumber \\
   & = & \int {\bm \Upsilon} {\bm \Pi}_s(t_1-t_3) {\bm \Upsilon}^\dagger 
   {\bm K}_{xc}(t_3-t_4) 
    {\bm \Upsilon} {\bm \Pi}(t_4-t_2) \, dt_3 dt_4 \, ,
  \label{eq:local.12}
\end{eqnarray}
which Fourier transforms to remove all the integrations,
\begin{equation}
    {\bm \Upsilon} {\bm \Pi}_s(\omega) {\bm \Upsilon}^\dagger
  {\bm f}_{xc}(\omega) {\bm \Upsilon} {\bm \chi}(\omega) 
   {\bm \Upsilon}^\dagger  
    =  {\bm \Upsilon} {\bm \Pi}_s(\omega) {\bm \Upsilon}^\dagger 
   {\bm K}_{xc}(\omega) 
    {\bm \Upsilon} {\bm \Pi}(\omega)  \, .
  \label{eq:local.13}
\end{equation}

\paragraph{Localizer}

Evidently,
\begin{equation}
   {\bm f}_{xc}(\omega) = {\bm \Lambda}_s(\omega) {\bm K}_{xc}(\omega)
                          {\bm \Lambda}(\omega) \, ,
   \label{eq:local.14}
\end{equation}
where we have introduced the notion of noninteracting (${\bm \Lambda}_s$) and
interacting (${\bm \Lambda}$) localizers,
\begin{eqnarray}
  {\bm \Lambda}_s(\omega) & = & \left( {\bm \Upsilon} {\bm \Pi}_s(\omega) {\bm \Upsilon}^\dagger \right)^{-1} {\bm \Upsilon} {\bm \Pi}_s(\omega) {\bm \Upsilon}^\dagger \nonumber \\
  {\bm \Lambda}(\omega) & = & \left( {\bm \Upsilon} {\bm \Pi}(\omega) {\bm \Upsilon}^\dagger \right)^{-1} {\bm \Upsilon} {\bm \Pi}(\omega) {\bm \Upsilon}^\dagger \, .
  \label{eq:local.15}
\end{eqnarray}
The localizer arises quite naturally in the context of the time-dependent
OEP problem.  According to the Runge-Gross theory \cite{RG84}, the exact
time-dependent xc-potential $v_{xc}(t)$, is not only a functional of the 
density $\rho(t)$, but also of an initial condition which can be taken as
the wavefunction $\Psi(t_0)$ at some prior time $t_0$.  On the other hand,
linear response theory begins with the static ground state case where the 
first Hohenberg-Kohn theorm tells us that the wavefunction is a functional
of the denisty $\Psi(t_0) = \Psi[\rho_{t_0}]$.  G\"orling has pointed out
that this greatly simplifies the problem \cite{G98} because we can then
show that,
\begin{equation}
  \int \Pi_s(1,1;2,2;\omega) v_x(2;\omega) \, d2 
  = \int \Pi_s(1,1;2,3;\omega) \Sigma_x(2,3) \, d2 d3 \, ,
  \label{eq:local.16}
\end{equation}
where $\Sigma_x$ is the Hartree-Fock exchange operator.  Equivalently this
may be written as,
\begin{equation}
  {\bm \Upsilon} {\bm \Pi}_s(\omega) {\bm \Upsilon}^\dagger {\bm v}_x
  = {\bm \Upsilon} {\bm \Pi}_s(\omega) {\bm \Sigma_x} \, ,
  \label{eq:local.17}
\end{equation}
or,
$\Sigma_x$,
\begin{equation}
  {\bm v}_x(\omega) = {\bm \Lambda}_s(\omega) {\bm \Sigma}_x \, .
  \label{eq:local.18}
\end{equation}
Equations~(\ref{eq:local.14}) and (\ref{eq:local.18}) are telling us 
something of fundamental importance, namely that the very act of spatially 
localizing the xc-coupling matrix involves introducing additional 
frequency dependence.  

For the special case of the non-interacting susceptibility, we can
easily derive an expression for the dynamic localizer.  Since,
\begin{eqnarray}
  \Pi_s(1,2;3,4;\omega) & = & \sum_i^{occ} \sum_a^{virt}
  \frac{\psi_i(1) \psi_a^*(2) \psi_i^*(3) \psi_a(4)} {\omega -
    \epsilon_{a,i} } \nonumber \\ & - & \sum_i^{occ}
  \sum_a^{virt} \frac{\psi_a(1) \psi_i^*(2) \psi_a^*(3) \psi_i(4)}
      {\omega + \epsilon_{a,i} } \, , \nonumber \\
  \label{eq:local.19}
\end{eqnarray}
we can express the kernel of ${\bm \Upsilon} {\bm \Pi}_s(\omega)$ as
\begin{eqnarray}
  \left( \Upsilon \Pi_s \right)(1;2,3;\omega) & = & \sum_i^{occ}
  \sum_a^{virt} \frac{\psi_i(1) \psi_a^*(1) \psi_i^*(2) \psi_a(3)}
      {\omega - \epsilon_{a,i}} \nonumber \\ & - &
      \sum_i^{occ} \sum_a^{virt} \frac{\psi_a(1) \psi_i^*(1)
        \psi_a^*(2) \psi_i(3)} {\omega + \epsilon_{a,i}} \,
      .  \nonumber \\
  \label{eq:local.20}
\end{eqnarray}
Also, the kernel of ${\bm \Upsilon} {\bm \Pi}_s(\omega) {\bm
  \Upsilon}^\dagger$ is just,
\begin{eqnarray}
   \left(\Upsilon \Pi_s \Upsilon^\dagger\right)(1;2;\omega) & = &
   \sum_i^{occ} \sum_a^{virt} \frac{\psi_i(1) \psi_a^*(1) \psi_i^*(2)
     \psi_a(2)} {\omega - \epsilon_{a,i} } \nonumber \\ & -
   & \sum_i^{occ} \sum_a^{virt} \frac{\psi_a(1) \psi_i^*(1)
     \psi_a^*(2) \psi_i(2)} {\omega + \epsilon_{a,i} } \, .
   \nonumber \\
  \label{eq:local.21}
\end{eqnarray}
Like the susceptibility, The two operators have poles at the
independent particle excitation energies $\omega = \pm \epsilon_{a,i} = \pm (\epsilon_a-\epsilon_i)$.

In order to construct the dynamic localizer, the kernel~(\ref{eq:local.17}) 
has to be inverted.  This is not generally possible to do this analytically, 
though it can be done in a finite-basis representation with great care.  
However Gonze and Scheffler have noted that exact inversion is possible in 
the special case of a frequency, $\omega=\epsilon_{b,j}$, of a pole well 
separated from the other poles.~\cite{GS99}  Near this pole, the kernels,
${\bm \Upsilon \Pi}_s(\omega)$ and 
${\bm \Upsilon \Pi}_s(\omega) {\bm \Upsilon}^\dagger$, are
each dominated by single terms,
\begin{eqnarray}
  \left( \Upsilon \Pi_s \right) & \approx & \frac{\psi_j(1) \psi_b^*(1) \psi_j^*(2) \psi_b(3)}
      {\omega - \epsilon_{b,j}} \nonumber \\
  \left(\Upsilon \Pi_s \Upsilon^\dagger\right)(1;2;\omega) & \approx & 
         \frac{\psi_j(1) \psi_b^*(1) \psi_j^*(2) \psi_b(2)} {\omega - \epsilon_{b,j} }
    \, .
  \label{eq:local.22}
\end{eqnarray}
Thus Eq.~(\ref{eq:local.17}) becomes,
\begin{equation}
    \frac{\psi_j(1) \psi_b^*(1)}{\omega - \epsilon_{b,j}} 
     \langle \psi_b \vert v_x(\epsilon_{b,j}) \vert \psi_j \rangle
     \approx  \frac{\psi_j(1) \psi_b^*(1)}{\omega - \epsilon_{b,j}} 
      \langle \psi_b \vert \hat{\Sigma}_x \vert \psi_j \rangle
      \, ,
     \label{eq:local.23}
\end{equation}
with the approximation becoming increasingly exact as $\omega$ approaches $\epsilon_{b,j}$.
Hence,
\begin{equation}
   \langle \psi_b \vert v_x(\epsilon_{b,j}) \vert \psi_j \rangle = 
   \langle \psi_b \vert \hat{\Sigma}_x \vert \psi_j \rangle \, .
   \label{eq:local.24}
\end{equation}

More generally for an arbitrary dynamic kernel, $K(1,2;\omega)$,
\begin{equation}
  (\psi_b\psi^*_j \vert \Lambda(\epsilon_{b,j}) K(\epsilon_{b,j} )) = (\psi_j\vert K (\epsilon_{b,j})
  \vert \psi_b) \, ,
  \label{eq:local.24.1}
\end{equation}
and we can do the same for $-\epsilon_{b,j}$, obtaining
\begin{equation}
  (\psi_j\psi^*_b \vert \Lambda(-\epsilon_{b,j}) K(-\epsilon_{b,j}) ) = (\psi_j\vert K(-\epsilon_{b,j})
  \vert \psi_b)
  \label{eq:local.24.2}
\end{equation}
We refer to these last two equations as Gonze-Scheffler (GS) relations, since
they were first derived by these authors~\cite{GS99} and because we will
want to use them again.
These GS relations show that the dynamic localizer, ${\bm \Lambda}_s(\omega)$, is pole free if
the excitation energies, $\epsilon_{a,i}$, are discrete and nondegenerate and suggests that
the dynamic localizer maybe a smoother function of $\omega$ than might at first be suspected.
Equation~(\ref{eq:local.24})
is also very significant because we see that, at a particular frequency,
the matrix element of a local operator is the same as the matrix element
of a nonlocal operator.  Generalization to the xc-kernel will require
an approximation.

\paragraph{First approximation}

Equation~(\ref{eq:local.14}) is difficult to solve because of the need
to invert an expression involving the correlated PP.  However it may
be removed by instead using the approximate expression,
\begin{equation}
   {\bm f}_{xc}(\omega) = {\bm \Lambda}_s(\omega) {\bm K}_{xc}(\omega)
                          {\bm \Lambda}_{1/2}(\omega) \, ,
   \label{eq:local.25}
\end{equation}
where a localizer is used which is half-way between the noninteracting and
fully interacting form,
\begin{equation}
  {\bm \Lambda}_{1/2}(\omega) = \left( {\bm \Upsilon} {\bm \Pi}_s(\omega) {\bm \Upsilon}^\dagger \right)^{-1} {\bm \Upsilon} {\bm \Pi}(\omega) {\bm \Upsilon}^\dagger \, .
  \label{eq:local.26}
\end{equation}
Equation~(\ref{eq:local.25}) then becomes,
\begin{equation}
  {\bm f}_{xc}(\omega) = \left( {\bm \Upsilon} {\bm \Pi}_s(\omega) {\bm \Upsilon}^\dagger \right)^{-1} \left( {\bm \Pi}(\omega) - {\bm \Pi}_s(\omega) \right)
  \left( {\bm \Upsilon} {\bm \Pi}_s(\omega) {\bm \Upsilon}^\dagger \right)^{-1}
  \, .
  \label{eq:local.27}
\end{equation}
Such an approximation is expected to work well in the off-resonant regime.
As we shall see, it does give G\"orling's exact exchange (EXX) kernel for 
TD-DFT \cite{G98}.  On the other hand, the poles of the kernel in this 
approximation are {\em a priori} the poles of the exact and independent
particle PPs --- that is, the true and single-particle excitation energies ---
unless well-balanced approximations lead to fortuitous cancellations.

We can now return to a particular aspect of Casida's original PP approach~\cite{C05}
which was failure to take proper account of the localizer.  
This problem is rectified here.  The importance of
the localizer is made particularly clear by the GS relations in the
case of charge transfer excitations.  The single-pole approximation to the
$i \rightarrow a$ excitation energy is,
\begin{eqnarray}
  \omega & = & \epsilon_{a,i} + (ia \vert \Lambda(\epsilon_{a,i}) K_{xc}(\epsilon_{a,i}) \Lambda^\dagger(\epsilon_{a,i}) 
  \vert ai ) 
  \nonumber \\
  & = & \epsilon_{a,i} + ( aa \vert \Pi_s^{-1}(\epsilon_{a,i}) -
  \Pi^{-1}(\epsilon_{ai}) \vert ii) \, .
  \label{eq:local.28}
\end{eqnarray}
Thus once again we see that the frequency dependence of the localizer has 
transformed the matrix element of a  spatially-local frequency-dependent 
operator into the matrix element of a spatially-nonlocal operator.
Had the localizer been neglected, then we would have found incorrectly
that,
\begin{equation}
  \omega = \epsilon_{a,i} + ( ia \vert \Pi_s^{-1}(\epsilon_{ai}) -
  \Pi^{-1}(\epsilon_{a,i}) \vert ai) \, .
  \label{eq:local.29}
\end{equation}
While the latter reduces to just $\epsilon_{ai}$ for charge transfer
excitations at a distance (because $\psi_i \psi_a = 0$), the former
does not.~\cite{HIG09} However, for most excitations the overlap is non-zero. 
In such cases and around a pole well-separated pole the localizer can be
completely neglected.

\paragraph{Exchange-only case}

In order to apply Eq.~(\ref{eq:local.27}), we need only the 
previously derived terms represented by the diagrams in Fig.~\ref{fig:SOPPA1}.
The resultant expressions agree perfectly with the expanded
expressions of the TD-EXX kernel obtained by Hirata {\em et
 al.}~\cite{HIBG05}, which are equivalent to the more condensed form
given by G\"orling~\cite{G98}.

Use of the GS relation then leads to,
\begin{eqnarray}  
  \omega & = & \epsilon_{a,i}^{KS} + f_{xc}(\epsilon_{a,i}^{KS})
  \nonumber \\ & = & \epsilon_{a,i}^{KS} + \langle a \vert \hat{M}_{xc}
  \vert a \rangle - \langle i \vert \hat{M}_{xc} \vert i \rangle +
  (ai\vert \vert ia) \nonumber \\ & = & \epsilon_{a,i}^{HF} + (ai\vert
  \vert ia) \, ,
  \label{eq:local.30}
\end{eqnarray}
which is exactly the configuration interaction singles (CIS, i.e.,
TDHF Tamm-Dancoff approximation) expression evaluated using Kohn-Sham
orbitals.  This agrees with a previous exact result obtained using
G\"orling-Levy perturbation theory~\cite{GS99,FUG97,G96}.

\paragraph{Second approximation}

A second approximation, equivalent to the PP Born approximation,
\begin{equation}
  {\bm \Pi}(\omega)  =  {\bm \Pi}_s(\omega) + {\bm \Pi}_s(\omega) {\bm K}_{Hxc}(\omega) {\bm \Pi}_s(\omega)
  \, .
  \label{eq:local.31}
\end{equation}
is useful because of its potential for preserving as much as possible
of the basic algebraic structure of the exact equation 
[Eq.~(\ref{eq:local.14})] while still remaining computationally tractable.
This is our second approximation,
\begin{equation}
  {\bm f}_{Hxc}(\omega) =
  {\bm \Lambda}_s(\omega) \left( {\bm \Pi}_s^{-1}(\omega) - {\bm
    \Pi}^{-1}(\omega) \right) {\bm \Lambda}_s^\dagger(\omega) \, .
  \label{eq:local.32}
\end{equation}
Equation~(\ref{eq:local.32}) simply reads that ${\bm f}_{Hxc}(\omega)$ 
is a spatially localized form of  ${\bm K}_{Hxc}(\omega)$.
This is nothing but the PP analogue of the
basic approximation (\ref{eq:local.9}) used in the BSE approach on
the way to the Nanoquanta approximation 
\cite{ORR02,RORO02,SOR03,STP04,BDLS05,MSR03}.

%
%
%
\section{Conclusion and Perspectives}
\label{sec:conclude}

Time-dependent DFT has become part of the photochemical modelers toolbox, at least in the
FC region.  However extensions of TD-DFT are being made to answer the photochemical
challenge of describing photochemical funnel regions where double and possibly higher
excitations often need to be taken into account.  This article has presented the dressed
TD-DFT approach of using MBPT corrections to LR-TD-DFT in order to help address problems
which are particularly hard for conventional TD-DFT.  Illustrations have been given
for the dissociation of H$_2$ and for {\em cis}/{\em trans} isomerisation of ethylene.
We have also included a section deriving the form of the TD-DFT xc-kernel from MBPT.
This derivation makes it clear that localization in space is compensated in the exact
kernel by including additional frequency dependences.  In the short run, it may be that
such additional frequency dependences will be easier to model with hybrid MBPT/LR-TD-DFT
approaches.  Let us mention in closing the very similar ``configuration interaction-corrected
Tamm-Dancoff approximation'' of Truhlar and coworkers \cite{LMXT14}.  Yet another approach,
similar in spirit, but different in details is multiconfiguration  TD-DFT based upon range
separation \cite{FKJ13}.  In the future, if progress continues to be made at the current
rate, we may very well be using some combination of these, including elements of dressed 
LR-TD-DFT, as well as other tricks such as a Maitra-Tempel form of the 
xc-kernel \cite{MT06},
constricted variational DFT for double excitations \cite{SKZ14},
DFT multi-reference configuration interaction (DFT-MRCI) \cite{BTK+09},
spin-flip theory \cite{MG09,HNI+10,RVA10,MG11,C12,HFG+13,M14,HKZ+14,NGT+14,GMV+14,ZH14}, and restricted open-shell or spin-restricted ensemble-referenced Kohn-Sham 
theory \cite{FD07,FDF08,HFG+13,NGT+14,GMV+14,F15} to attack difficult photochemical 
problems on a routine basis.  Key elements to make this happen will be the right balance
between rigor and practicality, ease of automation, and last but not least ease of use
if many users are going to try these techniques and if they can be routinely applied at
every time step of a photochemical dynamics simulation.


%
%
\begin{acknowledgement}
We thank Andrei Ipatov, Valerio Olevano, Giovanni Onida, Lucia Reining, 
Pina Romaniello, Angel Rubio, Davide Sangalli, Jochen Schirmer, Eric Shirley,
and Hemanadhan Myneni
for useful discussions.  M.\ H.\ R.\ would like to acknowledge an {\em Allocation 
de Recherche} from the French Ministry of Education.  Over the years, this work has 
been carried out in the context of several programs: the French Rh\^one-Alpes 
{\em R\'eseau th\'ematique de recherche avanc\'ee (RTRA): Nanosciences aux 
limites de la nano\'electronique}, the Rh\^one-Alpes Associated Node of the 
European Theoretical Spectroscopy Facility (ETSF), and most recently
the grant ANR-12-MONU-0014-02 from the French {\em Agence
Nationale de la Recherche} for the ORGAVOLT project (ORGAnic solar cell VOLTage
by numerical computation).
\end{acknowledgement}
\section*{Appendix: Order Analysis}
\addcontentsline{toc}{section}{Appendix}
\label{app:order}

We have presented the superoperator PP procedure as if we simply 
manipulated Feynman diagrams.  In reality we expanded the matrices 
using Wick's theorem with the help of a home-made {\sc Fortran} program.  
The result was a series of algebraic expressions which were subsequently 
analyzed by drawing the corresponding Feynman diagrams.  This leads to 
about 200 diagrams which we ultimately resum to give a more compact 
expression.  It is the generation of this expression that we now wish to 
discuss.

Let us analyze this expression for the PP according to the order of excitation operator. Following Casida,~\cite{C05} we partition the space as,
\begin{equation}
  -\Pi_{sr,qp}(\omega) = \left( ({\hat p}^\dagger {\hat q} \vert {\bf
    T}_1^{\dagger})\ ({\hat p}^\dagger {\hat q} \vert {\bf
    T}_{2+}^{\dagger}) \right) {\bm \Gamma}^{-1}(\omega)
  \left(\begin{array}{c} ( {\bf T}_{1}^{\dagger} \vert {\hat r}^\dagger {\hat
      s}) \\ ({\bf T}_{2+}^{\dagger} \vert {\hat r}^\dagger {\hat s})
  \end{array}\right) \, ,
  \label{eq:order.1} 
\end{equation}
where ${\bm T}^\dagger_{2+}$ corresponds to the operator space of two-electron and higher excitations and
\begin{equation}
  {\bm \Gamma}^{-1}(\omega) = \left[
    \begin{array}{ll}
      {\bm \Gamma}_{1,1}(\omega) & {\bm \Gamma}_{1,2+} \\
      {\bm \Gamma}_{2+,1} & {\bm \Gamma}_{2+,2+}(\omega) \\
    \end{array}\right]^{-1} \, ,
  \label{eq:order.2} 
\end{equation}
has been blocked,
\begin{equation}
  {\bm \Gamma}_{i,j}(\omega) = ({\bf T}_i^\dagger|\omega{\breve
    1}+{\breve H}|{\bf T}_j^\dagger) \, .
  \label{eq:order.3} 
\end{equation}
Using the well-known expression for the inverse of a two-by-two block 
matrix allows us to transform Eq.~(\ref{eq:order.1}) into,
\begin{eqnarray}
  & -& \Pi_{sr,qp}(\omega) = [ ( {\hat p}^\dagger {\hat q} \vert {\bm
      T}_1^\dagger ) - ( {\hat p}^\dagger {\hat q} \vert {\bm
      T}_{2+}^\dagger ) {\bm \Gamma}_{2+,2+}^{-1}(\omega) {\bm
      \Gamma}_{2+,1} ] \nonumber \\ & & {\bm P}^{-1}(\omega) [ ( {\bm
      T}_1^\dagger \vert {\hat r}^\dagger {\hat s} ) - {\bm
      \Gamma}_{1,2+} {\bm \Gamma}^{-1}_{2+,2+}(\omega) ({\bm
      T}_{2+}^\dagger \vert {\hat r}^\dagger {\hat s} ) ] \nonumber
  \\ & + & ( {\hat p}^\dagger {\hat q} \vert {\bm T}_{2+}^\dagger )
     {\bm \Gamma}_{2+,2+}^{-1}(\omega) ( {\bm T}_{2+}^\dagger \vert
     {\hat r}^\dagger {\hat s} ) \, ,
  \label{eq:order.4} 
\end{eqnarray}
where,
\begin{equation}
  {\bm P}(\omega)={\bm \Gamma}_{1,1}(\omega) - {\bm \Gamma}_{1,2+}
  {\bm \Gamma}_{2+,2+}^{-1}(\omega) {\bm \Gamma}_{2+,1} \, .
  \label{eq:order.5} 
\end{equation}
Although Eq.~(\ref{eq:order.4}) is somewhat complicated, it turns out 
that ${\bm P}(\omega)$ plays much the same role in the smaller 
${\bm T}_1^\dagger$ space that ${\bm \Gamma}(\omega)$ plays in the 
full ${\bm T}^\dagger$ space.
To see how this comes about, it is necessary to introduce the concept of order
in the fluctuation operator [Eq.~(\ref{eq:MBPT.30})] and in $M_{xc}$ 
[Eq.~(\ref{eq:MBPT.31})]. 
We can now perform an order-by-order expansion of Eq.~(\ref{eq:order.4}).  
Through second order only the ${\bm T}_2^\dagger$ part of 
${\bm T}_{2+}^\dagger$ contributes, so we need not consider higher than 
double excitation operators.  However we shall make some additional 
approximations.  In particular, we will follow the usual practice and 
drop the last term in Eq.~(\ref{eq:order.4}) because it contributes 
only at second order
and appears to be small when calculating excitation energies and 
transitions moments using the Hartree-Fock approximation as 
zero-order~\cite{SRM73,JOR75,S82,OS83,OJY84}. For response functions 
such as dynamic polarizabilities, their inclusion is more critical, 
improving the agreement with experiments \cite{NJO80}. 
We will also have no need to consider the second term in
\begin{equation}
  ( {\hat p}^\dagger {\hat q} \vert {\bf T}_1^\dagger ) - ( {\hat
    p}^\dagger {\hat q} \vert {\bf T}_{2+}^\dagger ) {\bm
    \Gamma}_{2+,2+}^{-1}(\omega) {\bm \Gamma}_{2+,1} \, .
  \label{eq:order.6} 
\end{equation}
This means that for the purposes of this paper we can treat the PP in the present work as given by,
\begin{equation}
  -\Pi_{sr,qp}(\omega) = ({\hat p}^\dagger {\hat q} \vert {\bm
    T}_1^\dagger){\bf P}^{-1}(\omega)({\bm T}_1^\dagger \vert {\hat
    r}^\dagger {\hat s}) \, .
  \label{eq:order.7} 
\end{equation}
Comparing with Eq.~(\ref{eq:MBPT.43}) substantiates our earlier claim 
that ${\bm P}(\omega)$ plays the same role in the ${\bm T}^\dagger_1$ 
space that ${\bm \Gamma}(\omega)$ plays over the full ${\bm T}^\dagger$ space.

\paragraph{First-Order Exchange-Correlation Kernel}

We now turn to the first-order exchange-correlation kernel.  Our main 
motivation here is to verify that we obtain the same terms as in exact 
exchange (EXX) calculations when we evaluate 
${\bm \Pi} - {\bm \Pi}_s$~\cite{G98,HIBG05}. Since our approach is in 
some ways more general than previous approaches to the EXX kernel, 
this subsection may also provide some new insight into the meaning of the EXX equations.

Since we are limited to first order, only zero- and first-order wavefunction 
terms need be considered.  This implies that all the contributions due to 
the ${\bf T}^\dagger_{2+}$ space (the space of double- and higher-excitations) 
are zero and substantiates our claim that Eq.~(\ref{eq:order.7}) is exact to 
first-order. An order-by-order expansion gives,
\begin{eqnarray}
  -\Pi^{(0+1)}_{sr,qp}(\omega)&=&({\hat p}^\dagger {\hat q} \vert {\bm
    T}_1^\dagger)^{(1)} {\bm P}^{(0),-1}(\omega) ({\bm T}_1^\dagger
  \vert {\hat r}^\dagger {\hat s})^{(0)} \nonumber +({\hat
    p}^\dagger{\hat q}\vert {\bm T}_1^\dagger)^{(0)}{\bm
    P}^{(0),-1}(\omega)({\bm T}_1^\dagger \vert {\hat r}^\dagger {\hat
    s})^{(1)} \nonumber 
   \\ &+& ({\hat p}^\dagger {\hat q} \vert {\bm
    T}_1^\dagger)^{(0)} {\bm P}^{(1),-1}(\omega) ({\bm T}_1^\dagger
  \vert {\hat r}^\dagger{\hat s})^{(0)} 
   -\Pi^s_{sr,qp}(\omega) \, , 
  \label{eq:order.8} 
\end{eqnarray}
where,
\begin{equation}
  -\Pi^s_{sr,qp}(\omega) = ( {\hat p}^\dagger {\hat q} \vert {\bm
    T}_1^\dagger )^{(0)} ( {\bm T}_1^\dagger \vert \omega{\breve 1} +
  {\breve h}_{KS} \vert {\bm T}^\dagger_1 )^{(0),-1} ({\bm
    T}_1^\dagger \vert {\hat r}^\dagger {\hat s} )^{(0)} \, .
  \label{eq:order.9} 
\end{equation}
The evaluation of each of first-order blocks is straightforward using the basic definitions and Wick's theorem.

Let us first consider the ${\bm P}$ parts.  The zeroth-order contribution is,
\begin{eqnarray}
  \label{eq:order.10} 
  P_{kc,ia}^{(0)}(\omega) &=&
  (\omega-\epsilon_{i,a})\delta_{ik}\delta_{ac} \\
  \label{eq:order.11} 
  P_{ck,ia}^{(0)}(\omega) &=& 0 \, ,
\end{eqnarray}
and the first-order contribution gives
\begin{eqnarray}
  \label{eq:order.12} 
  P_{kc,ia}^{(1)} &=& (ai\vert\vert kc) + M_{ac}\delta_{ik} -
  M_{ik}\delta_{ac} \\
  \label{eq:order.13} 
  P_{ck,ia}^{(1)} &=& (ci\vert\vert ak) \, .
\end{eqnarray}
(Note that $P_{kc,ia}$ is part of the ${\bm A}$ block, while $P_{ck,ia}$
is part of the ${\bm B}$ block.)
The sum of ${\bf P}^{(0)}+{\bf P}^{(1)}$ gives the exact pole structure up to first-order in the SOPPA approach.

The zero-order contribution,
\begin{equation}
  ({\hat p}^\dagger {\hat q} \vert {\bf T}_1^\dagger )^{(0)} = ({\bf
    T}^\dagger_1 \vert {\bf T}^\dagger_1 ) \, ,
  \label{eq:order.14} 
\end{equation}
and the first-order contributions are given by,
\begin{eqnarray}
  \label{eq:order.15} 
  [({\hat p}^\dagger {\hat q} \vert {\bf T}^\dagger_1 )]^{(1)}_{kc,ji}
  &=& -\frac{M_{jc}}{\epsilon_{j,c}}\delta_{ik} \\
  \label{eq:order.16} \label{eq:EOM.32}
  [({\hat p}^\dagger {\hat q} \vert {\bf T}^\dagger_1 )]^{(1)}_{ck,ji}
  &=& \ \ \frac{M_{ic}}{\epsilon_{i,c}}\delta_{kj} \\
  \label{eq:order.17} 
  [({\hat p}^\dagger {\hat q} \vert {\bf T}^\dagger_1 )]^{(1)}_{kc,ba}
  &=& \ \ \frac{M_{ka}}{\epsilon_{k,a}}\delta_{bc} \\
  \label{eq:order.18} 
  [({\hat p}^\dagger {\hat q} \vert {\bf T}^\dagger_1 )]^{(1)}_{ck,ba}
  &=& -\frac{M_{kb}}{\epsilon_{k,b}}\delta_{ca} \, .
\end{eqnarray}

The PP ${\bm \Pi}(\omega)$ is now easily constructed by simple matrix 
multiplication according to Eq.~(\ref{eq:order.8}).  Applying the 1st 
approximation from Sec.~\ref{sec:local} and expanding 
${\bm \Pi}_s(\omega)-{\bm \Pi}(\omega)$ through first order allows 
us to recover G\"orling's TD-EXX kernel.~\cite{G98} The most convenient 
way to do this is to expand ${\bm P}^{(1),-1}$ using,
\begin{eqnarray}
 && ({\bf T}_1^\dagger \vert \omega{\breve 1} + {\breve H} \vert {\bf
    T}_1^\dagger)^{-1} \approx ({\bf T}_1^\dagger \vert \omega{\breve 1}
  +{\breve H}^{(0)} \vert {\bf T}_1^\dagger)^{-1} \nonumber \\ &+&
  ({\bf T}_1^\dagger \vert \omega{\breve 1} + {\breve H}^{(0)} \vert
  {\bf T}_1^\dagger)^{-1} ({\bf T}_1^\dagger \vert {\breve H}^{(1)}
  \vert {\bf T}_1^\dagger) ({\bf T}_1^\dagger \vert \omega{\breve 1} +
        {\breve H}^{(0)} \vert {\bf T}_1^\dagger)^{-1} \, .
  \label{eq:order.19} 
\end{eqnarray}
The result is represented diagrammatically in Fig.~\ref{fig:SOPPA1}.  
The corresponding expressions agree perfectly with the expanded expressions 
of the TD-EXX kernel obtained by Hirata {\em et al},~\cite{HIBG05} which 
are equivalent to the more condensed form given by G\"orling.~\cite{G98} 
The diagrammatic treatment makes clear the connection with the BSE approach.  
There are in fact just three time-unordered diagrams shown in 
Fig.~\ref{fig:tdia} whose various time orderings generate the diagrams in 
Fig.~\ref{fig:SOPPA1}.  However the ``hanging parts'' above and below the 
horizontal dotted lines now have the physical interpretation of 
initial and final state wave function correlation. 
Had we applied the 2nd approximation of sec.~\ref{sec:local}, then only 
diagrams (a-f) of Fig.~\ref{fig:SOPPA1} would have survived.

Use of the Gonze-Scheffler relation (see further Sec.~\ref{sec:local})
then leads to,
\begin{eqnarray}  
  \omega & = & \epsilon_{a,i}^{KS} + f_{xc}(\epsilon_{a,i}^{KS})
  \nonumber \\ & = & \epsilon_{a,i}^{KS} + \langle a \vert \hat{M}_{xc}
  \vert a \rangle - \langle i \vert \hat{M}_{xc} \vert i \rangle +
  (ai\vert \vert ia) \nonumber \\ & = & \epsilon_{a,i}^{HF} + (ai\vert
  \vert ia) \, ,
  \label{eq:order.20} 
\end{eqnarray}
which is exactly the configuration interaction singles (CIS, i.e., TDHF 
Tamm-Dancoff approximation) expression evaluated using Kohn-Sham orbitals.  
This agrees with a previous exact result obtained using G\"orling-Levy 
perturbation theory.~\cite{GS99,FUG97,G96}

\paragraph{Second-Order Exchange-Correlation Kernel}

Having verified some known results, let us go on to do the MBPT necessary 
to obtain the pole structure of the xc-kernel through second order in the 
2nd approximation.  That is, we need to evaluate 
${\bm \Pi}_s^{-1}(\omega)-{\bm \Pi}^{-1}(\omega)$ through second order 
in such a way that its pole structure is evident.  The SOPPA/ADC strategy 
for this is to make a diagrammatic ${\bm \Pi}_s(\omega)-{\bm \Pi}(\omega)$ 
expansion of this quantity and then resum the expansion in an order-consistent 
way to have the form
\begin{equation}
  [{\bm \Pi}_s(\omega)-{\bm \Pi}(\omega)]^{(0+1+...+n)}_{rs,qp} =
  \nonumber \sum_{k=0}^n{\sum_{i=0}^{k} {\sum_{j=0}^{k-i} {
        ({\hat p}^\dagger {\hat q} \vert {\bf T}_1^\dagger)^{(i)} {\bf
          P}^{(j),-1}(\omega) ({\bf T}_1^\dagger \vert {\hat
          r}^\dagger {\hat s})^{(k-i-j)} }}} \, , 
  \label{eq:order.21} 
\end{equation} 
when the Born approximation is applied to the ${\bf P}(\omega)$, in the same 
fashion as in Sec.~\ref{sec:local}. The number of diagrams contributing to 
this expansion is large and, for the sake of simplicity, we will only give 
the resumed expressions for each block.  
Evidently, after the calculation of each block there will be an additional step matrix inversion in order to apply the 2nd approximation to the xc-kernel.

It should be emphasized that although the treatment below may seem simple, 
application of Wick's theorem is complicated and has been carried out using 
an in-house {\sc Fortran} program written specifically for the purpose.  
The result before resummation is roughly 200 diagrams which have been 
included as supplementary material.

It can be shown that the operator space may be truncated without loss of generality in a second-order treatment to only 1- and 2-electron excitation operators.~\cite{S82} The wavefunction may also be truncated at second-order.  This truncation breaks the orthonormality of the ${\bf T}^\dagger_1$ space:
\begin{equation}
  ({\bf T}_1^\dagger \vert {\bf T}_1^\dagger) \approx ({\bf
    T}^\dagger_1 \vert {\bf T}^\dagger_1)^{(0)} + ({\bf T}^\dagger_1
  \vert {\bf T}^\dagger_1)^{(2)} \ne \left(\begin{array}{cc} 1 & 0
    \\ 0 & -1 \\
  \end{array}
  \right) \, .
  \label{eq:order.22} 
\end{equation}
This complication is dealt with by orthonormalizing our operator space.  
The new operator set expressed in terms of the original set contains only 
second-order corrections,
\begin{eqnarray}
  [{\hat a}^\dagger{\hat i}]^{(2)} &=& \sum_b{ \left(\frac{1}{4}
    \sum_{kld} { \frac{(kd \vert\vert lb)(dk \vert\vert al )}{
        \epsilon_{kl,bd} \epsilon_{kl,da}}}
    + \sum_{k}{\frac{M_{kb}M_{ka}} {\epsilon_{k,b}\epsilon_{k,a}}}
    \right) {\hat b}^\dagger{\hat i}} \nonumber \\ &+&
  \sum_j{\left(\frac{1}{4}\sum_{mcd}{\frac{(md\vert\vert
        jc)(ci\vert\vert dm)}{\epsilon_{mj,cd}
        \epsilon_{im,cd}}}+ \sum_{d}{\frac{M_{jd}M_{di}}
      {\epsilon_{j,d}\epsilon_{i,d}}} \right){\hat a}^\dagger{\hat j}}
  \, . 
  \label{eq:order.23} 
\end{eqnarray}
(Note that we have used the linked-cluster theorem to eliminate contributions from disconnected diagrams. For a proof for the EOM of the one- and two-particle the Green's function see Ref.~\cite{K66})

We may now proceed to calculate,
\begin{eqnarray}
  -{\Pi}^{(2)}_{sr,qp}(\omega) &=& ({\hat p}^\dagger{\hat q}\vert{\bf
    T}_1^\dagger)^{(1)}{\bf P}^{(1),-1}(\omega)( {\bf
    T}^\dagger_1\vert{\hat r}^\dagger{\hat s})^{(0)} \nonumber \\ 
   &+& ({\hat p}^\dagger{\hat q}\vert{\bf T}_1^\dagger)^{(0)}{\bf
    P}^{(1),-1}(\omega)({\bf T}^\dagger_1\vert{\hat r}^\dagger{\hat s})^{(1)}
  \nonumber \\ 
   &+& ({\hat p}^\dagger{\hat q}\vert{\bf
    T}_1^\dagger)^{(1)}{\bf P}^{(0),-1}(\omega)({\bf
    T}^\dagger_1\vert{\hat r}^\dagger{\hat s})^{(1)} \nonumber \\ 
   &+&
  ({\hat p}^\dagger{\hat q}\vert{\bf T}_1^\dagger)^{(0)} {\bf
    P}^{(2),-1}(\omega) ({\bf T}^\dagger_1\vert{\hat r}^\dagger{\hat
    s})^{(0)} \, . 
  \label{eq:order.24} 
\end{eqnarray}
The only new contributions that arise at this level are due to the block ${\bf P}^{(2)}$, which is given by
\begin{equation}
  {\bf P}^{(2)} = {\bm \Gamma}_{1,1}^{(2)} - {\bm
    \Gamma}_{1,2}^{(1)} {\bm \Gamma}_{2,2}^{(0),-1}(\omega) {\bm
    \Gamma}_{2,1}^{(1)} \, .
  \label{eq:order.25} 
\end{equation}
(We are anticipating the $\omega$-dependence of the various 
$\bm \Gamma$-blocks which will be derived below.) Since the block 
${\bm \Gamma}_{1,1}^{(2)}$ is affected by the orthonormalization procedure, 
it may be useful to provide a few more details.  Expanding order-by-order,
\begin{eqnarray}
  {\bm \Gamma}_{1,1}^{(2)} &=& \langle 0^{(1)} \vert [{\bf
      T}_1^\dagger,[\omega {\breve 1} + {\breve H}^{(0)},{\bf
        T}^\dagger_1]] \vert 0^{(1)} \rangle \nonumber \\ &+& \langle
  0^{(0)} \vert[{\bf T}_1^\dagger,[ \omega {\breve 1} + {\breve
        H}^{(0)},{\bf T}_1^\dagger]] \vert 0^{(2)} \rangle \nonumber
  \\ &+& \langle 0^{(2)} \vert [{\bf T}_1^\dagger,[ \omega {\breve 1}
      + {\breve H}^{(0)},{\bf T}_1^\dagger ]] \vert 0^{(0)} \rangle
  \nonumber \\ &+& \langle 0^{(0)} \vert [{\bf
      T}^{\dagger(2)}_1,[\omega{\breve 1} + {\breve H}^{(0)},{\bf
        T}_1^\dagger]] \vert 0^{(0)} \rangle \, , \nonumber \\ &+&
  \langle 0^{(0)} \vert [{\bf T}_1^\dagger,[\omega{\breve 1} + {\breve
        H}^{(0)},{\bf T}^{\dagger(2)}_1]] \vert 0^{(0)} \rangle \, ,
  \nonumber \\ &+& \langle 0^{(1)} \vert [{\bf T}_1^\dagger,[{\hat
        H}^{(1)},{\bf T}^\dagger_1] \vert 0 ^{(0)} \rangle \nonumber
    \\ &+& \langle 0^{(0)} \vert [{\bf T}_1^\dagger,[{\hat
          H}^{(1)},{\bf T}^\dagger_1] \vert 0 ^{(1)} \rangle \, ,
  \label{eq:order.26} 
\end{eqnarray}
where ${\bf T}^{\dagger(2)}_1$ is the vector of second-order operators 
defined in Eq.~(\ref{eq:order.23}). It is easily shown that the first term 
cancels with the contributions coming from the second-order operators, and 
that the contributions from second-order wave function are exactly zero. 
Hence, that block is simply
\begin{eqnarray}
  {\bm \Gamma}_{1,1}^{(2)} &=& \langle 0^{(1)} \vert [{\bf
      T}_1^\dagger,[{\hat H}^{(1)},{\bf T}^\dagger_1] \vert 0 ^{(0)}
    \rangle \nonumber \\ &+& \langle 0^{(0)} \vert [{\bf
        T}_1^\dagger,[{\hat H}^{(1)},{\bf T}^\dagger_1] \vert 0 ^{(1)}
      \rangle \, , 
  \label{eq:order.27} 
\end{eqnarray}
which makes it frequency-independent. Its calculation gives
\begin{eqnarray}
  [\Gamma_{1,1}^{(2)}]_{kc,ia} &=&
  \delta_{ac}\sum_{d}{\frac{M_{kd}M_{di}}{\epsilon_{i,d}}}
  +\delta_{ik}\sum_{l}{\frac{M_{la}M_{lc}}{\epsilon_{l,a}}} \nonumber
  \\ 
  \\ &+&\frac{\delta_{ac}}{2}\sum_{lde}{\frac{(le\vert\vert
      kd)(dl\vert\vert ei)}{\epsilon_{im,de}}} 
   - \frac{\delta_{ik}}{2}\sum_{lmd}{\frac{(ld\vert\vert
      mc)(dl\vert\vert
      ma)}{\epsilon_{lm,ad}}} \label{eq:order.28} 
  \\ {[\Gamma_{1,1}^{(2)}]}_{ck,ia} &=&
  \frac{M_{ak}M_{id}}{\epsilon_{i,d}} +
  \frac{M_{ci}M_{ka}}{\epsilon_{k,a}} \nonumber \\ &+&
  2\sum_{d}{\frac{M_{dk}(ad\vert\vert ci)}{\epsilon_{k,d}}} \nonumber
  + 2\sum_{l}{\frac{M_{lc}(lk\vert\vert ai)}{\epsilon_{l,c}}}
  \nonumber \\ &-&\sum_{md}{\frac{(ce\vert\vert ad)(di\vert\vert
      em)}{\epsilon_{im,de}}} \nonumber
  -\sum_{me}{\frac{(ce\vert\vert mi)(ak \vert \vert
      me)}{\epsilon_{km,ae}}} \nonumber
  \\ &-&\frac{1}{2}\sum_{de}{\frac{(ce\vert \vert ad)(dk \vert \vert
      ei)}{\epsilon_{ik,de}}} \nonumber
  -\frac{1}{2}\sum_{ml}{\frac{(ik\vert \vert ml)(ac \vert \vert
      ml)}{\epsilon_{lm,ac}}} \, . 
  \label{eq:order.29} 
\end{eqnarray}

The block ${\bm \Gamma}_{1,2}$ and its adjoint is of at least first-order, due to the fact that the space is orthonormal. For that reason, it is not affected by the orthonormalization at this level of approximation. Its calculation gives
\begin{eqnarray}
  {[{\Gamma^{(2)}_{2,1}}]}_{kc,jbia} =& -&\delta_{ik}(bc \vert \vert
  aj) +\delta_{jk}(bc \vert \vert ai) \nonumber \\ & -& \delta_{bc}(ai
  \vert \vert kj) +\delta_{ac}(bi \vert \vert kj) \nonumber
  \\ {[{\Gamma^{(2)}_{2,1}}]}_{ck,jbia} =& 0& \, .
  \label{eq:order.30} 
\end{eqnarray}
Finally, the block ${\bm \Gamma}_{2,2}(\omega)$ gives
\begin{eqnarray}
  {[{\Gamma}_{2,2}^{(2)}(\omega)]_{ldkc,jbia}} &=& (\omega -
  \epsilon_{ij,ab})\delta_{jl}\delta_{ik}\delta_{ca}\delta_{db}
  \nonumber \\ {[{\Gamma}_{2,2}^{(2)}(\omega)]_{ckdl,jbia}} &=& 0
  \, \label{eq:order.31a} 
\end{eqnarray}
Notice that double excitations are treated only to zeroth-order in a 
second-order approach. To obtain a consistent theory with first-order 
corrections to double excitations, one should go at least to third order.  
This however becomes computationally quite heavy.

It is interesting to speculate what would happen if we were to include the 
first-order doubles correction within the present second-order theory.  
There are, in fact, indications that this can lead to improved agreement 
between calculated and experimental double excitations, though the quality 
of the single excitations is simultaneously decreased due to an imbalanced 
treatment~\cite{OJB78,TS95}.

We can now construct the PP necessary to construct the 2nd approximation 
of the xc-kernel Eq.~(\ref{eq:local.32}) according to 
Eq.~(\ref{eq:order.7}).  Since the the localizers of both left- and 
right-side are constructed from the non-interacting KS PP, we are only 
concerned with ph and hp contributions.  This means that the blocks 
involving pp or hh indices, corresponding to density shift operators, 
can be ignored at this level of approximation. This simplifies the 
construction of ${\bm P}(\omega)$ in Eq.~(\ref{eq:order.7}), which 
up to second-order gives
\begin{equation}
  {\bm \Pi}^{(0+1+2),-1}(\omega) = ({\bm T}_1^\dagger \vert {\bm
    T}_1^\dagger )^{-1} {\bm P}^{(0+1+2)}(\omega) ({\bm T}_1^\dagger \vert
  {\bm T}_1^\dagger )^{-1} \, .
  \label{eq:order.31} 
\end{equation}
Separating ph and hp contributions, the PP takes the form of a 
2 $\times$ 2 block-matrix in the same spirit as the LR-TD-DFT 
formulation of Casida,
\begin{eqnarray}
  &&{\bm \Pi}^{(0+1+2),-1}(\omega) = \nonumber
  \\ &&\left(\begin{array}{cc} {\bm 1} & {\bm 0} \\ {\bm 0} & -{\bm 1} \\
  \end{array}
  \right) \left(\begin{array}{cc} {\bm P}^{(0+1+2)}(\omega) &
    {\bm P}^{(0+1+2)}(\omega) \\ {\bm P}^{(0+1+2)}(\omega) &
    {\bm P}^{(0+1+2)}(\omega) \\
  \end{array}
  \right) \left(\begin{array}{cc} {\bm 1} & {\bm 0} \\ {\bm 0} & -{\bm 1} \\
  \end{array}
  \right) \nonumber \\
  & &  = \left(\begin{array}{rr}
    {\bm P}^{(0+1+2)}(\omega) & -{\bm P}^{(0+1+2)}(\omega)
    \\ -{\bm P}^{(0+1+2)}(\omega) & {\bm P}^{(0+1+2)}(\omega) \\
  \end{array}
  \right)\, .
  \label{eq:order.32} 
\end{eqnarray}
It follows that,
\begin{eqnarray}
  & & 
  {\bm \Pi}^{-1}_s(\omega) - {\bm \Pi}^{(0+1+2),-1}(\omega) =
  \nonumber \\
  & & \left(\begin{array}{rr} {\bm P}^{(1+2)}(\omega) &
    -{\bm \Gamma}^{(1+2)}_{1,1} \\ -{\bm \Gamma}^{(1+2)}_{1,1}
    & {\bm P}^{(1+2)}(\omega) \\
  \end{array}
  \right) \, .
  \label{eq:order.33} 
\end{eqnarray}
Note that the off-diagonal (ph,hp)- and (hp,ph)-blocks are frequency-independent and that the diagonal blocks are given by Eq.~(\ref{eq:order.25}).  
Ignoring localization for the moment, we may now cast the present 
Kohn-Sham based second-order polarization propagator approximation 
(SOPPA/KS) into the familiar form of Eq.~(\ref{eq:review.25}) with,
\begin{eqnarray}
  A_{ia,jb}(\omega) & = & \delta_{i,j} \delta_{a,b}\epsilon_{a,i} + P^{(1+2)}_{ia,jb}(\omega) \nonumber \\
  B_{ia,bj}(\omega) & = & -\left( \Gamma_{1,1}^{(1+2)} \right)_{ia,bj} \, .
  \label{eq:order.34} 
\end{eqnarray}

Localization [Eq.~(\ref{eq:local.32})] will complicate these formulae 
by mixing the ${\bm P}^{(1+2)}(\omega)$ and ${\bm \Gamma}^{(1+2)}_{1,1}$ terms,
\begin{eqnarray}
  A_{ia,jb}(\omega) & = & \delta_{i,j} \delta_{a,b} (\epsilon_a - \epsilon_i)  \nonumber \\ 
  & + &  \left[ ({\bm \Lambda}_s)_{hp,hp}(\omega) {\bm P}^{(1+2)}(\omega) ({\bm \Lambda}_s^\dagger)_{hp,hp}(\omega) \right]_{ia,jb}
  \nonumber \\
   & + & \left[ ({\bm \Lambda}_s)_{hp,ph}(\omega) {\bm P}^{(1+2)}(\omega) ({\bm \Lambda}_s^\dagger)_{ph,hp}(\omega) \right]_{ia,jb}
   \nonumber \\
   & - & \left[ ({\bm \Lambda}_s)_{hp,ph}(\omega) {\bm \Gamma }^{(1+2)} ({\bm \Lambda}_s^\dagger)_{hp,hp}(\omega) \right]_{ia,jb}
    \nonumber \\
   & - & \left[ ({\bm \Lambda}_s)_{hp,hp}(\omega) {\bm \Gamma}^{(1+2)} ({\bm \Lambda}_s^\dagger)_{ph,hp}(\omega) \right]_{ia,jb}
   \nonumber \\
  B_{ia,bj}(\omega) & = &  
  \left[ ({\bm \Lambda}_s)_{hp,hp} {\bm P}^{(1+2)}(\omega) ({\bm \Lambda}_s^\dagger)_{hp,ph} \right]_{ia,bj}
  \nonumber \\
   & + & \left[ ({\bm \Lambda}_s)_{hp,ph} {\bm P}^{(1+2)}(\omega) ({\bm \Lambda}_s^\dagger)_{ph,ph} \right]_{ia,bj}
   \nonumber \\
  & - & \left[ ({\bm \Lambda}_s)_{hp,ph}(\omega) {\bm \Gamma }^{(1+2)} ({\bm \Lambda}_s^\dagger)_{hp,ph}(\omega) \right]_{ia,bj}
  \nonumber \\
  & - & \left[ ({\bm \Lambda}_s)_{hp,hp}(\omega) {\bm \Gamma }^{(1+2)} ({\bm \Lambda}_s^\dagger)_{ph,ph}(\omega) \right]_{ia,bj}
         \, . \nonumber \\
  \label{eq:order.35} \label{eq:EOM.50b}
\end{eqnarray}
Of course this extra complication is unnecessary if all we want to do is to calculate improved excitation energies and transition amplitudes by doing DFT-based many-body perturbation theory.  It is only needed when our goal is to study the effect of localization on purely TDDFT quantities such as the xc-kernel and the TDDFT vectors ${\bm X}$ and ${\bm Y}$.

%


\bibliographystyle{spphys}
\bibliography{refs}

\begin{thebibliography}{100}
\providecommand{\url}[1]{{#1}}
\providecommand{\urlprefix}{URL }
\expandafter\ifx\csname urlstyle\endcsname\relax
  \providecommand{\doi}[1]{DOI \discretionary{}{}{}#1}\else
  \providecommand{\doi}{DOI \discretionary{}{}{}\begingroup
  \urlstyle{rm}\Url}\fi

\bibitem{R09}
J.S. Rowlinson, Bull. Hist. Chem. \textbf{34}, 1 (2009).
\newblock {\color{blue} \sf The border between physics and chemistry}

\bibitem{CJCS98}
M.E. Casida, C.~Jamorski, K.C. Casida, D.R. Salahub, J. Chem. Phys.
  \textbf{108}, 4439 (1998).
\newblock {\color{blue} \sf Molecular excitation energies to high-lying bound
  states from time-dependent density-functional response theory:
  Characterization and correction of the time-dependent local density
  approximation ionization threshold}

\bibitem{C02}
M.E. Casida, in \emph{Accurate Description of Low-Lying Molecular States and
  Potential Energy Surfaces}, ed. by M.R.H. Hoffmann, K.G. Dyall (ACS Press,
  Washington, D.C., 2002), p. 199.
\newblock {\color{blue} Jacob's ladder for time-dependent density-functional
  theory: {S}ome rungs on the way to photochemical heaven}

\bibitem{DM02}
N.L. Doltsinis, D.~Marx, J. Theo. Comput. Chem. \textbf{1}, 319 (2002).
\newblock {\color{blue} \sf First Principles Molecular Dynamics Involving
  Excited States and Nonadiabatic Transitions}

\bibitem{CJI+07}
F.~Cordova, L.J. Doriol, A.~Ipatov, M.E. Casida, C.~Filippi, A.~Vela, J. Chem.
  Phys. \textbf{127}, 164111 (2007).
\newblock {\color{blue} \sf Troubleshooting Time-Dependent Density-Functional
  Theory for Photochemical Applications: {O}xirane}

\bibitem{TTR+08}
E.~Tapavicza, I.~Tavernelli, U.~Rothlisberger, C.~Filippi, M.E. Casida, J.
  Chem. Phys. \textbf{129}(12), 124108 (2008).
\newblock {\color{blue} \sf Mixed time-dependent density-functional
  theory/classical trajectory surface hopping study of oxirane photochemistry}

\bibitem{CND11}
M.E. Casida, B.~Natarajan, T.~Deutsch, in \emph{Fundamentals of Time-Dependent
  Density-Functional Theory}, \emph{Lecture Notes in Physics}, vol. 837, ed. by
  M.~Marques, N.~Maitra, F.~Noguiera, E.K.U. Gross, A.~Rubio (Springer Verlag,
  2011), p. 279.
\newblock {\color{blue} \sf Non-{B}orn-{O}ppenheimer dynamics and conical
  intersections}

\bibitem{CH12}
M.E. Casida, M.~{Huix-Rotllant}, Annu. Rev. Phys. Chem. \textbf{63}, 287
  (2012).
\newblock {\color{blue} \sf Progress in time-dependent density-functional
  theory}

\bibitem{HK64}
P.~Hohenberg, W.~Kohn, Phys. Rev. \textbf{136}, B864 (1964).
\newblock {\color{blue}\sf Inhomogenous electron gas}

\bibitem{KS65}
W.~Kohn, L.J. Sham, Phys. Rev. \textbf{140}, A1133 (1965).
\newblock {\color{blue} \sf Self-consistent equations including exchange and
  correlation effects}

\bibitem{PY89}
R.G. Parr, W.~Yang, \emph{Density-Functional Theory of Atoms and Molecules}
  (Oxford University Press, New York, 1989)

\bibitem{DG90}
D.M. Dreizler, E.K.U. Gross, \emph{Density Functional Theory, An Approach to
  the Quantum Many-Body Problem} (Springer-Verlag, New York, 1990)

\bibitem{KH00}
W.~Koch, M.C. Holthausen, \emph{A Chemist's Guide to Density Funcational
  Theory} (Wiley-VCH, New York, 2000)

\bibitem{PS01}
J.P. Perdew, K.~Schmidt, in \emph{Density Functional Theory and its
  Applications to Materials}, ed. by V.E.V. Doren, K.V. Alseoy, P.~Geerlings
  (American Institute of Physics, Melville, New York, 2001), p.~1.
\newblock {\color{blue} \sf Jacob's ladder of density functional approximations
  for the exchange-correlation energy}

\bibitem{PRC+09}
J.P. Perdew, A.~Ruzsinsky, L.A. Constantin, J.~Sun, G.I. Csonka, J. Chem.
  Theor. Comput. \textbf{5}, 902 (2009).
\newblock {\color{blue} \sf Some fundamental issues in ground-state density
  functional theory: a guide for the perplexed}

\bibitem{PC07}
J.P. Perdew, L.A. Constantin, Phys. Rev. B \textbf{75}, 155109 (2007).
\newblock {\color{blue} \sf Laplacian-level density functionals for the kinetic
  energy density and exchange-correlation energy}

\bibitem{G01}
P.M. Gill, Aust. J. Chem. \textbf{54}, 661 (2001).
\newblock {\color{blue} \sf Obituary: {D}ensity-functional theory (1927-1993)}

\bibitem{B93}
A.~Becke, J. Chem. Phys. \textbf{98}, 1372 (1993).
\newblock {\color{blue} \sf A new mixing of Hartree–Fock and local
  density‐functional theories}

\bibitem{PEB96}
J.P. Perdew, M.~Ernzerhof, K.~Burke, J. Chem. Phys. \textbf{105}, 9982 (1996).
\newblock {\color{blue} \sf Rationale for mixing exact exchange with density
  functional approximations}

\bibitem{S95}
A.~Savin, in \emph{Recent Advances in Density Functional Theory}, ed. by D.P.
  Chong (World Scientific, Singapore, 1995), p. 129.
\newblock {\color{blue} \sf Beyond the {K}ohn-{S}ham determinant}

\bibitem{BLS10}
R.~Baer, E.~Livshits, U.~Salzner, Annu. Rev. Phys. Chem. \textbf{61}, 85
  (2010).
\newblock {\color{blue} \sf Tuned range-separated hybrids in density functional
  theory}

\bibitem{MUN+06}
M.A.L. Marques, C.~Ullrich, F.~Nogueira, A.~Rubio, E.K.U. Gross (eds.),
  \emph{Time-Dependent Density-Functional Theory}, \emph{Lecture Notes in
  Physics}, vol. 706 (Springer, Berlin, 2006)

\bibitem{MNN+11}
M.~Marques, N.~Maitra, F.~Noguiera, E.K.U. Gross, A.~Rubio, \emph{Fundamentals
  of Time-Dependent Density-Functional Theory}, \emph{Lecture Notes in
  Physics}, vol. 837 (Springer Verlag, 2011)

\bibitem{U12}
C.A. Ullrich, \emph{Time-Dependent Density-Functional Theory: Concepts and
  Applications} (Oxford University Press, 2012)

\bibitem{RG84}
E.~Runge, E.K.U. Gross, Phys. Rev. Lett. \textbf{52}, 997 (1984).
\newblock {\color{blue} \sf Density functional theory for time-dependent
  systems}

\bibitem{L99}
R.~van Leeuwen, Phys. Rev. Lett. \textbf{82}, 3863 (1999).
\newblock {\color{blue} \sf Mapping from Densities to Potentials in
  Time-Dependent Density-Functional Theory}

\bibitem{MTWB10}
N.T. Maitra, T.N. Todorov, C.~Woodward, K.~Burke, Phys. Rev. A \textbf{81},
  042525 (2010).
\newblock {\color{blue} \sf Density-potential mapping in time-dependent
  density-functional theory}

\bibitem{RL11}
M.~Ruggenthaler, R.~van Leeuwen, Europhys. Lett. \textbf{95}, 13001 (2011).
\newblock {\color{blue} \sf Global fixed-point proof of time-dependent
  density-functional theory}

\bibitem{RGPL12}
M.~Ruggenthaler, K.J.H. Glesbertz, M.~Penz, R.~van Leeuwen, Phys. Rev. A
  \textbf{85}, 052504 (2012).
\newblock {\color{blue} \sf Density-potential mappings in quantum dynamics}

\bibitem{RNL13}
M.~Ruggenthaler, S.E.B. Nlelsen, R.~van Leeuwen, Phys. Rev. A \textbf{88},
  022512 (2013).
\newblock {\color{blue} \sf Analytic density functionals with initial-state
  dependence}

\bibitem{V08}
G.~Vignale, Physical Review A \textbf{77}(6), 1 (2008).
\newblock \doi{10.1103/PhysRevA.77.062511}.
\newblock {\color{blue} \sf Real-time resolution of the causality paradox of
  time-dependent density-functional theory}

\bibitem{MDRS11}
J.~Messud, P.M. Dinh, P.~Reinhard, E.~Suraud, Ann. Phys. (Berlin) \textbf{523},
  270 (2011).
\newblock {\color{blue} \sf The generalized {SIC-OEP} formalism and the
  generalized {SIC}-Slater approximation (stationary and time-dependent cases)}

\bibitem{R96}
A.K. Rajagopal, Phys. Rev. A \textbf{54}, 3916 (1996).
\newblock {\color{blue} \sf Time-dependent variational principle and the
  effective action in density-functional theory and Berry’s phase}

\bibitem{L98}
R.~van Leeuwen, Phys. Rev. Lett. \textbf{80}, 1280 (1998).
\newblock {\color{blue} \sf Causality and symmetry in time-dependent
  density-functional theory}

\bibitem{L01}
R.~van Leeuwen, Int. J. Mod. Phys. \textbf{15}, 1969 (2001).
\newblock {\color{blue} \sf Key concepts in time-dependent density-functional
  theory}

\bibitem{M05}
S.~Mukamel, Phys. Rev. A \textbf{024503} (2005).
\newblock {\color{blue} \sf Generalized time-dependent
  density-functional-theory response functions for spontaneous density
  fluctuations and nonlinear response: {R}esolving the causality paradox}

\bibitem{M13}
M.A. Mosquera, Phys. Rev. B \textbf{88}, 022515 (2013).
\newblock {\color{blue} \sf Action formalism in time-dependent
  density-functional theory}

\bibitem{C95}
M.E. Casida, in \emph{Recent Advances in Density Functional Methods, Part I},
  ed. by D.P. Chong (World Scientific, Singapore, 1995), p. 155.
\newblock {\color{blue} Time-dependent density-functional response theory for
  molecules}

\bibitem{C96}
M.E. Casida, in \emph{Recent Developments and Applications of Modern Density
  Functional Theory}, ed. by J.~Seminario (Elsevier, Amsterdam, 1996), p. 391.
\newblock {\color{blue} \sf Time-Dependent Density Functional Response Theory
  of Molecular Systems: Theory, Computational Methods, and Functionals}

\bibitem{L64}
P.O. L\"owdin, J. Mol. Spectr. \textbf{14}, 112 (1964).
\newblock {\color{blue} Studies in Perturbation Theory. Part VI. Contraction of
  Secular Equations}

\bibitem{ORR02}
G.~Onida, L.~Reining, A.~Rubio, Rev. Mod. Phys. \textbf{74}, 601 (2002).
\newblock {\color{blue} \sf Electronic excitations: density-functional versus
  many-body {G}reen’s-function approaches}

\bibitem{RORO02}
L.~Reining, V.~Olevano, A.~Rubio, G.~Onida, Phys. Rev. Lett. \textbf{88},
  066404 (2002).
\newblock {\color{blue} \sf Excitonic Effects in Solids Described by
  Time-Dependent Density-Functional Theory}

\bibitem{SOR03}
F.~Sottile, V.~Olevano, L.~Reining, Phys. Rev. Lett. \textbf{91}, 056402
  (2003).
\newblock {\color{blue} \sf Parameter-Free Calculation of Response Functions in
  Time-Dependent Density-Functional Theory}

\bibitem{MSR03}
A.~Marini, R.D. Sole, A.~Rubio, Phys. Rev. Lett. \textbf{91}, 256402 (2003).
\newblock {\color{blue} \sf Bound Excitons in Time-Dependent Density-Functional
  Theory: {O}ptical and Energy-Loss Spectra}

\bibitem{STP04}
R.~Stubner, I.V. Tokatly, O.~Pankratov, Phys. Rev. B \textbf{70}, 245119
  (2004).
\newblock {\color{blue} \sf Excitonic effects in time-dependent
  density-functional theory: {A}n analytically solvable model}

\bibitem{BDLS05}
U.~von Barth, N.E. Dahlen, R.~van Leeuwen, G.~Stefanucci, Phys. Rev. B
  \textbf{72}, 235109 (2005).
\newblock {\color{blue} \sf Conserving approximations in time-dependent density
  functional theory}

\bibitem{RSB+09}
P.~Romaniello, D.~Sangalli, J.A. Berger, F.~Sottile, L.G. Molinari, L.~Reining,
  G.~Onida, J. Chem. Phys. \textbf{130}, 044108 (2009).
\newblock {\color{blue} \sf Double excitations in finite systems}

\bibitem{OJ77}
J.~Oddershede, P.~J{\o}rgensen, J. Chem. Phys. \textbf{66}, 1541 (1977).
\newblock {\color{blue} An order analysis of the particle-hole propagator}

\bibitem{NJO80}
E.S. Nielsen, P.~J{\o}rgensen, J.~Oddershede, J. Chem. Phys. \textbf{73}, 6238
  (1980).
\newblock {\color{blue} \sf Transition moments and dynamic polarizabilities in
  a second order polarization propagator approach}

\bibitem{NJO81}
E.S. Nielsen, P.~J{\o}rgensen, J.~Oddershede, J. Chem. Phys. \textbf{75}, 499
  (1981).
\newblock {\color{blue} \sf Erratum {J. Chem. Phys.}, {\bf 73}, 6238 (1980)}

\bibitem{JS81}
P.~J{\o}rgensen, J.~Simons, \emph{Second Quantization-Based Methods in Quantum
  Chemistry} (Academic Press, New York, 1981)

\bibitem{S82}
J.~Schirmer, Phys. Rev. A \textbf{26}, 2395 (1982).
\newblock {\color{blue} Beyond the random phase approximation: {A} new
  approximation scheme for the polarization propagator}

\bibitem{TSS99}
A.B. Trofimov, G.~Stelter, J.~Schirmer, J. Chem. Phys. \textbf{111}, 9982
  (1999).
\newblock {\color{blue} \sf A consistent third-order propagator method for
  electronic excitation}

\bibitem{FW71}
A.L. Fetter, J.D. Walecka, \emph{Quantum Theory of Many-Particle Systems}
  (McGraw-Hill, New York, 1971)

\bibitem{K66}
D.H. Kobe, Journal of Mathematical Physics \textbf{7}(10), 1806 (1966).
\newblock {\color{blue} Linked Cluster Theorem and the Green's Function
  Equations of Motion for a Many-Fermion System}

\bibitem{W84}
S.~Wilson, \emph{Electron correlation in Molecules} (Clarendon Press, 1984)

\bibitem{SRC+11}
D.~Sangalli, P.~Romaniello, G.~Col\`o, A.~Marini, G.~Onida, J. Chem. Phys.
  \textbf{134}, 034115 (2011).
\newblock {\color{blue} \sf {D}ouble excitation in correlated systems: {A}
  many-body approach}

\bibitem{C05}
M.E. Casida, J. Chem. Phys. \textbf{122}, 054111 (2005).
\newblock {\color{blue} \sf Propagator corrections to adiabatic time-dependent
  density-functional theory linear response theory}

\bibitem{HIBG05}
S.~Hirata, S.~Ivanov, R.J. Bartlett, I.~Grabowski, Phys. Rew. A \textbf{71},
  032507 (2005).
\newblock {\color{blue} \sf Exact-exchange time-dependent density-functional
  theory for static and dynamic polarizabilities}

\bibitem{G98}
A.~G\"orling, Int. J. Quant. Chem. \textbf{69}, 265 (1998).
\newblock {\color{blue} \sf Exact exchange kernel for time-dependent
  density-functional theory}

\bibitem{MZCB04}
N.T. Maitra, F.~Zhang, R.J. Cave, K.~Burke, J. Chem. Phys. \textbf{120}, 5932
  (2004).
\newblock {\color{blue} \sf Double excitations within time-dependent density
  functional theory linear response theory}

\bibitem{CZMB04}
R.J. Cave, F.~Zhang, N.T. Maitra, K.~Burke, Chem. Phys. Lett. \textbf{389}, 39
  (2004).
\newblock {\color{blue} \sf A dressed {TDDFT} treatment of the $^1A_g$ states
  of butadiene and hexatriene}

\bibitem{MW09}
G.~Mazur, R.~W{\l}odarczyk, J. Comput. Chem. \textbf{30}, 811 (2009).
\newblock {\color{blue} \sf Application of the dressed time-dependent density
  functional theory for the excited states of linear polyenes}

\bibitem{GB09}
O.V. Gritsenko, E.J. Baerends, Phys. Chem. Chem. Phys. \textbf{11}, 4640
  (2009).
\newblock {\color{blue} \sf Double excitation effect in non-adiabatic
  time-dependent density functional theory with an analytic construction of the
  exchange-correlation kernel in the common energy denominator approximation}

\bibitem{HIRC11}
M.~Huix-Rotllant, A.~Ipatov, A.~Rubio, M.E. Casida, Chem. Phys. \textbf{391},
  120 (2011).
\newblock {\color{blue} \sf Assessment of Dressed Time-Dependent
  Density-Functional Theory for the Low-Lying Valence States of 28 Organic
  Chromophores}

\bibitem{SSST08a}
M.~Schreiber, M.R. Silva-Junior, S.P.A. Sauer, W.~Thiel, J. Chem. Phys.
  \textbf{128}, 134110 (2008).
\newblock {\color{blue} \sf Benchmarks for electronically excited states:
  {CASPT2, CC2, CCSD, and CC3}}

\bibitem{HHH01}
C.P. Hsu, S.~Hirata, M.~Head-Gordon, J. Phys. Chem. A \textbf{105}, 451 (2001).
\newblock {\color{blue} \sf Excitation Energies from Time-Dependent Density
  Functional Theory for Linear Polyene Oligomers: {B}utadiene to
  Decapentaene}

\bibitem{MT06}
N.T. Maitra, D.G. Tempel, J. Chem. Phys. \textbf{125}, 184111 (2006).
\newblock {\color{blue} \sf Long-range excitations in time-dependent density
  functional theory}

\bibitem{H11}
M.~Huix-Rotllant, Improved correlation kernels for linear-response
  time-dependent density-functional theory.
\newblock Ph.D. thesis, Universit\'e de Grenoble (2011)

\bibitem{BSB05}
D.~Bokhan, I.G. Schweigert, R.J. Bartlett, Mol. Phys. \textbf{103}, 2299
  (2005).
\newblock {\color{blue} \sf Interconnection between functional derivative and
  effective operator approaches in {\em ab initio} density functional theory}

\bibitem{BB06}
D.~Bokhan, R.J. Bartlett, Phys. Rev. A \textbf{73}, 022502 (2006).
\newblock {\color{blue} \sf Adiabatic {\em ab initio} time-dependent
  density-functional theory emplying optimized-effective-potential many-body
  perturbation theory potentials}

\bibitem{TS76}
J.D. Talman, W.F. Shadwick, Phys. Rev. A \textbf{14}, 36 (1976).
\newblock {\color{blue} \sf Optimized effective atomic central potential}

\bibitem{T89}
J.D. Talman, Comp. Phys. Commun. \textbf{54}, 85 (1989).
\newblock {\color{blue} \sf {A} program to compute variationally optimized
  effective atomic potentials}

\bibitem{G99}
A.~G\"orling, Phys. Rev. Lett. \textbf{83}, 5459 (1999).
\newblock {\color{blue} \sf New {KS} Method for Molecules Based on an Exchange
  Charge Density Generating the Exact Local {KS} Exchange Potential}

\bibitem{IHB99}
S.~Ivanov, S.~Hirata, R.J. Bartlett, Phys. Rev. Lett. \textbf{83}, 5455 (1999).
\newblock {\color{blue} \sf Exact exchange treatment for molecules in
  finite-basis-set {K}ohn-{S}ham theory}

\bibitem{C95a}
M.E. Casida, Phys. Rev. A \textbf{51}, 2505 (1995).
\newblock {\color{blue} \sf Generalization of the Optimized Effective Potential
  Model to Include Electron Correlation: A Variational Derivation of the
  {S}ham--{S}chl\"uter Equation for the Exact Exchange-Correlation Potential}

\bibitem{C99a}
M.E. Casida, Phys. Rev. B \textbf{59}, 4694 (1999).
\newblock {\color{blue} \sf Correlated optimized effective potential treatment
  of the derivative discontinuity and of the highest occupied {K}ohn-{S}ham
  eigenvalue: {A} {J}anak-type theorem for the optimized effective potential
  method}

\bibitem{HIGB02}
S.~Hirata, S.~Ivanov, I.~Grabowski, R.J. Bartlett, J. Chem. Phys. \textbf{116},
  6468 (2002).
\newblock {\color{blue} \sf Time-dependent density functional theory employing
  optimized effective potentials}

\bibitem{BB07}
D.~Bokhan, R.J. Barlett, J. Chem. Phys. \textbf{127}, 174102 (2007).
\newblock {\color{blue} \sf Exact-exchange density functional theory for
  hyperpolarizabilities}

\bibitem{TP01}
I.V. Tokatly, O.~Pankratov, Phys. Rev. Lett. \textbf{86}, 2078 (2001).
\newblock {\color{blue} \sf Many-body diagrammatic expansion in a {K}ohn-{S}ham
  basis: {I}mplications for time-dependent density functional theory of excited
  states}

\bibitem{TSP02}
I.V. Tokatly, R.~Stubner, O.~Pankratov, Phys. Rev. B \textbf{65}, 113107
  (2002).
\newblock {\color{blue} \sf Many-body diagrammatic expansion of the
  exchange-correlation kernel in time-dependent density-functional theory}

\bibitem{GS99}
X.~Gonze, M.~Scheffler, Phys. Rev. Lett. \textbf{82}, 4416 (1999).
\newblock {\color{blue} \sf Exchange and correlation kernels at the resonance
  frequency: {I}mplications for excitation energies in density-functional
  theory}

\bibitem{H83}
J.E. Harriman, Phys. Rev. A \textbf{27}, 632 (1983).
\newblock {\color{blue} \sf Geometry of Density-Matrices. 4. {T}he relationship
  between density-matrices and densities}

\bibitem{H86}
J.E. Harriman, Phys. Rev. A \textbf{34}, 29 (1986).
\newblock {\color{blue} \sf {D}ensities, operators, and basis sets}

\bibitem{HIG09}
A.~He{\ss}elmann, A.~Ipatov, A.~G\"orling, Phys. Rev. A \textbf{80}, 012507
  (2009).
\newblock {\color{blue} \sf Charge-transfer excitation energies with a
  time-dependent density-functional method suitable for orbital-dependent
  exchange-correlation functionals}

\bibitem{FUG97}
C.~Filippi, C.J. Umrigar, X.~Gonze, The Journal of Chemical Physics
  \textbf{107}(23), 9994 (1997).
\newblock {\color{blue} \sf Excitation energies from density functional
  perturbation theory}

\bibitem{G96}
A.~G\"orling, Phys. Rev. A \textbf{54}(5), 3912 (1996).
\newblock {\color{blue} \sf Density-functional theory for excited states}

\bibitem{LMXT14}
S.L. Li, A.V. Marenich, X.~Xu, D.G. Truhlar, J. Chem. Phys. Lett. \textbf{5},
  322 (2014).
\newblock {\color{blue} \sf Configuration interaction-corrected
  {T}amm-{D}ancoff approximation: {A} time-dependent density functional method
  with the correct dimensionality of conical intersections}

\bibitem{FKJ13}
E.~Fromager, S.~Knecht, H.J.A. Jensen, J. Chem. Phys. \textbf{138}, 084101
  (2013).
\newblock {\color{blue} \sf Multi-configuration time-dependent
  density-functional theory based upon range separation}

\bibitem{SKZ14}
I.~Seidu, M.~Krykunov, T.~Ziegler, Mol. Phys. \textbf{112}, 661 (2014).
\newblock {\color{blue} \sf The formulation of a constricted variational
  density functional theory for double excitations}

\bibitem{BTK+09}
M.~B\"ohm, J.~Tatchen, D.~Kr\"ugler, K.~Kleinermanns, M.G.D. Nix, T.A.
  {LaGreve}, T.S. Zwier, M.~Schmitt, J. Phys. Chem. A \textbf{113}, 2456
  (2009).
\newblock {\color{blue} \sf High-Resolution and Dispersed Fluorescence
  Examination of Vibronic Bands of Tryptamine: {S}pectroscopic Signatures for
  $L_a/L_b$ Mixing near a Conical Intersection}

\bibitem{MG09}
N.~Minezawa, M.S. Gordon, J. Phys. Chem. A \textbf{113}, 12749 (2009).
\newblock {\color{blue} \sf Optimizing conical intersections by spin-flip
  density-functional theory: {A}pplication to ethylene}

\bibitem{HNI+10}
M.~{Huix-Rotllant}, B.~Natarajan, A.~Ipatov, C.M.W.T. Deutsch, M.E. Casida,
  Phys. Chem. Chem. Phys. \textbf{12}, 12811 (2010).
\newblock {\color{blue} \sf Assessment of Noncollinear Spin-Flip
  {T}amm-{D}ancoff Approximation Time-Dependent Density-Functional Theory for
  the Photochemical Ring-Opening of Oxirane}

\bibitem{RVA10}
Z.~Rinkevicius, O.~Vahtras, H.~{\AA}gren, J. Chem. Phys. \textbf{133}, 114104
  (2010).
\newblock {\color{blue} \sf Spin-flip time dependent density functional theory
  applied to excited states with single, double, or mixed electron excitation
  character}

\bibitem{MG11}
N.~Minezawa, M.S. Gordon, J. Phys. Chem. A \textbf{115}, 7901 (2011).
\newblock {\color{blue} \sf Photoisomerization of stilbene: {A} spin-flip
  density functional theory approach}

\bibitem{C12}
D.~Casanova, J. Chem. Phys. \textbf{137}, 084105 (2012).
\newblock {\color{blue} \sf Avoided crossings, conical intersections, and
  low-lying excited states with a single reference method: {T}he restricted
  active space spin-flip configuration interaction approach}

\bibitem{HFG+13}
M.~{Huix-Rotllant}, F.~Filatov, S.~Gozem, I.~Schapiro, M.~Olivucci, N.~Ferr\'e,
  J. Chem. Theory Comput. \textbf{9}, 3917 (2013).
\newblock {\color{blue} \sf Assessment of Density Functional Theory for
  Describing the Correlation Effects on the Ground and Excited State Potential
  Energy Surfaces of a Retinal Chromophore Model}

\bibitem{M14}
N.~Minezawa, J. Chem. Phys. \textbf{141}, 164118 (2014).
\newblock {\color{blue} \sf Optimizing minimum free-energy crossing points in
  solution: {L}inear-response free energy/spin-flip density functional theory
  approach}

\bibitem{HKZ+14}
Y.~Harabuchi, K.~Keipert, F.~Zahariev, T.~Taketsugu, M.S. Gordon, J. Phys.
  Chem. A \textbf{118}, 11987 (2014).
\newblock {\color{blue} \sf Dynamics Simulations with Spin-Flip Time-Dependent
  Density Functional Theory: {P}hotoisomerization and Photocyclization
  Mechanisms of cis-Stilbene in $(\pi,\pi^*)$ States}

\bibitem{NGT+14}
A.~Nikiforov, J.A. Gamez, W.~Thiel, M.~{Huix-Rotllant}, M.~Filatov, J. Chem.
  Phys. \textbf{141}, 124122 (2014).
\newblock {\color{blue} \sf Assessment of approximate computational methods for
  conical intersections and branching plane vectors in organic molecules}

\bibitem{GMV+14}
S.~Gozem, F.~Melaccio, A.~Valentini, M.~Filatov, M.~{Huix-Rotllant},
  N.~Ferr\'e, L.M. Frutos, C.~Angeli, A.I. Krylov, A.A. Granovsky, R.~Lindh,
  M.~Olivucci, J. Chem. Theory Comput. \textbf{10}, 3074 (2014).
\newblock {\color{blue} \sf Shape of Multireference, Equation-of-Motion
  Coupled-Cluster, and Density Functional Theory Potential Energy Surfaces at a
  Conical Intersection}

\bibitem{ZH14}
X.~Zhang, J.M. Herbert, J. Chem. Phys. \textbf{141}, 064104 (2014).
\newblock {\color{blue} \sf Analytic derivative couplings for spin-flip
  configuration interaction singles and spin-flip time-dependent density
  functional theory}i

\bibitem{FD07}
I.~Frank, K.~Damianos, J. Chem. Phys. \textbf{126}, 125105 (2007).
\newblock {\color{blue} \sf Restricted open-shell {K}ohn-{S}ham theory:
  {S}imulation}

\bibitem{FDF08}
J.~Friedrichs, K.~Darnianos, I.~Frank, J. Chem. Phys. \textbf{347}, 17 (2008).
\newblock {\color{blue} \sf Solving restricted open-shell equations in excited
  state molecular dynamics simulations}

\bibitem{F15}
M.~Filatov, Comput. Molec. Sci. \textbf{5}, 146 (2015).
\newblock {\color{blue} \sf Spin-restricted ensemble-referenced {K}ohn-{S}ham
  method: basic principles and application to strongly correlated ground and
  excited states of molecules}

\bibitem{SRM73}
T.~Shibuya, J.~Rose, V.~McKoy, J. Chem. Phys.  (1973).
\newblock {\color{blue} \sf Equations-of-motion method including
  renormalization and double-excitation mixing}

\bibitem{JOR75}
P.~J{\o}rgensen, J.~Oddershede, M.A. Ratner, Chem. Phys. Lett. \textbf{32}, 111
  (1975).
\newblock {\color{blue} \sf Two-Particle, Two-Hole Corrections to a
  Self-Consistent Time-Dependent Hartree-Fock Scheme}

\bibitem{OS83}
J.~Oddershede, J.R. Sabin, J. Chem. Phys.  (1983).
\newblock {\color{blue} \sf The use of modified virtual orbitals in
  perturbative polarization propagator calculations}

\bibitem{OJY84}
J.~Oddershede, P.~J{\o}rgensen, D.L. Yeager, Comp. Phys. Rep. \textbf{2}, 33
  (1984).
\newblock {\color{blue} \sf Polarization Propagator Methods in Atomic and
  Molecular Calculations}

\bibitem{OJB78}
J.~Oddershede, P.~J{\o}rgensen, N.H.F. Beebe, J. Phys. B: Atom. Molec. Phys.
  \textbf{11}, 1 (1978).
\newblock {\color{blue} \sf Analysis of excitation energies and transition
  moments}

\bibitem{TS95}
A.B. Trofimov, J.~Schirmer, J. Phys. B: At. Mol. Opt. Phys. \textbf{28}, 2299
  (1995).
\newblock {\color{blue} \sf An efficient polarization propagator approach to
  valence electron excitation spectra}

\end{thebibliography}
\end{document}